\documentclass[11pt,journal]{IEEEtran}

\usepackage{graphicx}
\graphicspath{{./fig/}}
\usepackage[caption=false,font=footnotesize,subrefformat=parens,labelformat=parens]{subfig}
\usepackage{amsmath,amssymb,bm}
\usepackage[ruled]{algorithm2e}
\usepackage{array}
\usepackage{mathtools} 
\usepackage{datetime}
\usepackage{tikz,pgfplots}

\newcommand{\calA}{\mathcal{A}}

\newcommand{\bbR}{\mathbb{R}}
\newcommand{\bbC}{\mathbb{C}}
\newcommand{\bbE}{\mathbb{E}}
\newcommand{\bbP}{\mathbb{P}}
\newcommand{\rmF}{\mathrm{F}}
\newcommand{\rms}{\mathrm{s}}
\newcommand{\rmn}{\mathrm{n}}
\newcommand{\calCN}{\mathcal{CN}}
\newcommand{\norm}[1]{\|{#1}\|}

\newcommand{\T}{\top}
\newcommand{\vect}{\mathrm{vec}}
\newcommand{\diag}{\mathrm{diag}}
\newcommand{\supp}{\mathrm{supp}}
\newcommand{\re}{\mathrm{Re}}
\newcommand{\im}{\mathrm{Im}}

\usepackage{umoline}

\newtheorem{theorem}{Theorem}[section]
\newtheorem{lemma}[theorem]{Lemma}

\newtheorem{remark}[theorem]{Remark}

\newtheorem{assumption}[theorem]{Assumption}

\begin{document}

\title{Blind Gain and Phase Calibration \\via Sparse Spectral Methods}

\author{Yanjun~Li,
        Kiryung~Lee,~\IEEEmembership{Member,~IEEE},
        and~Yoram~Bresler,~\IEEEmembership{Fellow,~IEEE}
\thanks{This work was supported in part by the National Science Foundation under Grant IIS 14-47879. This paper was presented in part at SampTA 2017 \cite{Li2017}.}
\thanks{Y. Li and Y. Bresler are with the Coordinated Science Laboratory and the Department
of Electrical and Computer Engineering, University of Illinois at
Urbana-Champaign, Urbana, IL 61801, USA (e-mail: yli145@illinois.edu, ybresler@illinois.edu). 
K. Lee is with the School of Electrical and Computer Engineering, Georgia Institute of Technology, Atlanta, GA 30332, USA (e-mail: kiryung@ece.gatech.edu).}}

\markboth{Last revised: \currenttime,~\today}%
{Last revised: \currenttime,~\today}

\maketitle

\begin{abstract}
Blind gain and phase calibration (BGPC) is a bilinear inverse problem involving the determination of unknown gains and phases of the sensing system, and the unknown signal, jointly. BGPC arises in numerous applications, e.g., blind albedo estimation in inverse rendering, synthetic aperture radar autofocus, and sensor array auto-calibration. In some cases, sparse structure in the unknown signal alleviates the ill-posedness of BGPC. Recently there has been renewed interest in solutions to BGPC with careful analysis of error bounds. In this paper, we formulate BGPC as an eigenvalue/eigenvector problem, and propose to solve it via power iteration, or in the sparsity or joint sparsity case, via truncated power iteration. Under certain assumptions, the unknown gains, phases, and the unknown signal can be recovered simultaneously. Numerical experiments show that power iteration algorithms work not only in the regime predicted by our main results, but also in regimes where theoretical analysis is limited. We also show that our power iteration algorithms for BGPC compare favorably with competing algorithms in adversarial conditions, e.g., with noisy measurement or with a bad initial estimate.
\end{abstract}


\IEEEpeerreviewmaketitle

\section{Introduction}
Blind gain and phase calibration (BGPC), the joint recovery of the unknown gains and phases in the sensing system and the unknown signal, is a bilinear inverse problem that arises in many applications: the joint estimation of albedo and lighting conditions in inverse rendering \cite{Nguyen2013}; the joint recovery of phase error and radar image in synthetic aperture radar (SAR) autofocus \cite{Morrison2009}; and auto-calibration of sensor gains and phases in array processing \cite{Paulraj1985}. 
There exists a long line of research regarding the solutions for each application. However, theoretical analysis of the problem and error bounds for its solutions have been established only recently \cite{Li2015,Li2015e,Ling2016,Wang2016}.

In this paper, we reformulate the BGPC problem as an eigenvalue/eigenvector problem. In the subspace case, we use algorithms that find principal eigenvectors such as the power iteration algorithm (also known as the power method) \cite[Section 8.2.1]{Golub1996}, to find  the concatenation of the gain and phase vector and the vectorized signal matrix in the form of the principal component of a structured matrix. In the sparsity case, the problem resembles sparse principal component analysis (sparse PCA) \cite{Moghaddam2006}. We then propose to solve the sparse eigenvector problem using truncated power iteration \cite{Yuan2013}.
The main contribution of this paper is the theoretical analysis of the error bounds of power iteration and truncated power iteration for BGPC in the subspace and joint sparsity cases, respectively. When the measurement matrix is random, and the signals and the noise are adversarial, our algorithms stably recover the unknown gains and phases, and the unknown signals with high probability under near optimal sample complexities. Since truncated power iteration relies on a good initial estimate, we also propose a simple initialization algorithm, and prove that the output is sufficiently good under certain technical conditions.

We complement the theoretical results with numerical experiments, which show that the algorithms can indeed solve BGPC in the optimal regime. We also demonstrate that the algorithms are robust against noise and an inaccurate initial estimate. Experiments with different initialization schemes show that our initialization algorithm significantly outperforms the baseline. Then we apply the power iteration algorithm to inverse rendering, and showcase its effectiveness in real-world applications.

The rest of the paper is organized as follows. In this section, we introduce the formulation of the BGPC problem, and related work. We then introduce the power iteration algorithms and our main theoretical results in Sections \ref{sec:algorithm} and \ref{sec:main}, respectively. Sections \ref{sec:fun_est} and \ref{sec:proof} give some fundamental estimates regarding the structured matrix in our BGPC formulation, and the proofs for our main results. We conduct some numerical experiments in Section \ref{sec:experiment}, and conclude the paper with some discussion in Section \ref{sec:conclusion}. 

\subsection{Notations} \label{sec:notation}

We use $A^\T$, $\overline{A}$, and $A^*$ to denote the transpose, the complex conjugate, and the conjugate transpose of a matrix $A$, respectively. The $k$-th entry of a vector $\lambda$ is denoted by $\lambda_k$. The $j$-th column, the $k$-th row (in a column vector form), and the $(k,j)$-th entry of a matrix $A$ are denoted by $a_{\cdot j}$, $a_{k \cdot}$, and $a_{k j}$, respectively. Upper script $t$ in a vector $\eta^{(t)}$ denotes the iteration number in an iterative algorithm.
We use $I_n$ to denote the identity matrix of size $n\times n$, and $\bm{1}_{n,m}$ and $\bm{0}_{n,m}$ to denote the matrices of all ones and all zeros of size $n\times m$, respectively. The $i$-th standard basis vector is denoted by $e_i$, whose ambient dimension is clear in the context.
The $\ell_p$ norm and $\ell_0$ ``norm'' of a vector $x$ are denoted by $\norm{x}_p$ and $\norm{x}_0$, respectively. The Frobenius norm and the spectral norm of a matrix $A$ are denoted by $\norm{A}_\rmF$ and $\norm{A}$, respectively. The support of a sparse vector $x$ is denoted by $\supp(x)$. The vector $\vect(X)$ denotes the concatenation of the columns of $X=[x_{\cdot 1},x_{\cdot 2},\dots, x_{\cdot N}]$, i.e., $\vect(X) = [x_{\cdot 1}^\T,x_{\cdot 2}^\T,\dots,x_{\cdot N}^\T]^\T$. A diagonal matrix with the entries of vector $x$ on the diagonal is denoted by $\diag(x)$. The Kronecker product is denoted by $\otimes$. We use $\gtrsim$ to denote the relation greater than up to log factors.
We use $[n]$ to denote the set $\{1,2,\dots,n\}$. 
For an index set $T$, the projection operator onto $T$ is denoted by $\Pi_T$, and the operator that restricts onto $T$ is denoted by $\Omega_T$. We use these operator notations for different spaces, and the ambient dimensions will be clarified in the context.


\subsection{Problem Formulation} \label{sec:formulation}
In this section, we introduce the BGPC problem with a subspace constraint or a sparsity constraint. Suppose $A\in\bbC^{n\times m}$ is the known measurement matrix, and $\lambda \in\bbC^{n}$ is the vector of unknown gains and phases, the $k$-th entry of which is $\lambda_k = |\lambda_k|e^{\sqrt{-1}\varphi_k}$. Here, $|\lambda_k|$ and $\varphi_k$ denote the gain and phase of the $k$-th sensor, respectively. The BGPC problem is the simultaneous recovery of $\lambda$ and the unknown signal matrix $X\in\bbC^{m\times N}$ from the following measurement:
\begin{equation}
\label{eq:bgpc}
Y = \diag(\lambda) A X + W,
\end{equation}
where $W\in\bbC^{n\times N}$ is the measurement noise. The $(k,j)$-th entry in the measurement $y_{kj}$ has the following expression:
\[
y_{kj} = \lambda_k \, a_{k \cdot}^\T \, x_{\cdot j} + w_{kj}.
\]

Clearly, BGPC is a bilinear inverse problem. The solution $(\lambda,X)$ suffers from scaling ambiguity, i.e., $(\lambda/\sigma,\sigma X)$ generates the same measurements as $(\lambda,X)$, and therefore cannot be distinguished from it.
Despite the fact that the solution can have other ambiguity issues, in this paper, we consider the generic setting where the solution suffers only from scaling ambiguity \cite{Li2015e}.\footnote{An example of another ambiguity is a shift ambiguity when $A$ is the discrete Fourier transform matrix \cite{Li2015,Wang2016}. For a generic matrix $A$, the solution to BGPC does not suffer from shift ambiguity.} Even in this setting, the solution is not unique, unless we exploit the structure of the signals. In this paper, we solve the BGPC problem under two scenarios -- BGPC with a subspace structure, and BGPC with sparsity.

1) \textbf{Subspace case:} Suppose that the known matrix $A$ is tall ($n>m$) and has full column rank. Then the columns of $AX$ reside in the low-dimensional subspace spanned by the columns of $A$. The problem is effectively unconstrained with respect to $X$.

2) \textbf{Sparsity case:} Suppose that $A$ is a known dictionary with $m \geq n$, while the columns of $X$ are $s_0$-sparse, i.e., $\norm{x_{\cdot j}}_0 \leq s_0$ for all $j\in [N]$. A variation of this setting is that the columns of $X$ are jointly $s_0$-sparse, i.e., there are at most $s_0$ nonzero rows in $X$. In this case, the subspace constraint on $AX$ no longer applies, and one must solve the problem with a sparsity (or joint sparsity) constraint.

The BGPC problem arises in applications including inverse rending, sensor array processing, multichannel blind deconvolution, and SAR autofocus. We refer the reader to our previous work \cite[Section II.C]{Li2015e} for a detailed account of applications of BGPC. For consistency, from now on, we use the convention in sensor array processing, and refer to $n$ and $N$ as the numbers of sensors and snapshots, respectively.

\begin{table}%
\renewcommand{\arraystretch}{1.2}
\caption{Comparison of Sample Complexities with Prior Work}
\label{tab:compare}
\begin{tabular}{|c||c|c|c|}
\hline
& Subspace & Joint Sparsity & Sparsity \\
\hline
\hline
Unique Recovery \cite{Li2015e} & \begin{tabular}[c]{@{}c@{}}$n > m$\\ $N \geq \frac{n-1}{n-m}$\end{tabular} & \begin{tabular}[c]{@{}c@{}}$n > 2s_0$\\ $N \geq \frac{n-1}{n-2s_0}$\end{tabular} & -- \\
\hline
Least Squares \cite{Ling2016} & \begin{tabular}[c]{@{}c@{}} $n\gtrsim m$ \\ $N \gtrsim 1$ \end{tabular} & -- & -- \\
\hline
$\ell_1$ Minimization \cite{Wang2016} & -- & -- & \begin{tabular}[c]{@{}c@{}} $n\gtrsim s_0$ \\ $N \gtrsim n$ \end{tabular} \\
\hline
This Paper & \begin{tabular}[c]{@{}c@{}} $n\gtrsim m$ \\ $N \gtrsim 1$ \end{tabular} & \begin{tabular}[c]{@{}c@{}} $n\gtrsim s_0$ \\ $N \gtrsim \sqrt{s_0}$ \end{tabular} & -- \\
\hline
\end{tabular} \vspace{0.1in}

{\scriptsize \textbf{Note:} $n$, $N$, $m$ and $s_0$ represent the number of sensors, the number of snapshots, the subspace dimension, and the sparsity level, respectively.}
\end{table}

\subsection{Our Contributions} \label{sec:contribution}

We reformulate BGPC as the problem of finding the principal eigenvector of a matrix (or operator). In the subspace case, this can be solved using any eigen-solver, e.g., power iteration (Algorithm \ref{alg:pi}). In the sparsity case, we propose to solve this problem using truncated power iteration (Algorithm \ref{alg:tpi}). Our main results can be summarized as follows:
\begin{theorem} \label{thm:summary}
Under certain assumptions on $A$, $\lambda$, $X$, and $W$, one can solve the BGPC problem with high probability using:

1) \textbf{Subspace case:} algorithms that find the principal eigenvector of a certain matrix, e.g., power iteration, if $n\gtrsim m$ and $N\gtrsim 1$.

2) \textbf{Joint sparsity case:} truncated power iteration with a good initialization, if $n\gtrsim s_0$ and $N\gtrsim \sqrt{s_0}$. 
\end{theorem}

In Table \ref{tab:compare}, we compare the above results with the sample complexities for unique recovery in BGPC \cite{Li2015e}, and previous guaranteed algorithms for BGPC in the subspace and sparsity case \cite{Ling2016,Wang2016}. In the subspace case, power iteration solves BGPC using optimal (up to log factors) numbers of sensors and snapshots. These sample complexities are comparable to the least squares method in \cite{Ling2016}. Moreover, we show that power iteration is empirically more robust against noise than least squares.

Truncated power iteration solves BGPC with a joint sparsity structure, with an optimal (up to log factors) number of sensors, and a slightly suboptimal (within a factor of $\sqrt{s_0}$ and log factors) number of snapshots. In comparison, the $\ell_1$ minimization method for the sparsity case of BGPC uses a similar number of sensors, but a much larger number of snapshots. Numerical experiments show that truncated power iteration empirically succeed, in both the joint sparsity case and the more general sparsity case, in the optimal regime.

The success of truncated power iteration relies on a good initial estimate of $X$ and $\lambda$. We propose a simple initialization algorithm (Algorithm \ref{alg:init}) with the following guarantee:
\begin{theorem} \label{thm:summary_init}
Under additional assumptions on the absolute values of the nonzero entries in $X$, our initialization algorithm produces a sufficiently good estimate of $\lambda$ and $X$ if $n\gtrsim s_0^2$. (We not require any additional assumption on the number $N$ of snapshots.)
\end{theorem}
Despite the above scaling law predicted by theory, numerical experiments suggest that our initialization scheme is effective when $n\gtrsim s_0$.

\subsection{Related Work} \label{sec:related}

BGPC arises in many real-world scenarios, and previous solutions have mostly been tailored to specific applications such as sensor array processing \cite{Paulraj1985,Rockah1988,Weiss1990}, sensor network calibration \cite{Balzano2007,Lipor2014}, synthetic aperture radar autofocus \cite{Morrison2009}, and computational relighting \cite{Nguyen2013}. 
A special case for BGPC is multichannel blind deconvolution (MBD) with a circular convolution model. Most previous works on MBD consider linear convolution with a finite impulse response (FIR) filter model (see \cite{Tong1991,Moulines1995,Harikumar1998,Harikumar1999}, and a recent stabilized method \cite{Lee2017a,Lee2016}). In comparison, the BGPC problem discussed in this paper involves a more general subspace or sparsity model.

The idea of solving BGPC by reformulating it into a linear inverse problem, which is a key idea in this paper, has been proposed by many prior works \cite{Balzano2007,Morrison2009,Nguyen2013}. In particular, Bilen et al. \cite{Bilen2014} provided a solution to BGPC with high-dimensional but sparse signals using $\ell_1$ minimization. However, such methods have not been carefully analyzed until recently. 
Ling and Strohmer \cite{Ling2016} derived an error bound for the least squares solution in the subspace case of BGPC. In this paper, the power iteration method has sample complexities comparable to those of the least squares method \cite{Ling2016}, and is empirically more robust to noise than the latter. 
Wang and Chi \cite{Wang2016} gave a theoretical guarantee for $\ell_1$ minimization that solves BGPC in the sparsity case, where they assumed that $A$ is the discrete Fourier transform (DFT) matrix and $X$ is random following a Bernoulli-Subgaussian model. 
In this paper, we give a guarantee for truncated power iteration under the assumption that $A$ is a complex Gaussian random matrix, and $X$ is \emph{jointly} sparse, well-conditioned, and deterministic. In this sense, we consider an adversarial scenario for the signal $X$. Our sample complexity results require a near optimal number $n$ of sensors, and a much smaller number $N$ of snapshots. Moreover, truncated power iteration is more robust against noise and inaccurate initial estimate of phases. 
Very recently, Eldar et al. \cite{Eldar2017} proposed new methods for BGPC with signals whose sparse components may lie off the grid. Similar to earlier work on blind calibration of sensor arrays \cite{Paulraj1985}, these methods rely on empirical covariance matrices of the measurements and therefore need a relatively large number of snapshots.

To position BGPC in a more broad context, it is a special bilinear inverse problem \cite{Li2015}, which in turn is a special case of low-rank matrix recovery from incomplete measurements \cite{Davenport2016,Li2016,Kech2016,Lee2013}. A resurgence of interest in bilinear inverse problems was pioneered by the recent studies in single-channel blind deconvolution of signals with subspace or sparsity structures, where both the signal and the filter are structured \cite{Ahmed2014,Ling2015,Chi2015,Lee2015a,XLi2016}. 

Another related bilinear inverse problem is blind calibration via repeated measurements from multiple sensing operators \cite{Bahmani2015a,Tang2014,Cambareri2016,Cambareri2017,Ahmed2016a,Cosse2017}. Since blind calibration with repeated measurements is in principle an easier problem than BGPC \cite{Ling2016}, we believe our methods for BGPC and our theoretical analysis can be extended to this scenario.

Also related is the phase retrieval problem \cite{Fienup1982}, where there only exists uncertainty in the phases (and not the gains) of the sensing system. An active line of work solves phase retrieval with guaranteed algorithms (see \cite{Candes2013a,Netrapalli2013,Candes2015,Cai2016,Bahmani2016,Goldstein2016} and \cite{Shechtman2015} for a recent review).

The error bounds of power iteration and truncated power iteration have been analyzed in general settings, e.g., in \cite[Section 8.2.1]{Golub1996} and \cite{Yuan2013}. These previous results hinge on spectral properties of matrices such as gaps between eigenvalues, which do not translate directly to sample complexity requirements. This paper undertakes analysis specific to BGPC. We relate spectral properties in BGPC to some technical conditions on $\lambda$, $A$, $X$, and $W$, and derive recovery error under near optimal sample complexities. We also adapt the analysis of sparse PCA \cite{Yuan2013} to accommodate a structured sparsity constraint in BGPC.

BGPC and our proposed methods are non-convex in nature. In particular, our truncated power iteration algorithm can be interpreted as projected gradient descent for a non-convex optimization problem. There have been rapid developments in guaranteed non-convex methods \cite{Sun2015} in a variety of domains such as matrix completion \cite{Keshavan2010,Jain2013,RSun2016}, dictionary learning \cite{Sun2017,Sun2017a}, blind deconvolution \cite{Lee2015a,XLi2016}, and phase retrieval \cite{Candes2015,Netrapalli2013,Sun2016}. It is a common theme that carefully crafted non-convex methods have better theoretical guarantees in terms of sample complexity than their convex counterparts, and often have faster implementations and better empirical performance. This paper is a new example of such superiority of non-convex methods.


\section{Power Iteration Algorithms for BGPC} \label{sec:algorithm}

Next, we describe the algorithms we use to solve BGPC. In Section \ref{sec:linear}, we introduce a simple trick that turns the bilinear inverse problem in BGPC to a linear inverse problem. In Sections \ref{sec:pi} and \ref{sec:tpi}, we introduce the power iteration algorithm we use to solve BGPC with a subspace structure, and the truncated (or sparse) power iteration algorithm we use to solve BGPC with sparsity, respectively.  

\subsection{From Bilinearity to Linearity}\label{sec:linear}
We use a simple trick to turn BGPC into a linear inverse problem \cite{Balzano2007}. Without loss of generality, assume that $\lambda_k\neq 0$ for $k\in[n]$. Indeed, if any sensor has zero gain, then the corresponding row in $Y$ is all zero or contains only noise, and we can simply remove the corresponding row in \eqref{eq:bgpc}. Let $\gamma$ denote the entrywise inverse of $\lambda$, i.e., $\gamma_k = 1/\lambda_k$ for $k\in[n]$. We have
\begin{equation}
\label{eq:bgpc_linear}
\diag(\gamma) Y_\rms = AX,
\end{equation}
where $Y_\rms = \diag(\lambda) AX$ is the noiseless measurement. Equation \eqref{eq:bgpc_linear} is linear in all the entries of $\gamma$ and $X$. The bilinear inverse problem in $(\lambda, X)$ now becomes a linear inverse problem in $(\gamma, X)$. In practice, since only the noisy measurement $Y$ is available, one can solve $\diag(\gamma) Y \approx AX$.

This technique was widely used to solve BGPC with a subspace structure, in applications such as sensor network calibration \cite{Balzano2007}, synthetic aperture radar autofocus \cite{Morrison2009}, and computational relighting \cite{Nguyen2013}. Recently, Ling and Strohmer \cite{Ling2016} analyzed the least squares solution to \eqref{eq:bgpc_linear}. Wang and Chi \cite{Wang2016} considered a special case where $A$ is the DFT matrix, and analyzed the solution of a sparse $X$ by minimizing the $\ell_1$ norm of $A^{-1}\diag(\gamma)Y$. 

We use the same trick in our algorithms. Define
\begin{equation}
\label{eq:D}
D \coloneqq
\begin{bmatrix}
I_N \otimes a_{1\cdot}^\T \\
\vdots \\
I_N \otimes a_{n\cdot}^\T
\end{bmatrix},
\end{equation}
\begin{equation}
\label{eq:E}
E \coloneqq
\begin{bmatrix}
y_{1\cdot} & & \\
& \ddots & \\
& & y_{n\cdot} 
\end{bmatrix}.
\end{equation}
We can decompose $E$ into $E = E_\rms + E_\rmn$, where
\[
E_\rms \coloneqq
\begin{bmatrix}
\lambda_1 X^\T a_{1\cdot} & & \\
& \ddots & \\
& & \lambda_n X^\T a_{n\cdot}
\end{bmatrix},
\]
\[
E_\rmn \coloneqq
\begin{bmatrix}
w_{1\cdot} & & \\
& \ddots & \\
& & w_{n\cdot}
\end{bmatrix}.
\]
Define also
\begin{equation}
\label{eq:B}
B \coloneqq
\begin{bmatrix}
D^*D & \alpha D^*E \\
\alpha E^*D & \alpha^2 E^*E
\end{bmatrix},
\end{equation}
\[
B_\rms \coloneqq
\begin{bmatrix}
D^*D & \alpha D^*E_\rms \\
\alpha E_\rms^*D & \alpha^2 E_\rms^*E_\rms
\end{bmatrix},
\]
where $\alpha$ is a nonzero constant specified later. 

Clearly, \eqref{eq:bgpc_linear} can be rewritten as
\[
D x - E_\rms \gamma = 0,
\]
where $x=\vect(X)$.
Equivalently, $\eta = [x^\T, -\gamma^\T/\alpha]^\T$ is a null vector of $B_\rms$. When certain sufficient conditions are satisfied, $\eta$ is the unique null vector of $B_\rms$. For example, if $\lambda$, $A$, and $X$ are in general positions in $\bbC^n$, $\bbC^{n\times m}$, and $\bbC^{m\times N}$, respectively, then $N \geq \frac{n-1}{n-m}$ snapshots are sufficient to guarantee uniqueness of the solution to BGPC in the subspace case. We refer readers to our work on the identifiability in BGPC for more details \cite{Li2015,Li2015e}. 

Since only the noisy matrix $B$ is accessible in practice, one can instead find the minor eigenvector, i.e., the eigenvector corresponding to the smallest eigenvalue of $B$. The rest of this section focuses on algorithms that find such an eigenvector of $B$, with no constraint (in the subspace case), or with a sparsity constraint (in the sparsity case).

\subsection{Power Iteration for BGPC with a Subspace Structure}\label{sec:pi}
In the subspace case ($n>m$), we solve for the minor eigenvector of the positive definite matrix $B$. In Section \ref{sec:main}, we derive an upper bound on the error between this eigenvector and the true solution $\eta$. 

The minor eigenvector of $B$ can be computed by a variety of methods. Here, we propose an algorithm that remains computationally efficient for  large scale problems.
By eigenvalue decomposition, the null vector of $B$ is identical to the principal eigenvector of
\begin{equation}
\label{eq:G}
G = \beta I_{mN+n}-B,
\end{equation}
for a large enough constant $\beta$. This eigenvector can be computed using the power iteration algorithm (see Algorithm \ref{alg:pi}). 

The size of $G$ is $(Nm+n)\times (Nm+n)$.
An advantage of Algorithm \ref{alg:pi} over an eigen-solver that decomposes $G$, is that one does not need to explicitly compute the entries of $G$ to iteratively apply it to a vector. Furthermore, rather than $O((Nm+n)^2)$, by the structure of $D$ and $E$, the per iteration time complexity of applying the operator $G$ to a vector is only $O(mnN)$. This can be further reduced if $A$ and $A^*$ are linear operators with implementations faster than $O(mn)$.

The rule of thumb for selecting parameter $\alpha$ is that the $\ell_2$ norms of the columns of $D$ be close to those of $\alpha E$ so that $G$ in \eqref{eq:G} exhibits good spectral properties for power iterations. A safe choice for $\beta$ is $\norm{B}$, which may be conservatively large in some cases, but works well in practice.
In Section \ref{sec:main}, we discuss our choice of parameters $\alpha,\beta$ under certain normalization assumptions (see Remark \ref{rem:parameter}). 

Algorithm \ref{alg:pi} converges to the principal eigenvector of $G$, as long as the initial estimate $\eta^{(0)}$ is not orthogonal to that eigenvector. This insensitivity to initialization is a privilege not shared by the sparsity case (see Section \ref{sec:tpi}).

\begin{algorithm}
\DontPrintSemicolon
\SetKwInput{KwPara}{Parameters}
\caption{Power Iteration for BGPC}
\label{alg:pi}
\KwIn{$A\in\bbC^{n\times m}$, $Y\in\bbC^{n\times N}$, initial estimate $\eta^{(0)}\in\bbC^{Nm+n}$}
\KwOut{$\eta^{(t)} \in \bbC^{Nm+n}$}
\KwPara{$\alpha$, $\beta$}
Compute operator $G: \bbC^{Nm+n} \rightarrow \bbC^{Nm+n}$ by \eqref{eq:D}, \eqref{eq:E}, \eqref{eq:B}, \eqref{eq:G} \;
$t \leftarrow 1$ \;
\Repeat{convergence criterion is met}{
Compute $\eta^{(t)} = G \eta^{(t-1)}/\norm{G \eta^{(t-1)}}_2$ \;
$t \leftarrow t+1$ \;
}
\end{algorithm}

\subsection{Truncated Power Iteration for BGPC with Sparsity}\label{sec:tpi}
When $2\leq n\leq m$, $[D,\alpha E]\in\bbC^{Nn\times (Nm+n)}$ is a fat matrix, and the null space of $B$ has dimension at least $2$. Therefore, there exist at least two linearly independent eigenvectors corresponding to the largest eigenvalue of $G$. To overcome the ill-posedness, one can leverage the sparsity structure in $X$ to make the solution to the eigenvector problem unique.

Let $\Pi_s(x)$ denote the projection of a vector $x$ onto the set of $s$-sparse vectors. It is computed by setting to zero all but the $s$ entries of $x$ of the largest absolute values. Let $\Pi'_s(X)$ denote the projection of a matrix $X$ onto the set of matrices whose columns are jointly $s$-sparse. This projection is computed by setting to zero all but the $s$ rows of $X$ of the largest $\ell_2$ norms.
We define two projection operators on $\eta = [x^\T, -\gamma^\T/\alpha]^\T$ that will be used repeatedly in the rest of this paper:
\begin{align*}
& \widetilde{\Pi}_s(\eta) \coloneqq [\Pi_s(x_{\cdot 1})^\T,\Pi_s(x_{\cdot 2})^\T,\dots,\Pi_s(x_{\cdot N})^\T,-\gamma^\T/\alpha]^\T ,\\
& \widetilde{\Pi}'_s(\eta) \coloneqq [\vect\bigl(\Pi'_s(X) \bigr)^\T,-\gamma^\T/\alpha]^\T .
\end{align*} 

For the sparsity case of BGPC, we adapt the eigenvector problem in Section \ref{sec:pi} by adding a sparsity constraint:
\begin{align}
\label{eq:sparse_pca}
\begin{split}
\max_{\eta} \quad & \eta^* G \eta\\
\mathrm{s.t.} \quad & \norm{\eta}_2 = 1,\\
& \widetilde{\Pi}_{s_0}(\eta) = \eta.
\end{split}
\end{align}
This nonconvex optimization is very similar to the sparse PCA problem. The only difference lies in the structure of the sparsity constraint. In sparse PCA, the principal component is $s_0$-sparse. In \eqref{eq:sparse_pca}, the vector $\eta$ consists of $s_0$-sparse vectors $x_{\cdot 1},x_{\cdot 2},\dots,x_{\cdot N}$, and a dense vector $-\gamma/\alpha$.

To solve \eqref{eq:sparse_pca}, we adopt a sparse PCA algorithm called truncated power iteration \cite{Yuan2013}, and revise it to adapt to the sparsity structure of BGPC (see Algorithm \ref{alg:tpi}). One can choose parameters $\alpha$ and $\beta$ using the same rules as in Section \ref{sec:pi}. Note that we use a sparsity level $s_1\geq s_0$ in this algorithm, for two reasons: (a) in practice, it is easier to obtain an upper bound on the sparsity level instead of the exact number of nonzero entries in the signal; and (b) the ratio $s_0/s_1$ is an important constant in the main results, controlling the trade-off between the number of measurements and the rate of convergence. 

For the joint sparsity case, we use essentially the same algorithm, with $\widetilde{\Pi}_{s_1}$ replaced by $\widetilde{\Pi}'_{s_1}$. 

Since \eqref{eq:sparse_pca} is a nonconvex optimization problem, a good initialization $\eta^{(0)}$ is crucial to the success of Algorithm \ref{alg:tpi}. Algorithm \ref{alg:init} outlines one such initialization. We denote by $\Pi_{T_x}$ the projection onto the support set $T_x$, which sets to zero all rows of $D^*E$ but the $s_1$ rows of the largest $\ell_2$ norms in each block. (The $j$-th block of $D^*E$ consists of $m$ contiguous rows indexed by $\{(j-1)m+\ell\}_{\ell \in [m]}$.) Then the normalized left and right singular vectors $u$ and $v$ of $\Pi_{T_x} D^*E$ are computed as initial estimates for $x$ and $\lambda$. We use $1./v$ to denote the entrywise inverse of $v$ except for zero entries, which are kept zero. In Section \ref{sec:main}, we further comment on how to choose a proper initial estimate $\eta^{(0)}$ (see Remark \ref{rem:initialize}).

\begin{algorithm}
\DontPrintSemicolon
\SetKwInput{KwPara}{Parameters}
\caption{Truncated Power Iteration for BGPC with Sparsity}
\label{alg:tpi}
\KwIn{$A\in\bbC^{n\times m}$, $Y\in\bbC^{n\times N}$, initial estimate $\eta^{(0)}\in\bbC^{Nm+n}$}
\KwOut{$\eta^{(t)} \in\bbC^{Nm+n}$}
\KwPara{$\alpha$, $\beta$, $s_1$}
Compute operator $G: \bbC^{Nm+n}\rightarrow \bbC^{Nm+n}$ by \eqref{eq:D}, \eqref{eq:E}, \eqref{eq:B}, \eqref{eq:G} \;
$t \leftarrow 1$ \;
\Repeat{convergence criterion is met}{
Compute $\tilde{\eta}^{(t)} = G \eta^{(t-1)}/\norm{G \eta^{(t-1)}}_2$ \;
Compute $\eta^{(t)} = \widetilde{\Pi}_{s_1}(\tilde{\eta}^{(t)})/\norm{\widetilde{\Pi}_{s_1}(\tilde{\eta}^{(t)})}_2$ \;
$t \leftarrow t+1$ \;
}
\end{algorithm}

\begin{algorithm}
\DontPrintSemicolon
\SetKwInput{KwPara}{Parameters}
\caption{Initialization for Truncated Power Iteration}
\label{alg:init}
\KwIn{$A\in\bbC^{n\times m}$, $Y\in\bbC^{n\times N}$}
\KwOut{initial estimate $\eta^{(0)}\in\bbC^{Nm+n}$}
\KwPara{$s_1$}
Compute matrix $D^*E\in\bbC^{Nm \times n}$ by \eqref{eq:D}, \eqref{eq:E}\;
$T_x \leftarrow \emptyset$ \; 
\For{$j\in [N]$}{
Compute the row norms $\norm{d_{\cdot((j-1)m+\ell)}^*E}_2$ for $\ell\in [m]$\;
Find subset $T_j\subset [m]$ ($|T_j|=s_1$) s.t. for $\ell \in T_j$ and $\ell' \in [m]\backslash T_j$:
\[
\norm{d_{\cdot((j-1)m+\ell)}^*E}_2 \geq \norm{d_{\cdot((j-1)m+\ell')}^*E}_2  
\] \vspace{-0.1in}
\;
Merge support $T_x \leftarrow T_x\bigcup \left(T_j + \{(j-1)m\} \right)$ \;
}
Compute the principal left and right singular vectors $u$, $v$ of $\Pi_{T_x} D^*E$\;
$\eta^{(0)} \leftarrow [u^\T, -(1./v^\T)/n]^\T$\;
$\eta^{(0)} \leftarrow \eta^{(0)} / \norm{\eta^{(0)}}_2$\;
\end{algorithm}

\subsection{Alternative Interpretation as Projected Gradient Descent}
Algorithms \ref{alg:pi} and \ref{alg:tpi} can be interpreted as gradient descent and projected gradient descent, respectively. Next, we explain such equivalence using the sparsity case as an example.

Recall that BGPC is linearized as $\begin{bmatrix}D & \alpha E\end{bmatrix}\eta = 0$. Relaxing the sparsity level from $s_0$ to $s_1$, the optimization in \eqref{eq:sparse_pca} is equivalent to:
\begin{align*}
\min_{\eta} \quad & \frac{1}{2}\left\| \begin{bmatrix}D & \alpha E\end{bmatrix} \eta \right\|_2^2\\
\mathrm{s.t.} \quad & \norm{\eta}_2 = 1,\\
& \widetilde{\Pi}_{s_1}(\eta) = \eta.
\end{align*}
The gradient of the objective function at $\eta^{(t-1)}$ is
\[
\begin{bmatrix}
D^*\\
\alpha E^*
\end{bmatrix}
\begin{bmatrix}
D & \alpha E
\end{bmatrix} \eta^{(t-1)} = B \eta^{(t-1)}.
\]
Each iteration of projected gradient descent consists of two steps: 

\noindent(i) \textbf{Gradient descent} with a step size of $1/\beta$:
\begin{align*}
\tilde{\eta}^{(t)} = \eta^{(t-1)} - \frac{1}{\beta} B\, \eta^{(t-1)} = \frac{1}{\beta} G \, \eta^{(t-1)}.
\end{align*}

\noindent(ii) \textbf{Projection} onto the constraint set, i.e., the intersection of a cone ($\widetilde{\Pi}_{s_1}(\eta) = \eta$) and a sphere ($\norm{\eta}_2 = 1$):
\[
\eta^{(t)} = \widetilde{\Pi}_{s_1}(\tilde{\eta}^{(t)})/\norm{\widetilde{\Pi}_{s_1}(\tilde{\eta}^{(t)})}_2.
\]

Clearly, the two steps are identical to those in each truncated power iteration except for a different scaling in Step (i), which, due to the normalization in Step (ii), is insignificant.

\section{Main Results} \label{sec:main}
In this section, we give theoretical guarantees for Algorithms \ref{alg:pi} and \ref{alg:tpi} in the subspace case and in the joint sparsity case, respectively. We also give a guarantee for the initialization by Algorithm \ref{alg:init}.

\subsection{Main Assumptions} \label{sec:assumption}
We start by stating the assumptions on $A$, $\lambda$, $X$ and $W$, which we use throughout this section. 

\begin{assumption} \label{ass:A}
$A$ is a complex Gaussian random matrix, whose entries are i.i.d. following $\calCN(0, \frac{1}{n})$. Equivalently, the vectors $\{a_{k\cdot}\}_{k=1}^{n}$ are i.i.d. following $\calCN(\bm{0}_{m,1}, \frac{1}{n}I_m)$.
\end{assumption}

\begin{assumption} \label{ass:lambda}
The vector $\lambda$ has ``flat'' gains in the sense that $1-\delta \leq |\lambda_k|^2 \leq 1+\delta$ for some $\delta \in (0,1)$.
\end{assumption}

\begin{assumption} \label{ass:X}
The matrix $X\in\bbC^{m\times N}$ is normalized and has good conditioning, i.e., $\norm{X}_\rmF = 1$, and for some $\theta \in (0,1)$ we have:
\begin{itemize}
	\item \textbf{Subspace case:}
	\[
	\min\{\norm{NX^*X-I_N}, \norm{mXX^*-I_m}\} \leq \theta,
	\]
	\item \textbf{Joint sparsity case:}
	\[
	\min\{\norm{NX^*X-I_N}, \norm{s_0\Omega_{T_0}XX^*\Omega_{T_0}^*-I_{s_0}}\} \leq \theta,
	\]
\end{itemize}
where $\Omega_T$ denotes the operator that restricts a matrix to the row support $T$, and $T_0 \coloneqq \{i\in[m] | \norm{e_i^\T X}_2 > 0 \}$ ($|T_0|=s_0$) is the row support of $X$. 
\end{assumption}

Assumptions \ref{ass:A} -- \ref{ass:X} can be relaxed in practice.
\begin{itemize}
	\item The complex Gaussian distribution in Assumption \ref{ass:A} can be relaxed to $\calCN(0,\sigma_A^2)$ for any $\sigma_A>0$. We choose the particular scaling $\sigma_A^2 = 1/n$, because then $A$ satisfies the restricted isometry property (RIP) \cite{Candes2005}, i.e., $(1-\delta_s)\norm{x}_2^2\leq \norm{Ax}_2^2 \leq (1+\delta_s)\norm{x}_2^2$ for some $\delta_s\in(0,1)$, when $n$ is large compared to the number $s$ of nonzero entries in $x$. 
	\item The gains can center around any $\sigma >0$, i.e., $\sigma(1-\delta) \leq |\lambda_k|^2 \leq \sigma(1+\delta)$. Due to bilinearity, we may assume that $\lambda_k$'s center around $1$ without loss of generality by solving for $(\lambda/\sigma, \sigma X)$.
	\item The Frobenius norm $\norm{X}_\rmF$ of matrix $X$ can be any positive number. If $\norm{X}_\rmF$ is known, one can scale $X$ to have unit Frobenius norm before solving BGPC. In practice, the norm of $X$ is generally unknown. However, due to Assumptions \ref{ass:A} (RIP) and \ref{ass:lambda} (``flat'' gains), we have
	\begin{align*}
	\sqrt{(1-\delta_s)(1-\delta)}\leq \frac{\norm{\diag(\lambda)AX}_\rmF}{\norm{X}_\rmF} \\
	\leq \sqrt{(1+\delta_s)(1+\delta)}.
	\end{align*}
	Hence $\norm{Y}_\rmF$ is a good surrogate for $\norm{X}_\rmF$ in noiseless or low noise settings, and one can scale $X$ by $1/\norm{Y}_\rmF$ to achieve the desired scaling. The slight deviation of $\norm{X}_\rmF/\norm{Y}_\rmF$ from $1$ does \emph{not} have any significant impact on our theoretical analysis. Therefore, we assume $\norm{X}_\rmF = 1$ to simply the constants in our derivation.
	\item The conditioning of $X$ can also be relaxed. When $N$ is large, one can choose a subset of $N'<N$ columns in $Y$, such that the matrix formed from the corresponding columns of $X$ has good conditioning. When noise amplification is not of concern (noiseless or low noise settings), one can choose a preconditioning matrix $H\in\bbC^{N\times N}$ such that $X' = XH$ is well conditioned, and then solve the BGPC with $Y' = YH$.   
\end{itemize}
In summary, we can manipulate the BGPC problem and make it approximately satisfy our assumptions. For example, \eqref{eq:bgpc} can be rewritten as:
\begin{align*}
\frac{1}{\norm{YH}_\rmF} YH  = &~ \diag\Bigl(\frac{\lambda}{\sigma}\Bigr) \Bigl(\frac{1}{\sqrt{n}\sigma_A}A \Bigr) \Bigl(\frac{\sqrt{n}\sigma\sigma_A}{\norm{YH}_\rmF}XH \Bigr) \\
& + \frac{1}{\norm{YH}_\rmF} WH.
\end{align*}
We can run Algorithms \ref{alg:pi} and \ref{alg:tpi} with input $\frac{1}{\sqrt{n}\sigma_A}A$ and $\frac{1}{\norm{YH}_\rmF} YH$, and solve for $\frac{\lambda}{\sigma}$ and $\frac{\sqrt{n}\sigma\sigma_A}{\norm{YH}_\rmF}XH$. 
The above manipulations do not have any significant impact on the solution, or on our theoretical analysis. However, by making these assumptions, we eliminate some tedious and unnecessary discussions.

We also need an assumption on the noise level.
\begin{assumption} \label{ass:W}
The noise term $W$ satisfies
\begin{itemize}
	\item \textbf{Subspace case:} $\max_{k\in [n],j\in[N]} |w_{kj}| \leq \frac{C_W}{\sqrt{nN}}$
	\item \textbf{Joint sparsity case:} $\max_{k\in [n],j\in[N]} |w_{kj}| \leq \frac{C_W}{\sqrt{n}N^2}$
\end{itemize}
for an absolute constant $C_W > 0$.
\end{assumption}
In the subspace case, the assumption on the noise level is very mild. Because under Assumptions \ref{ass:A} -- \ref{ass:X}, $\norm{\diag(\lambda)AX}_\rmF \leq \sqrt{(1+\delta_s)(1+\delta)}$, the noise term $W$, which satisfies $\norm{W}_\rmF \leq C_W$, can be on the same order in terms of Frobenius norm as the clean signal $\diag(\lambda)AX$.

Finally, the following assumption is required for a theoretical guarantee of the initialization.
\begin{assumption} \label{ass:X_flat}
For all $j\in [N]$, there exists $T_j'\subset \supp(x_{\cdot j}) \subset[m]$, such that for all $\ell \in T_j'$,
\[
\frac{|x_{\ell j}|^2}{\norm{x_{\cdot j}}_2^2} \geq \frac{\omega}{s_0},
\]
for some absolute constant $\omega$, and 
\[
\frac{\sum_{\ell'\in [m]\backslash T_j'}|x_{\ell' j}|^2}{\norm{x_{\cdot j}}_2^2} \leq \delta_X,
\]
for some small absolute constant $\delta_X \in (0,1)$.
\end{assumption}
Assumption \ref{ass:X_flat} says that the support of $x_{\cdot j}$ can be partitioned into two subsets. The absolute values of the entries in the first subset $T_j'$ are sufficiently large. Moreover, the total energy (sum of squares of the entries) in the second subset is small compared to the squared norm of $x_{\cdot j}$.
For example, the assumption is satisfied in the following special case: $T_j'=\supp(x_{\cdot j})$ (therefore $x_{\ell' j} = 0$ for $\ell'\in [m]\backslash T_j'$), and the absolute values of the nonzero entries are all comparable, e.g., $x_{\ell j} = \pm \frac{\norm{x_{\cdot j}}}{\sqrt{s_0}}$.

Before introducing our main results, we disclose the choice of parameters $\alpha$ and $\beta$ for our theoretical analysis of Algorithms \ref{alg:pi} and \ref{alg:tpi}. 
\begin{remark} \label{rem:parameter}
When Assumptions \ref{ass:A} -- \ref{ass:X} are satisfied, we choose $\alpha = \sqrt{n}$ and $\beta = 3/2$.
\end{remark}

\subsection{A Perturbation Bound for the Eigenvector Problem}

Next, we introduce a key result, a perturbation bound for the eigenvector problem, which is used to derive error bounds for power iteration algorithms.

Let $\{T_j\}_{j=1}^N$ denote subsets of $[m]$, such that $|T_j|= s$ and $\supp(x_{\cdot j}) \subset T_j$. We define $T_x \subset [Nm]$ and $T_\eta \subset [Nm + n]$ as follows:
\begin{align}
\label{eq:Tx} & T_x \coloneqq \bigcup_{j\in[N]} \bigl( T_j + \{(j-1)m\} \bigr), \\
\label{eq:Teta} & T_\eta \coloneqq T_x \bigcup \bigl( [n] + \{Nm\} \bigr).
\end{align}
Recall that $\Omega_{T}$ restricts a vector to the support $T$, and hence $\Omega_{T}^*\Omega_{T}$ is the projection operator onto the support $T$. Clearly, we have $x = \Omega_{T_x}^*\Omega_{T_x} x$, and $\eta = \Omega_{T_\eta}^*\Omega_{T_\eta}\eta$. In the subspace case discussed in Theorem \ref{thm:perturbation}, we have $s = m$, $T_j = [m]$, $T_x = [Nm]$, and $T_\eta = [Nm+n]$. In the \emph{joint} sparsity case, we have $T_1 = T_2 =\dots = T_N$. We set $|T_j| = s = s_0 + 2s_1$, which we justify later in the analysis of truncated power iteration.

Let 
\[
\dot{\eta} \coloneqq \frac{\eta}{\norm{\eta}_2}
\]
denote the normalized version of $\eta$, which is the eigenvector of $B_\rms$ and $\bbE B_\rms$ corresponding to eigenvalue $0$. 
Let $\hat{\eta}$ denote the principal eigenvector of $G$. In the joint sparsity case, let $\hat{\eta}_{T_\eta}$ denote the principal eigenvector of $\Omega_{T_\eta} G \Omega_{T_\eta}^*$, where $T = T_1 = \dots = T_N$, $|T|=s$, and the support of $\eta$ is a subset of $T_\eta$ defined in \eqref{eq:Teta}.

In Algorithms \ref{alg:pi} and \ref{alg:tpi} and in our analysis, vectors $\dot{\eta}$, $\hat{\eta}$, and $\eta^{(t)}$ are normalized to unit norm. However, multiplication by a scalar of unit modulus is a remaining ambiguity, i.e., the set $\{e^{\sqrt{-1}\varphi}\dot{\eta}: \varphi \in [0,2\pi)\}$ is an equivalence class for $\dot{\eta}$. Our main results use $d(\eta,\eta') \coloneqq \min_\varphi \norm{e^{\sqrt{-1}\varphi}\eta - \eta'}_2$ to denote the distance between $\eta$ and $\eta'$, which is a metric on the set of such equivalence classes.

\begin{theorem}[\textbf{Subspace Case}] \label{thm:perturbation}
Let $\alpha =\sqrt{n}$, and suppose Assumptions \ref{ass:A} -- \ref{ass:W} are satisfied with $\delta < 1/3$ and a sufficiently small absolute constant $C_W > 0$. 
Then there exist absolute constants $c,C,C' > 0$, such that if
\begin{align}
\max\Bigl\{\frac{m\log^2(Nm+n)}{n}, \frac{\log(Nm+n)}{N},  \nonumber\\
\frac{\log(Nm+n)}{m} \Bigr\} \leq C,  \label{eq:size_subspace}
\end{align}
then with probability at least $1-2n^{-c} - e^{-cm}$,
\[
d(\hat{\eta},~\dot{\eta}) \leq \Delta,
\]
where
\begin{align}
\Delta \coloneqq
\frac{8C'}{1-3\delta} \max\{\nu,~\nu^2\},  \label{eq:Delta}
\end{align}
and
\begin{align}
\nu \coloneqq \sqrt{nN} \max_{k\in[n], j\in[N]}|w_{kj}|. \label{eq:nu}
\end{align}
\end{theorem}


When $m$ is large (e.g., $m\geq n$), \eqref{eq:size_subspace} does not hold, hence the perturbation bound of the eigenvector $\hat{\eta}$ of $G$ in Theorem \ref{thm:perturbation} is no longer true. We can, however, bound the perturbation of the eigenvector of a submatrix of $G$.

\begin{theorem}[\textbf{Joint Sparsity Case}] \label{thm:perturbation2_alt}
Let $\alpha =\sqrt{n}$ and $s=s_0+2s_1$, and suppose Assumptions \ref{ass:A} -- \ref{ass:W} are satisfied with $\delta < 1/3$ and a sufficiently small absolute constant $C_W > 0$. Then there exist absolute constants $c,C,C' > 0$, such that if
\begin{align}
 \max\Bigl\{\frac{(s+N)\log^8 n \log^2(sN+m)}{n}, \nonumber\\
 \frac{\sqrt{s}\log^2 n \log(sN+m)}{N},  \nonumber\\
 \frac{\log^4 n \log^2(sN+m)}{s_0} \Bigr\} \leq C, \label{eq:size_sparsity_alt}
\end{align}
then with probability at least $1-2n^{-c} - m^{-cs}$,
\[
d(\hat{\eta}_{T_\eta},~\Omega_{T_\eta} \dot{\eta}) \leq \widetilde{\Delta},
\]
where
\begin{align}
\widetilde{\Delta} \coloneqq
\frac{8C'}{1-3\delta} \max \{N^{3/2} \nu,~\nu^2 \},  \label{eq:Delta_alt}
\end{align}
and $\nu$ is defined in \eqref{eq:nu}.
\end{theorem}

The error bounds for Algorithms \ref{alg:pi} and \ref{alg:tpi} in the next section rely on Theorems \ref{thm:perturbation} and \ref{thm:perturbation2_alt}, and existing analysis of power iteration \cite{Golub1996} and truncated power iteration \cite{Yuan2013}. Additionally, the perturbation bounds in this section are of independent interest. In particular, 
Theorem \ref{thm:perturbation} shows that if the assumptions and the prescribed sample complexities in \eqref{eq:size_subspace} are satisfied, then with high probability the principal eigenvector $\hat{\eta}$ of $G$ is an accurate estimate of the vector $\dot{\eta}$ that concatenates the unknown variables. It gives an error bound for any algorithm that finds the principal eigenvector of $G$.
Similarly, we believe that Theorem \ref{thm:perturbation2_alt} can be used to analyze other algorithms that find the sparse principal component of $G$.

\subsection{Error Bounds for the Power Iteration Algorithms}
In this section, we give performance guarantees for Algorithms \ref{alg:pi} and \ref{alg:tpi} under the assumptions stated in Section \ref{sec:assumption}. 

\begin{theorem}[\textbf{Subspace Case}] \label{thm:pi}
Suppose Assumptions \ref{ass:A} -- \ref{ass:W} are satisfied with $\delta < 1/4$ and a sufficiently small absolute constant $C_W > 0$. Let $\alpha =\sqrt{n}$, and $\beta = 3/2$. Assume that $\xi \coloneqq |\hat{\eta}^*\eta^{(0)}| > 0$. Then there exist absolute constants $c,C,C' > 0$, such that if \eqref{eq:size_subspace} is satisfied, then with probability at least $1-2n^{-c} - e^{-cm}$, the iterates in Algorithm \ref{alg:pi} satisfy
\begin{align*}
d(\eta^{(t)},~\dot{\eta}) \leq \rho^t
d(\eta^{(0)},~\dot{\eta}) + 2\Delta.
\end{align*}
where $\Delta$ is defined in \eqref{eq:Delta}, and
\begin{equation}
\label{eq:rho}
\rho \coloneqq \Bigl\{ 1 - \frac{1}{2} \Bigl[1-\Bigl(\frac{1+6\delta}{3-2\delta}\Bigr)^2\Bigr]\xi(1 + \xi) \Bigr\}^{1/2}.
\end{equation}
\end{theorem}

Theorem \ref{thm:pi} shows that the power iteration algorithm requires $n=O(m\log^2(Nm+n))$ sensors and $N=O(\log(Nm+n))$ snapshots to successfully recover $X$ and $\lambda$. This agrees, up to log factors, with the sample complexity required for the uniqueness of $(\lambda, X)$ in the subspace case, which is $n>m$ and $N\geq \frac{n-1}{n-m}$ \cite{Li2015e}.

Next, we compare Theorem \ref{thm:pi} with a similar error bound for the least squares approach by Ling and Strohmer \cite[Theorem 3.5]{Ling2016}. The sample complexity in Theorem \ref{thm:pi} matches the numbers required by the least squares approach $n=O(m\log^2(Nm+n))$ and $N=O(\log^2(Nm+n))$ (up to one log factor). One caveat in the least squares approach is that, apart from the linear equation \eqref{eq:bgpc_linear}, it needs an extra linear constraint to avoid the trivial solution $\gamma=0$, $X=0$. Unfortunately, in the noisy setting, the recovery error by the least squares approach is sensitive to this extra linear constraint. Our numerical experiments (Section \ref{sec:experiment}) show that power iteration outperforms least squares in the noisy setting.


\begin{theorem}[\textbf{Joint Sparsity Case}] \label{thm:tpi_alt}
Suppose Assumptions \ref{ass:A} -- \ref{ass:W} are satisfied with $\delta < 1/4$ and a sufficiently small absolute constant $C_W > 0$. Let $\alpha =\sqrt{n}$, $\beta = 3/2$, $s_1\geq s_0$ in Algorithm \ref{alg:tpi}, and define $s=s_0+2s_1$. Then there exist absolute constants $c,C,C' > 0$, such that if $|\dot{\eta}^*\eta^{(0)}|\geq \xi + \widetilde{\Delta}$ for some $\xi\in(0,1)$, and \eqref{eq:size_sparsity_alt} is satisfied, then with probability at least $1-2n^{-c} - m^{-cs}$, the iterates in Algorithm \ref{alg:tpi} for the \emph{joint} sparsity case satisfy
\begin{align*}
d(\eta^{(t)},~\dot{\eta}) \leq \tilde{\rho}^t
d(\eta^{(0)},~\dot{\eta}) + \frac{2\sqrt{5}\widetilde{\Delta}}{1-\tilde{\rho}}.
\end{align*}
where $\widetilde{\Delta}$ is defined in \eqref{eq:Delta_alt}, and $\tilde{\rho} < 1$ has the following expression:
\begin{align}
\tilde{\rho} \coloneqq \rho \cdot \Bigl(1 + 2\sqrt{\frac{s_0}{s_1}} + \frac{2s_0}{s_1}\Bigr)^{1/2}, \label{eq:rho_tilde}
\end{align}
and $\rho$ is defined in \eqref{eq:rho}.
\end{theorem}

Theorem \ref{thm:tpi_alt} is only valid when $\tilde{\rho} < 1$. With the choice $s_1 = 2s_0$, when $\delta$ approaches $0$, and $\xi$ approaches $1$, the convergence rate $\tilde{\rho}$ is roughly $\frac{1}{3}\sqrt{1 + \sqrt{2} + 2} \approx 0.62$. We discuss a more realistic scenario next. 

\begin{remark} \label{rem:initialize}
A good initialization for $\lambda$ alone is usually sufficient. Suppose one has a good initial estimate for the gains and phases, i.e., $\lambda$ satisfies $|\lambda_k - e^{\sqrt{-1}\varphi_k}| < \sqrt{1+\delta}-1$ for known phase estimates $\{\varphi_k\}_{k=1}^n$. One can initialize Algorithm \ref{alg:tpi} with $\eta^{(0)} = [\bm{0}_{Nm,1}^\T, e^{-\sqrt{-1}\varphi_1},\dots, e^{-\sqrt{-1}\varphi_n}]^\T$, then when $\Delta$ is negligible (noiseless or low noise settings), $\xi$ in Theorem \ref{thm:tpi_alt} can be set to $1/\sqrt{(1+\delta)(2+\delta)}$. For example, if $\delta = 0.05$ and $s_1 \geq 10 s_0$, then $\tilde{\rho}<1$.
Since we do not attempt to optimize the constants in this paper, the constants in this exemplary scenario are conservative. 
\end{remark}

Theorem \ref{thm:tpi_alt} states that for Algorithm \ref{alg:tpi} to recover $\lambda$ and a jointly sparse $X$, it is sufficient to have $n = O(s_0\log^8n\log^2(s_0N+m))$ sensors and $N=O(\sqrt{s_0}\log^2n \log(s_0N+m))$ snapshots. In comparison, the (up to a factor of $2$) optimal sample complexity for unique recovery in the joint sparsity case is $n > 2s_0$ and $N\geq \frac{n-1}{n-2s_0}$ \cite{Li2015e}. Hence, the number of sensors required in Theorem \ref{thm:tpi_alt} is (up to log factors) optimal, but the number of snapshots required is suboptimal.
Another drawback is that these results apply only to the joint sparsity case, and not to the more general sparsity case. However, we believe these drawbacks are due to artifacts of our analysis. 
For both the joint sparsity case and the sparsity case, we have $Nn$ complex-valued measurements, and $Ns_0 + n -1$ complex-valued unknowns. One may expect successful recovery when $n$ and $N$ are (up to log factors) on the order of $s_0$ and $1$, respectively. In fact, numerical experiments in Section \ref{sec:experiment} confirms that truncated power iteration successfully recovers $\lambda$ and $X$ in this regime for the more general sparsity case. 

Wang and Chi \cite{Wang2016} analyzed the performance of $\ell_1$ minimization for BGPC in the sparsity case, where they assumed that $A$ is the DFT matrix, and $X$ is a Bernoulli-Subgaussian random matrix. Their sample complexity for $\ell_1$ minimization is $n=O(s)$ and $N = O(n\log^4 n)$. The success of their algorithm relies on a restrictive assumption that $\lambda_k\approx 1$, which is analogous to the dependence of our algorithm on a good initialization of $\lambda_k$. In the next section, we show that such dependence can be relaxed under some additional conditions using the initialization provided by Algorithm \ref{alg:init}.


\subsection{A Theoretical Guarantee of the Initialization}

The next theorem shows that, under certain conditions, Algorithm \ref{alg:init} recovers the locations of the large entries in $X$ correctly, and yields an initial estimate $\eta^{(0)}$ that satisfies $|\dot{\eta}^* \eta^{(0)}| > 1-2\delta$ (close to $1$).

\begin{theorem}[Initialization] \label{thm:init}
Suppose Assumptions \ref{ass:A} -- \ref{ass:X_flat} are satisfied. Then there exist absolute constants $C'', c''>0$, such that if
\[
n > C'' s_0^2 \log^6 (nmN),
\]
then with probability at least $1-n^{-c''}$, for all $j\in [N]$ the set $T_j'$ in Assumption \ref{ass:X_flat} is a subset of $T_j$ in Algorithm \ref{alg:init}. 
Additionally, in the joint sparsity case, if
sample complexity \eqref{eq:size_sparsity_alt} is satisfied with a sufficiently large $C$, Assumption \ref{ass:W} is satisfied with a sufficiently small $C_W$, and Assumption \ref{ass:X_flat} is satisfied with a sufficiently small $\delta_X$, then $\eta_0$ produced by Algorithm \ref{alg:init} will satisfy that $|\dot{\eta}^* \eta^{(0)}|$ is arbitrarily close to 
\[
\frac{n^{3/2}+\norm{\lambda}_2\norm{\gamma}_2^2}{\sqrt{n^2+\norm{\lambda}_2^2\norm{\gamma}_2^2}\sqrt{n+\norm{\gamma}_2^2}} > 1-2\delta.
\]
\end{theorem}

By Theorem \ref{thm:init}, the constant $\xi$ in Theorem \ref{thm:tpi_alt} can be set to $1-2\delta$ in a low noise setting. For $\delta < 0.19$, this constant $\xi$ is larger than the one in Remark \ref{rem:initialize}, and allows $\tilde{\rho} < 1$ for more choices of $s_1$.

Our guarantee for the initialization requires that the number $n$ of sensors scales quadratically (up to log factors) in the sparsity $s_0$, which seems suboptimal. Since similar suboptimal sampling complexities show up in sparse PCA \cite{Berthet2013} and sparse phase retrieval \cite{Netrapalli2013,Cai2016,Jaganathan2017}, we conjecture that such a scaling law is intrinsic in the problem of sparse BGPC.

In the joint sparsity case, instead of estimating the supports of $x_{\cdot 1},x_{\cdot 2},\dots,x_{\cdot N}$ separately, one can estimate the row support of $X$ directly by sorting $\sum_{j\in[N]} \norm{d^*_{\cdot((j-1)m+\ell)}E}_2^2$ for $\ell\in [m]$ and finding the $s_1$ largest. In this case, Assumption \ref{ass:X_flat} can be changed to: There exists a subset $T'$ of large rows (in terms of $\ell_2$ norm), such that for all $\ell \in T'$,
\[
\frac{\sum_{j\in[N]} |x_{\ell j}|^2}{\norm{X}_\rmF^2} \geq \frac{\omega}{s_0},
\]
and
\[
\frac{\sum_{j\in [N], \ell' \in [m]\backslash T'} |x_{\ell' j}|^2}{\norm{X}_\rmF^2} \leq \delta_X.
\]
In this case, the subset $T'$ can be identified and an initialization $\eta^{(0)}$ can be computed under the same conditions as in Theorem \ref{thm:init}, which can be proved using the same arguments.

\section{Fundamental Estimates} \label{sec:fun_est}
To prove the main results, we must first establish some fundamental estimates specific to BGPC. Proofs of some lemmas in this section can be found in the appendix.

\subsection{A Gap in Eigenvalues} \label{sec:gap}

A key component in establishing a perturbation bound for an eigenvector problem (e.g., Theorem \ref{thm:perturbation}) is bounding the gap between eigenvalues. Lemma \ref{lem:expB} gives us such a bound.

\begin{lemma} \label{lem:expB}
Suppose Assumptions \ref{ass:A} -- \ref{ass:X} are satisfied and $\alpha =\sqrt{n}$. Then the smallest eigenvalue of $\bbE \Omega_{T_\eta} B_\rms \Omega_{T_\eta}^*$ is $0$, and the rest of the eigenvalues reside in the interval $[\frac{(1-\delta)^2}{1+\delta},\, 2(1+\delta)]$.
\end{lemma}

\subsection{Perturbation Due to Randomness in $A$} \label{sec:concentration}

Next, we show that $\Omega_{T_\eta} B_\rms \Omega_{T_\eta}^*$, whose randomness comes from $A$, is close to its mean $\bbE \Omega_{T_\eta} B_\rms \Omega_{T_\eta}^*$ under certain conditions. 

\begin{lemma} \label{lem:Bs}
Suppose Assumptions \ref{ass:A} -- \ref{ass:X} are satisfied, and $\alpha =\sqrt{n}$.
For any constant $\delta_B > 0$, there exist absolute constants $C,c>0$, such that:
\begin{itemize}
	\item \textbf{Subspace case:} If \eqref{eq:size_subspace} is satisfied with $C$, then
\[
\norm{B_\rms -\bbE B_\rms } \leq \delta_B
\]
with probability at least $1 - n^{-c} -e^{-cm}$.

	\item \textbf{Joint sparsity case:} If \eqref{eq:size_sparsity_alt} is satisfied with $C$, then
\[
\norm{\Omega_{T_\eta} B_\rms \Omega_{T_\eta}^* -\bbE \Omega_{T_\eta} B_\rms \Omega_{T_\eta}^*} \leq \delta_B
\]
for all $T_1=\dots = T_N$ and $T_\eta$ defined in \eqref{eq:Teta}, with probability at least $1 - n^{-c} -m^{-cs}$.
\end{itemize}
\end{lemma}

\begin{IEEEproof}[Proof of Lemma \ref{lem:Bs}]
Recall that
\[
\Omega_{T_\eta} B_\rms \Omega_{T_\eta}^* = \begin{bmatrix}
\Omega_{T_x} D^*D \Omega_{T_x}^* & \sqrt{n}\Omega_{T_x}D^*E_\rms \\
\sqrt{n}E_\rms^*D\Omega_{T_x}^* & n E_\rms^*E_\rms
\end{bmatrix}
\]
It follows that
\begin{align}
\nonumber & \norm{\Omega_{T_\eta} B_\rms \Omega_{T_\eta}^* -\bbE \Omega_{T_\eta} B_\rms \Omega_{T_\eta}^*}\\
\label{eq:sn1} & \leq \norm{\Omega_{T_x} D^*D \Omega_{T_x}^* -\bbE \Omega_{T_x} D^*D \Omega_{T_x}^*}\\
\label{eq:sn2} & + n \norm{E_\rms^*E_\rms - \bbE E_\rms^*E_\rms} \\
\label{eq:sn3} & + 2\sqrt{n} \norm{\Omega_{T_x}D^*E_\rms - \bbE \Omega_{T_x}D^*E_\rms}.
\end{align}
Lemma \ref{lem:Bs} follows from the bounds on the spectral norms in \eqref{eq:sn1} -- \eqref{eq:sn3} in Lemmas \ref{lem:DstarD} -- \ref{lem:DstarEs_alt}, respectively.
\end{IEEEproof}


\begin{lemma} \label{lem:DstarD}
Suppose Assumption \ref{ass:A} is satisfied, then there exist absolute constants $C_1, c_1 >0$, such that:
\begin{itemize}
	\item \textbf{Subspace case:}
\[
\norm{D^*D - \bbE D^*D} \leq C_1\sqrt{\frac{m}{n}},
\]
with probability at least $1-e^{-c_1 m}$.

	\item \textbf{Joint sparsity case:} For any $\{T_j\}_{j=1}^N$ and $T_x$ defined in \eqref{eq:Tx}, 
\[
\norm{\Omega_{T_x} D^*D \Omega_{T_x}^* - \bbE \Omega_{T_x} D^*D \Omega_{T_x}^*} \leq C_1\sqrt{\frac{s}{n}\log m},
\]
with probability at least $1-m^{-c_1 s}$.
\end{itemize}

\end{lemma}


\begin{lemma} \label{lem:EsstarEs}
Suppose Assumptions \ref{ass:A} -- \ref{ass:X} are satisfied, then there exist absolute constants $C_2, c_2>0$, such that
\begin{itemize}
	\item \textbf{Subspace case:}
	\begin{align*}
\norm{E_\rms^*E_\rms - \bbE E_\rms^*E_\rms} \leq \frac{C_2}{n}\max\Bigl\{\sqrt{\frac{\log n}{N}},\sqrt{\frac{\log n}{m}}, \\
\frac{\log n}{N},\frac{\log n}{m}\Bigr\}
\end{align*}
	
	\item \textbf{Joint sparsity case:}
	\begin{align*}
\norm{E_\rms^*E_\rms - \bbE E_\rms^*E_\rms} \leq \frac{C_2}{n}\max\Bigl\{\sqrt{\frac{\log n}{N}},\sqrt{\frac{\log n}{s_0}}, \\
\frac{\log n}{N},\frac{\log n}{s_0}\Bigr\}
\end{align*}
\end{itemize}
with probability at least $1-n^{-c_2}$.
\end{lemma}


\begin{lemma}[\textbf{Subspace Case}] \label{lem:DstarEs}
Suppose Assumptions \ref{ass:A} -- \ref{ass:X} are satisfied, and $\min\{N,m\}>\log n$, then there exist absolute constants $C_3, c_3>0$, such that 
\begin{align*}
\norm{D^*E_\rms-\bbE D^*E_\rms} \leq C_3 \max\Bigl\{ \sqrt{\frac{\log (Nm+n)}{nN}}, \\
\sqrt{\frac{\log(Nm+n)}{nm}},\frac{\sqrt{m}\log(Nm+n)}{n}\Bigr\}
\end{align*}
with probability at least $1-n^{-c_3}$.
\end{lemma}

\begin{lemma}[\textbf{Joint Sparsity Case}] \label{lem:DstarEs_alt}
Suppose Assumptions \ref{ass:A} -- \ref{ass:X} are satisfied, then there exist absolute constants $C_3, c_3>0$, such that for all $T_1 = \dots = T_N$,
\begin{align*}
& \norm{\Omega_{T_x}D^*E_\rms-\bbE \Omega_{T_x}D^*E_\rms} \\
& \leq  \frac{C_3 s_0^{1/4} (s+N)^{1/4} (\sqrt{n}+\sqrt{s+N})^{1/2}}{n \min\{\sqrt{s_0}, \sqrt{N}\}} \\
&\quad \log^3 n \log(sN+m),
\end{align*}
with probability at least $1-n^{-c_3}$.
\end{lemma}

\subsection{Perturbation Due to Noise} \label{sec:noise}

We established some fundamental estimates regarding $B_\rms$ in Sections \ref{sec:gap} and \ref{sec:concentration}. In this section, we turn to perturbation caused by noise. By the definitions of $B$, $B_\rms$, $E$, $E_\rms$, and $E_\rmn$, we have
\[
B = B_\rms + B_\rmn,
\]
where
\[
B_\rmn \coloneqq 
\begin{bmatrix}
0 & \alpha D^* E_\rmn \\
\alpha E_\rmn^*D & \alpha^2 (E_\rms^*E_\rmn + E_\rmn^* E_\rms + E_\rmn^* E_\rmn)
\end{bmatrix}.
\]
Therefore,
\begin{align*}
& \Omega_{T_\eta}B_\rmn \Omega_{T_\eta}^* \\
& = \begin{bmatrix}
0 & \alpha \Omega_{T_x} D^* E_\rmn \\
\alpha E_\rmn^*D \Omega_{T_x}^* & \alpha^2 (E_\rms^*E_\rmn + E_\rmn^* E_\rms + E_\rmn^* E_\rmn)
\end{bmatrix}.
\end{align*}

Lemma \ref{lem:Bn} gives an upper bound on the spectral norm of the perturbation from noise.

\begin{lemma} \label{lem:Bn}
Suppose Assumptions \ref{ass:A} -- \ref{ass:X} are satisfied. Let $\alpha =\sqrt{n}$ and let $\nu$ be defined by \eqref{eq:nu}. Then there exist absolute constants $c,C,C'>0$ such that:
\begin{itemize}
	\item \textbf{Subspace case:} If  \eqref{eq:size_subspace} is satisfied, then with probability at least $1 - n^{-c}$
\begin{align*}
\norm{B_\rmn} \leq C' \max \{\nu,~\nu^2 \}.
\end{align*}
Additionally, for any constant $\delta_W >0$, there exists an absolute constant $C_W>0$, if Assumption \ref{ass:W} is satisfied with $C_W$, then the above bound becomes
\[
\norm{B_\rmn} \leq \delta_W.
\]

	\item \textbf{Joint sparsity case:} If \eqref{eq:size_sparsity_alt} is satisfied, then with probability at least $1 - n^{-c}$
\begin{align*}
\norm{\Omega_{T_\eta} B_\rmn \Omega_{T_\eta}^*} \leq C' \max \{N^{3/2}\nu,~\nu^2 \}
\end{align*}
for all $T_1=\dots = T_N$ and $T_\eta$ defined in \eqref{eq:Teta}. Additionally, for any constant $\delta_W >0$, there exists an absolute constant $C_W>0$, if Assumption \ref{ass:W} is satisfied with $C_W$, then the above bound becomes
\[
\norm{\Omega_{T_\eta}B_\rmn \Omega_{T_\eta}^*} \leq \delta_W.
\]
\end{itemize}

\end{lemma}

\begin{IEEEproof}
To complete the proof, we bound the spectral norms of $\Omega_{T_x} D^* E_\rmn$, $E_\rms^*E_\rmn$, and $E_\rmn^* E_\rmn$ in Lemmas \ref{lem:DstarEn}, \ref{lem:EsstarEn}, and \ref{lem:EnstarEn}, respectively.
\end{IEEEproof}


\begin{lemma}[\textbf{Subspace Case}] \label{lem:DstarEn}
Suppose Assumption \ref{ass:A} is satisfied, and $m >\log n$, then there exist absolute constants $C_4, c_4>0$, such that
\begin{align*}
& \norm{D^*E_\rmn} \leq C_4 \max\Bigl\{\sqrt{\log (Nm+n)}, \\
& \qquad\qquad \sqrt{\frac{Nm}{n}} \log (Nm+n) \Bigr\} \max_{k\in[n], j\in[N]}|w_{kj}|,
\end{align*}
with probability at least $1-n^{-c_4}$.
\end{lemma}

\begin{lemma}[\textbf{Joint Sparsity Case}] \label{lem:DstarEn_alt}
Suppose Assumption \ref{ass:A} is satisfied, then there exist absolute constants $C_4, c_4>0$, such that for all $T_1 = \dots = T_N$,
\begin{align*}
\norm{\Omega_{T_x}D^*E_\rmn} \leq C_4 (\sqrt{s}N + \sqrt{sN\log m} + \sqrt{N\log^3 n}) \\
\times \sqrt{\log n}\max_{k\in[n], j\in[N]}|w_{kj}|,
\end{align*}
with probability at least $1-n^{-c_4}$.
\end{lemma}


\begin{lemma} \label{lem:EsstarEn}
Suppose Assumptions \ref{ass:A} -- \ref{ass:X} are satisfied, then there exist absolute constants $C_5, c_5>0$, such that
\begin{itemize}
	\item \textbf{Subspace case:}
	\begin{align*}
\norm{ E_\rms^*E_\rmn } \leq C_5\sqrt{\frac{N}{n}} \max\Bigl\{1, \sqrt{\frac{\log n}{N}}, \sqrt{\frac{\log n}{m}}\Bigr\} \\
\times \max_{k\in[n], j\in[N]}|w_{kj}|,
\end{align*}

	\item \textbf{Joint sparsity case:}
	\begin{align*}
\norm{ E_\rms^*E_\rmn } \leq C_5\sqrt{\frac{N}{n}} \max\Bigl\{1, \sqrt{\frac{\log n}{N}}, \sqrt{\frac{\log n}{s_0}}\Bigr\} \\
\times \max_{k\in[n], j\in[N]}|w_{kj}|,
\end{align*}
\end{itemize}

with probability at least $1-n^{-c_5}$.
\end{lemma}


\begin{lemma} \label{lem:EnstarEn}
\[
\norm{ E_\rmn^*E_\rmn } \leq N\max_{k\in[n], j\in[N]}|w_{kj}|^2,
\]
\end{lemma}

\subsection{Scalar Concentration} \label{sec:scalar}

We now introduce a few scalar concentration bounds that are useful in the proof of Theorem \ref{thm:init}.

\begin{lemma} \label{lem:ineq_square}
Suppose Assumptions \ref{ass:A} -- \ref{ass:W} is satisfied, then there exist absolute constants $C_6, c_6>0$, such that for all $j\in[N]$ and $\ell\in[m]$, we have
\begin{align}
& \left| \sum_{k\in[n]} \left( | \lambda_k \overline{a_{k\ell}} a_{k\cdot}^\T x_{\cdot j}|^2 - \bbE | \lambda_k \overline{a_{k\ell}} a_{k\cdot}^\T x_{\cdot j}|^2\right) \right|  \nonumber \\
& \leq \frac{C_6 \norm{x_{\cdot j}}_2^2 \log^3 (nmN)}{n^{3/2}},  \label{eq:ineq1}
\end{align}
\begin{align}
& \left| \sum_{k\in[n]} \lambda_k a_{k\ell} \overline{a_{k\ell}} a_{k\cdot}^\T x_{\cdot j} \overline{w_{kj}}\right|  \nonumber \\
& \leq \frac{C_6 \norm{x_{\cdot j}}_2 \log^2 (nmN)}{n} \max_{k\in[n],j\in[N]} |w_{kj}| \nonumber \\
& \leq \frac{C_6 C_W \norm{x_{\cdot j}}_2^2 \log^2 (nmN)}{\sqrt{1-\theta} n^{3/2}}, \label{eq:ineq2}
\end{align}
and
\begin{align}
& \left| \sum_{k\in[n]} \left( | \overline{a_{k\ell}} w_{kj} |^2 - \bbE | \overline{a_{k\ell}} w_{kj} |^2 \right) \right|  \nonumber \\
& \leq \frac{C_6 \log^2 (nmN)}{n^{1/2}} \max_{k\in[n],j\in[N]} |w_{kj}|^2  \nonumber\\
& \leq \frac{C_6 C_W^2 \norm{x_{\cdot j}}_2^2 \log^2 (nmN)}{(1-\theta) n^{3/2}},  \label{eq:ineq3}
\end{align}
with probability at least $1-n^{-c_6}$.
\end{lemma}

\section{Proofs of the Main Results} \label{sec:proof}

\subsection{Proof of the Perturbation Bound for the Eigenvector Problem}

In this section, we prove Theorem \ref{thm:perturbation}. Theorem \ref{thm:perturbation2_alt} can be proved similarly.

\begin{IEEEproof}[Proof of Theorem \ref{thm:perturbation}]
First, 
\begin{equation}
\label{eq:Gsum}
G = \beta I_{Nm+n} - B = (\beta I_{Nm+n} - \bbE B_\rms) - (B_\rms - \bbE B_\rms) - B_\rmn. 
\end{equation}
Lemma \ref{lem:expB} establishes a gap in the eigenvalues of the matrix $\bbE B_\rms$ -- the smallest and the second smallest eigenvalues of $\bbE B_\rms$ are separated by a gap of at least
\[
\frac{(1-\delta)^2}{1+\delta} \geq 1 - 3\delta > 0.
\]
Therefore, the gap between the largest and the second largest eigenvalues of $\beta I_{Nm+n} - \bbE B_\rms$ is at least $1-3\delta$. By Lemmas \ref{lem:Bs} and \ref{lem:Bn}, there exist absolute constants $c,C,C', C_W > 0$ such that if all the assumptions are satisfied, then with probability at least $1-2n^{-c} - e^{-cm}$,
\begin{equation}
\label{eq:perturb1}
\norm{(B_\rms - \bbE B_\rms) + B_\rmn} \leq \norm{B_\rms - \bbE B_\rms} + \norm{B_\rmn} \leq \frac{1-3\delta}{4},
\end{equation}
\begin{align}
\norm{B_\rmn} \leq C' \max \{\nu,~\nu^2 \}. \label{eq:perturb2}
\end{align}
Recall that $\dot{\eta}$ is the principal eigenvector of $\beta I_{Nm+n} - \bbE B_\rms$. By the Davis-Kahan $\sin\theta$ Theorem (\cite{Davis1970}; see also \cite[Theorem 8.1.12]{Golub1996}), \eqref{eq:perturb1} and \eqref{eq:perturb2} imply
\begin{align*}
& \sin\angle(\dot{\eta}, \hat{\eta}) \\
\leq & \frac{4}{1-3\delta} \norm{ (B_\rms - \bbE B_\rms + B_\rmn) \dot{\eta} }_2 \\
\leq & \frac{4}{1-3\delta} \norm{B_\rmn} \\
\leq & \frac{4C'}{1-3\delta} \max \{\nu,~\nu^2 \},
\end{align*}
where the second inequality is due to $B_\rms \dot{\eta} = \bbE B_\rms \dot{\eta} = 0$.

Theorem \ref{thm:perturbation} follows from the above bound, and the fact that
\begin{align*}
d(\dot{\eta}, \hat{\eta}) = \sqrt{2 - 2 \cos\angle(\dot{\eta},\hat{\eta})} = 2\sin\frac{\angle(\dot{\eta}, \hat{\eta})}{2} \\
\leq 2\sin\angle(\dot{\eta}, \hat{\eta}).
\end{align*}
\end{IEEEproof}

One can prove Theorem \ref{thm:perturbation2_alt} using the same steps as in the proof of Theorem \ref{thm:perturbation}, by restricting rows and columns of matrices to the support $T_\eta$ and applying the corresponding bounds on submatrices.


\subsection{Proof of the Error Bound for Algorithm \ref{alg:pi}}

\begin{IEEEproof}[Proof of Theorem \ref{thm:pi}]
Recall that the largest eigenvalue of $\beta I_{Nm+n} - \bbE B_\rms$ is $\beta - 0 = \frac{3}{2}$, and all other eigenvalues reside in the interval $[\frac{3}{2} - 2(1+\delta), \frac{3}{2} - \frac{(1-\delta)^2}{1+\delta}]$. By Lemmas \ref{lem:Bs} and \ref{lem:Bn}, there exist constants $c,C,C_W>0$ such that
\begin{align*}
\norm{(B_\rms - \bbE B_\rms) + B_\rmn} \leq \norm{B_\rms - \bbE B_\rms} + \norm{B_\rmn} \\
\leq \min\Bigl\{\delta, \frac{(1-\delta)^2}{1+\delta} +3\delta -1\Bigr\},
\end{align*}
with probability at least $1-2n^{-c}-e^{-cm}$. By \eqref{eq:Gsum}, the largest eigenvalue of $G$ is $\norm{G} \geq \frac{3}{2} - \delta$, the corresponding eigenvector is $\hat{\eta}$, and all the other eigenvalues of $G$ reside in the interval $[-\frac{1}{2}-3\delta, \frac{1}{2}+3\delta]$.
By the eigenvalue decomposition of $G$ and the Pythagorean theorem,
\[
G\hat{\eta} =\norm{G} \hat{\eta},
\]
\begin{align*}
&\norm{G\eta^{(t-1)}} \\
&\leq \sqrt{\norm{G}^2|\hat{\eta}^*\eta^{(t-1)}|^2+\Bigl(\frac{1}{2}+3\delta\Bigr)^2(1- |\hat{\eta}^*\eta^{(t-1)}|^2)}.
\end{align*}
Therefore,
\begin{align*}
& |\hat{\eta}^*\eta^{(t)}| \\
= & \frac{|\hat{\eta}^*G\eta^{(t-1)}|}{\norm{G\eta^{(t-1)}}_2} \\
\geq & \frac{\norm{G}|\hat{\eta}^*\eta^{(t-1)}|}{\sqrt{\norm{G}^2|\hat{\eta}^*\eta^{(t-1)}|^2+(\frac{1}{2}+3\delta)^2(1- |\hat{\eta}^*\eta^{(t-1)}|^2)}} \\
\geq & |\hat{\eta}^*\eta^{(t-1)}| \frac{1}{\sqrt{|\hat{\eta}^*\eta^{(t-1)}|^2+(\frac{1+6\delta}{3-2\delta})^2(1- |\hat{\eta}^*\eta^{(t-1)}|^2)}} \\
= & |\hat{\eta}^*\eta^{(t-1)}| \frac{1}{\sqrt{1-\bigl(1-(\frac{1+6\delta}{3-2\delta})^2\bigr)(1- |\hat{\eta}^*\eta^{(t-1)}|^2)}} \\
\geq & |\hat{\eta}^*\eta^{(t-1)}| \Big[ 1 + \frac{1}{2} \bigl(1-(\frac{1+6\delta}{3-2\delta})^2\bigr)(1- |\hat{\eta}^*\eta^{(t-1)}|^2) \Big],
\end{align*}
where the last inequality is due to $\frac{1}{\sqrt{1-z}} \geq 1 + \frac{1}{2}z$ for $z\in (0,1)$. It follows that
\begin{align}
& [1 - |\hat{\eta}^*\eta^{(t)}|] \leq [1 - |\hat{\eta}^*\eta^{(t-1)}|]  \nonumber\\
& \times \Big[ 1 - \frac{1}{2} \bigl(1-(\frac{1+6\delta}{3-2\delta})^2\bigr)|\hat{\eta}^*\eta^{(t-1)}|(1 + |\hat{\eta}^*\eta^{(t-1)}|) \Big]. \label{eq:converge}
\end{align}
Clearly, $\{ |\hat{\eta}^*\eta^{(\tau)}| \}_{\tau = 0}^t$ is monotonically increasing unless $|\hat{\eta}^*\eta^{(0)}| = 0$. By the definition $\xi \coloneqq |\hat{\eta}^*\eta^{(0)}| $, the convergence rate in \eqref{eq:converge} is bounded by $\rho^2 < 1$. It follows that
\begin{align*}
[1 - |\hat{\eta}^*\eta^{(t)}|] & \leq \rho^2[1 - |\hat{\eta}^*\eta^{(t-1)}|] \\
& \leq \rho^{2t} [1 - |\hat{\eta}^*\eta^{(0)}|]
\end{align*}
\begin{align*}
d(\hat{\eta}, \eta^{(t)}) \leq \rho^t
d(\hat{\eta}, \eta^{(0)})
\end{align*}

By Theorem \ref{thm:perturbation}, for $\tau = 0,\dots, t$
\[
d(\dot{\eta}, \hat{\eta}) \leq \Delta.
\]
It follows that
\begin{align*}
d(\dot{\eta}, \eta^{(t)}) \leq \rho^t
d(\dot{\eta}, \eta^{(0)}) + 2\Delta.
\end{align*}
\end{IEEEproof}


\subsection{Proof of the Error Bound for Algorithm \ref{alg:tpi}}

\begin{IEEEproof}[Proof of Theorem \ref{thm:tpi_alt}]
In the joint sparsity case, any iterate $\eta^{(\tau)} = [x^{(\tau)\T}, -\gamma^{(\tau)\T}/\alpha]^\T$ satisfies that $x^{(\tau)}$ is the concatenation of jointly sparse $\{x_{\cdot j}^{(\tau)}\}_{j=1}^N$. 
In the $t$-th iteration, we define a support set $T^{(t)}$ that has cardinality $s = s_0 + 2s_1$, and satisfies
\[
\supp(x_{\cdot j}) \bigcup \supp(x_{\cdot j}^{(t-1)}) \bigcup \supp(x_{\cdot j}^{(t)}) \subset T^{(t)},
\]
for all $j\in[N]$. Define $T_\eta^{(t)}$ using \eqref{eq:Tx} and \eqref{eq:Teta} with $T_1 = \dots = T_N = T^{(t)}$. Next, we focus on the submatrix $\Omega_{T_\eta^{(t)}} G \Omega_{T_\eta^{(t)}}^*$ and subvectors $\Omega_{T_\eta^{(t)}} \dot{\eta}$ and $\Omega_{T_\eta^{(t)}} \eta^{(t)}$, etc. Since the supports of $\eta^{(t)}$ and $\dot{\eta}$ are subsets of $T_\eta^{(t)}$, we have $|\dot{\eta}^*\Omega_{T_\eta^{(t)}}^*\Omega_{T_\eta^{(t)}}\eta^{(t)}| = |\dot{\eta}^*\eta^{(t)}|$.

We prove by induction that $\{|\dot{\eta}^*\eta^{(\tau)}|\}_{\tau=0}^t$ is monotonically increasing (until it crosses a threshold specified later in the proof). Suppose $\{|\dot{\eta}^*\eta^{(\tau)}|\}_{\tau=0}^{t-1}$ is monotonically increasing. Next, we prove 
\[
|\dot{\eta}^*\eta^{(t)}| > |\dot{\eta}^*\eta^{(t-1)}|.
\]

By the assumption that $|\dot{\eta}^*\eta^{(0)}|\geq \xi + \widetilde{\Delta}$ and Theorem \ref{thm:perturbation2_alt}, we have
\begin{align*}
& |\hat{\eta}_{T_\eta^{(t)}}^*\Omega_{T_\eta^{(t)}}\eta^{(t-1)}| \\
\geq & |\dot{\eta}^*\eta^{(t-1)}| - d(\Omega_{T_\eta^{(t)}}\dot{\eta},~\hat{\eta}_{T_\eta^{(t)}}) \\
\geq & \xi + \widetilde{\Delta} - \widetilde{\Delta} \\
= & \xi.
\end{align*}
Following the same steps in the proof of Theorem \ref{thm:pi}, we obtain a bound for $\tilde{\eta}^{(t)}$ similar to \eqref{eq:converge}:
\begin{align*}
& [1 - |\hat{\eta}_{T_\eta^{(t)}}^*\Omega_{T_\eta^{(t)}}\tilde{\eta}^{(t)}|] \\
\leq & [1 - |\hat{\eta}_{T_\eta^{(t)}}^*\Omega_{T_\eta^{(t)}}\eta^{(t-1)}|] \Big[ 1 - \frac{1}{2} \bigl(1-(\frac{1+6\delta}{3-2\delta})^2\bigr) \\
& \qquad\qquad |\hat{\eta}_{T_\eta^{(t)}}^*\Omega_{T_\eta^{(t)}}\eta^{(t-1)}|(1 + |\hat{\eta}_{T_\eta^{(t)}}^*\Omega_{T_\eta^{(t)}}\eta^{(t-1)}|) \Big] \\
\leq & [1 - |\hat{\eta}_{T_\eta^{(t)}}^*\Omega_{T_\eta^{(t)}}\eta^{(t-1)}|] \Big[ 1 - \frac{1}{2} \bigl(1-(\frac{1+6\delta}{3-2\delta})^2\bigr)\xi(1 + \xi) \Big] \\
= & \rho^2 [1 - |\hat{\eta}_{T_\eta^{(t)}}^*\Omega_{T_\eta^{(t)}}\eta^{(t-1)}|],
\end{align*}
where $\rho$ is defined in \eqref{eq:rho}. It follows that
\begin{align*}
d(\hat{\eta}_{T_\eta^{(t)}},~ \Omega_{T_\eta^{(t)}}\tilde{\eta}^{(t)}) \leq \rho \cdot d(\hat{\eta}_{T_\eta^{(t)}},~ \Omega_{T_\eta^{(t)}}\eta^{(t-1)}) 
\end{align*}
We use the perturbation bound in Theorem \ref{thm:perturbation2_alt} one more time:
\begin{align*}
d(\Omega_{T_\eta^{(t)}}\dot{\eta},~ \Omega_{T_\eta^{(t)}}\tilde{\eta}^{(t)}) \leq \rho \cdot d(\Omega_{T_\eta^{(t)}}\dot{\eta},~ \Omega_{T_\eta^{(t)}}\eta^{(t-1)}) + 2\widetilde{\Delta}.
\end{align*}
Equivalently,
\begin{equation}
\label{eq:before_proj}
\sqrt{1 - | \dot{\eta}^*\tilde{\eta}^{(t)} |} \leq \rho \sqrt{1 - | \dot{\eta}^* \eta^{(t-1)} |} + \sqrt{2}\widetilde{\Delta}.
\end{equation}

Next, we show that the truncation step amplifies the error only by a small factor. The vector $\widetilde{\Pi}_{s_1}(\tilde{\eta}^{(t)})$ is the projection of $\tilde{\eta}^{(t)}$ onto the set of structured sparse vectors, and $\eta^{(t)}$ is the normalized version. We define three index sets
\begin{align*}
& T_a = \supp(\dot{\eta})\backslash \supp(\eta^{(t)}), \\
& T_b = \supp(\dot{\eta})\bigcap \supp(\eta^{(t)}), \\
& T_c = \supp(\eta^{(t)})\backslash \supp(\dot{\eta}). 
\end{align*} 
By the Cauchy-Schwarz inequality,
\begin{align*}
& |\dot{\eta}^*\tilde{\eta}^{(t)}|^2 \\
\leq & \norm{\Omega_{T_a} \tilde{\eta}^{(t)}}_2^2 + \norm{\Omega_{T_b} \tilde{\eta}^{(t)}}_2^2 \\
\leq & 1 - \norm{\Omega_{T_c} \tilde{\eta}^{(t)}}_2^2 \\
\leq & 1 - \frac{|T_c|}{|T_a|} \norm{\Omega_{T_a} \tilde{\eta}^{(t)}}_2^2,
\end{align*}
where the last inequality is due to projection rule, i.e., $\widetilde{\Pi}_{s_1}(\tilde{\eta}^{(t)})$ keeps the largest entries of $\tilde{\eta}^{(t)}$ (in the part corresponding to $x$). Since ${|T_c|}/{|T_a|}\geq s_1/s_0$, we have
\begin{equation}
\label{eq:Tat}
\norm{\Omega_{T_a} \tilde{\eta}^{(t)}}_2 \leq \sqrt{\frac{s_0}{s_1} (1-|\dot{\eta}^*\tilde{\eta}^{(t)}|^2)}.
\end{equation}
Also by the Cauchy-Schwarz inequality,
\begin{align*}
& |\dot{\eta}^*\tilde{\eta}^{(t)}|^2 \\
& \leq (\norm{\Omega_{T_a} \tilde{\eta}^{(t)}}_2 \norm{\Omega_{T_a} \dot{\eta}}_2 + \norm{\Omega_{T_b} \tilde{\eta}^{(t)}}_2 \norm{\Omega_{T_b} \dot{\eta}}_2)^2 \\
& \leq \Bigl(\norm{\Omega_{T_a} \tilde{\eta}^{(t)}}_2 \norm{\Omega_{T_a} \dot{\eta}}_2 \\
& \qquad + \sqrt{1-\norm{\Omega_{T_a} \tilde{\eta}^{(t)}}_2^2} \sqrt{1-\norm{\Omega_{T_a} \dot{\eta}}_2^2} ~\Bigr)^2 \\
& \leq 1 - (\norm{\Omega_{T_a} \tilde{\eta}^{(t)}}_2 - \norm{\Omega_{T_a} \dot{\eta}}_2)^2.
\end{align*}
It follows that
\begin{equation}
\label{eq:Tad}
\norm{\Omega_{T_a} \dot{\eta}}_2 \leq \norm{\Omega_{T_a} \tilde{\eta}^{(t)}}_2 + \sqrt{1-|\dot{\eta}^*\tilde{\eta}^{(t)}|^2}.
\end{equation}
By \eqref{eq:Tat} and \eqref{eq:Tad},
\begin{align}
\nonumber & |\dot{\eta}^*\tilde{\eta}^{(t)}| - |\dot{\eta}^*\widetilde{\Pi}_{s_1}(\tilde{\eta}^{(t)})| \\
\nonumber \leq & |\dot{\eta}^*\bigl(\tilde{\eta}^{(t)}-\widetilde{\Pi}_{s_1}(\tilde{\eta}^{(t)})\bigr)| \\
\nonumber = & \norm{\Omega_{T_a} \tilde{\eta}^{(t)}}_2\norm{\Omega_{T_a} \dot{\eta}}_2 \\
\label{eq:proj} \leq & \Big(\sqrt{\frac{s_0}{s_1}} + \frac{s_0}{s_1}\Big) (1-|\dot{\eta}^*\tilde{\eta}^{(t)}|^2).
\end{align}

By \eqref{eq:before_proj} and \eqref{eq:proj},
\begin{align*}
& \sqrt{1-|\dot{\eta}^* \eta^{(t)}|} \\
\leq & \sqrt{1-|\dot{\eta}^* \widetilde{\Pi}_{s_1}(\tilde{\eta}^{(t)})|} \\
\leq & \sqrt{1-|\dot{\eta}^* \tilde{\eta}^{(t)}|} \sqrt{1 + \Big(\sqrt{\frac{s_0}{s_1}} + \frac{s_0}{s_1}\Big)(1 + |\dot{\eta}^* \tilde{\eta}^{(t)}|)} \\
\leq & \sqrt{1-|\dot{\eta}^* \tilde{\eta}^{(t)}|} \sqrt{1 + 2\Big(\sqrt{\frac{s_0}{s_1}} + \frac{s_0}{s_1}\Big)} \\
\leq & \rho \sqrt{1 + 2\sqrt{\frac{s_0}{s_1}} + \frac{2s_0}{s_1}} \sqrt{1 - | \dot{\eta}^* \eta^{(t-1)} |} + \sqrt{10}\widetilde{\Delta} \\
\leq & \tilde{\rho} \sqrt{1 - | \dot{\eta}^* \eta^{(t-1)} |} + \sqrt{10}\widetilde{\Delta}.
\end{align*}
Therefore, $\{|\dot{\eta}^* \eta^{(\tau)}|\}_{\tau=0}^t$ indeed monotonically increases unless $\sqrt{1-|\dot{\eta}^* \eta^{(\tau)}|}$ reaches $\sqrt{10}\widetilde{\Delta}/(1-\tilde{\rho})$ for some $\tau$. The proof by induction is complete.

It follows that
\[
\sqrt{1-|\dot{\eta}^* \eta^{(t)}|} \leq \tilde{\rho}^t \sqrt{1-|\dot{\eta}^* \eta^{(0)}|} + \frac{\sqrt{10}\widetilde{\Delta}}{1-\tilde{\rho}},
\]
or equivalently
\[
d(\dot{\eta}, \eta^{(t)}) \leq \tilde{\rho}^t
d(\dot{\eta}, \eta^{(0)}) + \frac{2\sqrt{5}\widetilde{\Delta}}{1-\tilde{\rho}}.
\]
\end{IEEEproof}


\subsection{Proof of the Guarantee for Algorithm \ref{alg:init}}

\begin{IEEEproof}[Proof of Theorem \ref{thm:init}]
We first show that, under the conditions in Theorem \ref{thm:init}, the support $T_j$ in Algorithm \ref{alg:init} contains $T_j'\subset \supp(x_{\cdot j})$ in Assumption \ref{ass:X_flat}. 
To this end, we prove that the norms of the rows of $D^*E$ indexed by $T_j'$ are larger than those outside $\supp(x_{\cdot j})$. For a fixed $j\in [N]$, the $j$-th block of $D^*E$ is indexed by the set $(j-1)m + [m]$. Therefore, the goal is to show that
\begin{align*}
& \min_{\ell \in T_j'}\norm{d^*_{\cdot ((j-1)m + \ell)}E}_2^2 \\
& > \max_{\ell' \in [m]\backslash \supp(x_{\cdot j})}\norm{d^*_{\cdot((j-1)m + \ell')}E}_2^2,
\end{align*}
or equivalently,
\begin{align*}
&\min_{\ell \in T_j'} \sum_{k\in[n]} |\overline{a_{k\ell}}y_{kj}|^2 \\
& > \max_{\ell' \in [m]\backslash \supp(x_{\cdot j})} \sum_{k\in[n]} |\overline{a_{k\ell'}}y_{kj}|^2.
\end{align*}
Since
\[
\bbE |\overline{a_{k\ell}}y_{kj}|^2 = \frac{1}{n^2} |\lambda_k|^2 (\norm{x_{\cdot j}}_2^2 + |x_{\ell j}|^2) + \frac{1}{n} |w_{kj}|^2,
\]
it suffices to show that for all $\ell\in T_j'$ and $\ell''\in [m]$,
\begin{align}
&\frac{1}{n^2} \sum_{k\in[n]} |\lambda_k|^2  |x_{\ell j}|^2  \nonumber \\
& > 2 \left| \sum_{k\in[n]} \left( |\overline{a_{k\ell''}}y_{kj}|^2 - \bbE |\overline{a_{k\ell''}}y_{kj}|^2\right) \right|. \label{eq:suffice}
\end{align}

Recall that
\[
y_{kj} = \lambda_k a_{k\cdot}^\T x_{\cdot j} + w_{kj}.
\]
By the triangle inequality and Lemma \ref{lem:ineq_square}, for all $j\in[N]$ and $\ell\in[m]$,
\begin{align*}
& \left| \sum_{k\in[n]} \left( |\overline{a_{k\ell}}y_{kj}|^2 - \bbE ||\overline{a_{k\ell}}y_{kj}|^2\right) \right| \\
& \leq \left| \sum_{k\in[n]} \left( | \lambda_k \overline{a_{k\ell}}a_{k\cdot}^\T x_{\cdot j}|^2 - \bbE | \lambda_k \overline{a_{k\ell}}a_{k\cdot}^\T x_{\cdot j}|^2\right) \right| \\
& + 2 \left| \sum_{k\in[n]} \mathrm{Re}\left(\lambda_k a_{k\ell} \overline{a_{k\ell}} a_{k\cdot}^\T x_{\cdot j} \overline{w_{kj}} \right) \right| \\
& + \left| \sum_{k\in[n]} \left( | \overline{a_{k\ell}} w_{kj} |^2 - \bbE | \overline{a_{k\ell}} w_{kj} |^2 \right) \right| \\
& \leq C_6 \left(1+\frac{C_W}{\sqrt{1-\theta}}\right)^2\frac{\norm{x_{\cdot j}}_2^2 \log^3 (nmN)}{n^{3/2}},
\end{align*}
with probability at least $1-n^{-c_6}$.

By Assumptions \ref{ass:lambda} and \ref{ass:X_flat}, if we plug the above result into \eqref{eq:suffice}, then the following sample complexity is sufficient for Algorithm \ref{alg:init} to correctly identify the subsets $T_j'$ ($j\in[N]$) with probability at least $1-n^{-c_6}$:
\[
n^{1/2} > \frac{2C_6}{\omega(1-\delta)} \left(1+\frac{C_W}{\sqrt{1-\theta}}\right)^2 s_0\log^3 (nmN).
\]
Thus the first half of Theorem \ref{thm:init} is proved.

Given that the support $T_j$ covers the large entries indexed by $T_j'$,
\begin{align}
& \norm{\bbE \Pi_{T_x} D^* E - \frac{1}{n} x \lambda^\T}  \nonumber\\
& = \norm{\frac{1}{n}\Pi_{T_x} x \lambda^\T - \frac{1}{n} x \lambda^\T}  \nonumber\\
& \leq \sqrt{\frac{1+\delta}{n} \sum_{j\in [N],\ell'\in[m]\backslash T_j'} |x_{\ell' j}|^2} \nonumber\\
& \leq \sqrt{\frac{(1+\delta)\delta_X}{n}}. \label{eq:small_energy}
\end{align}
We also have
\begin{align}
& \norm{ \Pi_{T_x} D^* E - \bbE \Pi_{T_x} D^* E } \nonumber\\
& \leq \norm{ \Omega_{T_x} D^* E_\rms - \bbE \Omega_{T_x} D^* E_\rms } + \norm{ \Omega_{T_x} D^* E_\rmn } \nonumber\\
& \leq \frac{1}{\alpha} (\norm{ \Omega_{T_\eta} B_\rms \Omega_{T_\eta}^* - \Omega_{T_\eta} \bbE B_\rms \Omega_{T_\eta}^*} + \norm{ \Omega_{T_\eta} B_\rmn \Omega_{T_\eta}^* } ) \nonumber\\
& \leq \frac{1}{\sqrt{n}} (\delta_B + \delta_W), \label{eq:use_perturbation}
\end{align}
where the last inequality follows from Lemmas \ref{lem:Bs} and \ref{lem:Bn}, give that the conditions of Theorem \ref{thm:tpi_alt} are satisfied. By the triangle inequality, and \eqref{eq:small_energy} and \eqref{eq:use_perturbation},
\[
\norm{ \Pi_{T_x} D^* E - \frac{1}{n} x \lambda^\T } \leq \frac{1}{\sqrt{n}} (\delta_B + \delta_W + \sqrt{(1+\delta)\delta_X}),
\]
where $\delta_B$ can be made arbitrarily small by a sufficiently large $C$ in \eqref{eq:size_sparsity_alt}, $\delta_W$ can be made arbitrarily small by a sufficiently small $C_W$ in Assumption \ref{ass:W}, and the last term can be made arbitrarily small by a sufficiently small $\delta_X$ in Assumption \ref{ass:X_flat}. Therefore, the first left and right singular vectors $u$ and $v$ can become arbitrarily close to $x$ and to $\lambda/\norm{\lambda}_2$ (up to a global phase factor, i.e., a constant of unit modulus), respectively, and $|\dot{\eta}^* \eta^{(0)}|$ approaches
\[
\frac{n^{3/2}+\norm{\lambda}_2\norm{\gamma}_2^2}{\sqrt{n^2+\norm{\lambda}_2^2\norm{\gamma}_2^2}\sqrt{n+\norm{\gamma}_2^2}} > 1-2\delta.
\]
The inequality follows from Assumption \ref{ass:lambda}, i.e., $\sqrt{1-\delta} \leq |\lambda_k| \leq \sqrt{1+\delta}$, and $1/\sqrt{1+\delta} \leq |\gamma_k| = 1 / |\lambda_k| \leq 1/\sqrt{1-\delta}$.
\end{IEEEproof}

\section{Numerical Experiments} \label{sec:experiment}

In this section, we test the empirical performance of Algorithm \ref{alg:pi} and Algorithm \ref{alg:tpi}.

\subsection{Subspace Case: Power Iteration vs. Least Squares} \label{sec:exp_pi}
In Algorithm \ref{alg:pi}, we choose $\alpha = \sqrt{n}$, and $\beta = \norm{B}$ (computed using another power iteration on $B$). We compare Algorithm \ref{alg:pi} with the least squares approach in \cite[Section 3.3]{Ling2016}, where $\gamma_1 = 1$ is used to avoid the trivial solution. 

\begin{figure}%
\centering
\includegraphics[width=0.5\columnwidth]{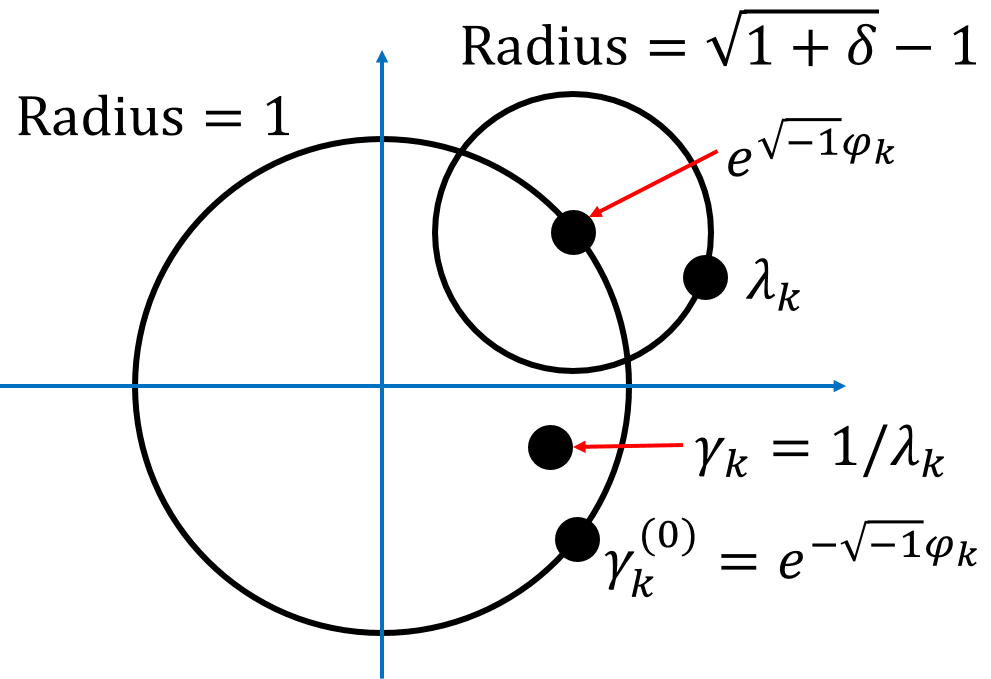}%
\caption{Illustration of $\lambda_k$ in the complex plane.}%
\label{fig:phase}%
\end{figure}

We generate $A\in\bbC^{n\times m}$ as a complex Gaussian random matrix, whose entries are drawn independently from $\calCN(0,\frac{1}{n})$, i.e., the real and imaginary part are drawn independently from $\mathcal{N}(0, \frac{1}{2n})$. The unknown gains and phases $\lambda_k$ are generated as follows:
\begin{equation}
\label{eq:lambda_exp}
\lambda_k = e^{\sqrt{-1}\varphi_k} \Big(1+(\sqrt{1+\delta}-1)e^{\sqrt{-1}\varphi_k'}\Big), \quad \forall k\in[n],
\end{equation}
such that $\lambda_k$ is on a small circle of radius $\sqrt{1+\delta}-1$ centered at a point on the unit circle, and $\varphi_k$ and $\varphi_k'$ are drawn independently from a uniform distribution on $[0,2\pi)$. Figure \ref{fig:phase} visualizes one such synthesized $\lambda_k$ in the complex plane. We set $\delta = 0.1$ in all the numerical experiments.
The entries of $X\in\bbC^{m\times N}$ are drawn independently from $\calCN(0,\frac{1}{Nm})$, so that the Frobenius norm of $X$ is approximately $1$. 
In the noisy setting, we generate complex white Gaussian noise $W\in\bbC^{n\times N}$, whose entries are drawn from $\calCN(0,\frac{\sigma_W^2}{Nn})$.
We define measurement signal-to-noise ratio (MSNR) and recovery signal-to-noise ratio (RSNR) as:
\[
\text{MSNR} \coloneqq 20\log_{10} \frac{\norm{\diag(\lambda)AX}_\rmF}{\norm{W}_\rmF},
\]
\[
\text{RSNR} \coloneqq -10\log_{10}(2-2|\dot{\eta}^*\eta^{(t)}|).
\]

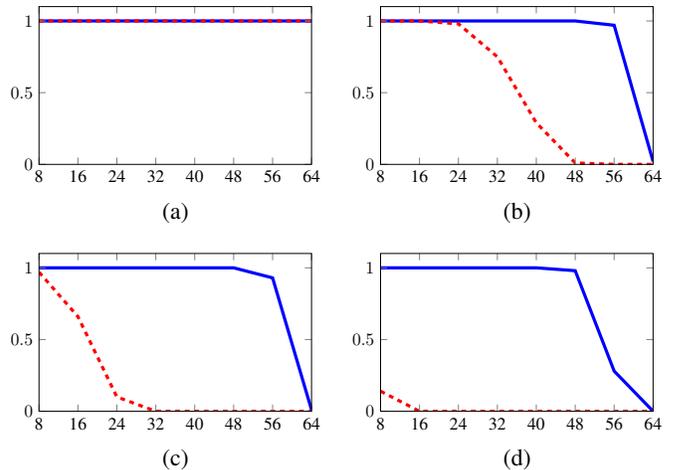
\begin{figure}[htbp]%
\centering
\subfloat[]{
\begin{tikzpicture}[scale=0.6]

\begin{axis}[
width=3in,
height=2in,
xmin=0, xmax=7,
ymin=0, ymax=1.1,
axis on top,
xtick={0,1,2,3,4,5,6,7},
xticklabels={8,16,24,32,40,48,56,64}
]
\addplot [color=blue,solid,line width=2.0pt]
table {%
0 1
1 1
2 1
3 1
4 1
5 1
6 1
7 1
};
\addplot [color=red,dashed,line width=2.0pt]
table {%
0 1
1 1
2 1
3 1
4 1
5 1
6 1
7 1
};
\end{axis}

\end{tikzpicture}
\label{fig:subspace_a}}
\subfloat[]{
\begin{tikzpicture}[scale=0.6]

\begin{axis}[
width=3in,
height=2in,
xmin=0, xmax=7,
ymin=0, ymax=1.1,
axis on top,
xtick={0,1,2,3,4,5,6,7},
xticklabels={8,16,24,32,40,48,56,64}
]
\addplot [color=blue,solid,line width=2.0pt]
table {%
0 1
1 1
2 1
3 1
4 1
5 1
6 0.97
7 0.02
};
\addplot [color=red,dashed,line width=2.0pt]
table {%
0 1
1 1
2 0.98
3 0.75
4 0.29
5 0.01
6 0
7 0
};
\end{axis}

\end{tikzpicture}
\label{fig:subspace_b}}
\\
\subfloat[]{
\begin{tikzpicture}[scale=0.6]

\begin{axis}[
width=3in,
height=2in,
xmin=0, xmax=7,
ymin=0, ymax=1.1,
axis on top,
xtick={0,1,2,3,4,5,6,7},
xticklabels={8,16,24,32,40,48,56,64}
]
\addplot [color=blue,solid,line width=2.0pt]
table {%
0 1
1 1
2 1
3 1
4 1
5 1
6 0.93
7 0.01
};
\addplot [color=red,dashed,line width=2.0pt]
table {%
0 0.97
1 0.66
2 0.1
3 0
4 0
5 0
6 0
7 0
};
\end{axis}

\end{tikzpicture}
\label{fig:subspace_c}}
\subfloat[]{
\begin{tikzpicture}[scale=0.6]

\begin{axis}[
width=3in,
height=2in,
xmin=0, xmax=7,
ymin=0, ymax=1.1,
axis on top,
xtick={0,1,2,3,4,5,6,7},
xticklabels={8,16,24,32,40,48,56,64}
]
\addplot [color=blue,solid,line width=2.0pt]
table {%
0 1
1 1
2 1
3 1
4 1
5 0.98
6 0.28
7 0
};
\addplot [color=red,dashed,line width=2.0pt]
table {%
0 0.14
1 0
2 0
3 0
4 0
5 0
6 0
7 0
};
\end{axis}

\end{tikzpicture}
\label{fig:subspace_d}}
\caption{Subspace case: The empirical success rates of power iteration (blue solid line) and least squares (red dashed line). The $x$-axis represents $m$, and the $y$-axis represents the empirical success rate. (a) -- (d) are the results with $\sigma_W=0$, $0.1$, $0.2$, and $0.5$, respectively.}%
\label{fig:subspace}%
\end{figure}

We test the two approaches at four noise levels: $\sigma_W=0$, $0.1$, $0.2$, and $0.5$, which roughly correspond to MSNR of $\infty$, $20$dB, $14$dB, and $6$dB. At these noise levels, we say the recovery is successful if the RSNR exceeds $30$dB, $20$dB, $14$dB, $6$dB, respectively. The success rates do not change dramatically as functions of these thresholds.
In the experiments, we set $n=128$, $N=16$, and $m= 8, 16, 24, \dots, 64$. For each $m$, we repeat the experiments $100$ times and compute the empirical success rates, which are shown in Figure \ref{fig:subspace}.

As seen in Figure \subref*{fig:subspace_a}, both power iteration and least squares achieve perfect recovery in the noiseless setting. However, as seen in Figures \subref*{fig:subspace_b} --  \subref*{fig:subspace_d}, power iteration is clearly more robust against noise than least squares, whose performance degrades more severely in the noisy settings. 

\begin{figure}[htbp]%
\centering
\subfloat[]{
\begin{tikzpicture}[scale=0.6]

\begin{axis}[
width=3in,
height=2in,
xmin=-0.5, xmax=7.5,
ymin=-0.5, ymax=3.5,
xtick={1,3,5,7},
xticklabels={64,128,192,256},
ytick={0,1,2,3},
yticklabels={64,128,192,256},
ultra thick
]
\addplot[plot graphics/node/.append style={yscale=-1,anchor=north west}] graphics [includegraphics cmd=\pgfimage,xmin=-0.5, xmax=7.5, ymin=3.5, ymax=-0.5] {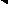};
\end{axis}

\end{tikzpicture}
\label{fig:pi_pt_n}}
\subfloat[]{
\begin{tikzpicture}[scale=0.6]

\begin{axis}[
width=3in,
height=2in,
xmin=-0.5, xmax=15.5,
ymin=-0.5, ymax=3.5,
xtick={3,7,11,15},
xticklabels={16,32,48,64},
ytick={0,1,2,3},
yticklabels={2,3,4,5},
ultra thick
]
\addplot[plot graphics/node/.append style={yscale=-1,anchor=north west}] graphics [includegraphics cmd=\pgfimage,xmin=-0.5, xmax=15.5, ymin=3.5, ymax=-0.5] {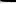};
\end{axis}

\end{tikzpicture}
\label{fig:pi_pt_N}}
\caption{The empirical phase transition of power iteration. Grayscale represents success rates, where white equals 1, and black equals 0. (a) The $x$-axis represents $m$, and the $y$-axis represents $n$. (b) The $x$-axis represents $m$, and the $y$-axis represents $N$.}%
\label{fig:pi_pt}%
\end{figure}

The empirical phase transitions of power iteration are shown in Figure \ref{fig:pi_pt}. We fix $N=16$ and plot the phase transition with respect to $n$ and $m$ (Figure \subref*{fig:pi_pt_n}); we then fix $n=2m$ and plot the phase transition with respect to $N$ and $m$ (Figure \subref*{fig:pi_pt_N}). Clearly, to achieve successful recovery, $n$ must scale linearly with $m$, but $N$ can be small compared to $m$ and $n$. This confirms the sample complexity in Theorem \ref{thm:pi}, of $n\gtrsim m$ and $N\gtrsim 1$. Careful readers may notice in Figure \subref*{fig:pi_pt_N} that for $N = 5$ the success rates at $m < 16$ are worse than those at $m \geq 16$. This seemingly peculiar phenomenon is caused by a small $n = 2m$, which does not belong to the large number regime associated with a high probability.

\subsection{Sparsity Case: Truncated Power Iteration vs. $\ell_1$ Minimization} \label{sec:exp_tpi}
In the sparsity case, we use the same setup described in the previous section, except for the signal $X$. The supports of the $s_0$-sparse columns of $X$ are chosen uniformly at random, and the nonzero entries follow $\calCN(0,\frac{1}{Ns_0})$. This unstructured
sparsity case is more challenging than the joint sparsity case in Theorem \ref{thm:tpi_alt}.

In Algorithm \ref{alg:tpi}, we choose $\alpha = \sqrt{n}$, and $\beta = \norm{B}$. In all the experiments, we assume that the sparsity level $s_0$ is known, and set $s_1 = 2s_0$ for convenience. A more sophisticated scheme that decreases $s_1$ as the iteration number increases may lead to better empirical performance \cite{Yuan2013}. 

For the experiment we suppose that the phases $\{\varphi_k\}_{k=1}^n$ in \eqref{eq:lambda_exp} are available, and let
\begin{equation}
\label{eq:gamma0}
\gamma^{(0)} \coloneqq [e^{-\sqrt{-1}\varphi_1},\dots, e^{-\sqrt{-1}\varphi_n}]^\T
\end{equation}
denote the initial estimate of $\gamma$, which is close to but different from the true $\gamma$, i.e., the entrywise inverse of $\lambda$ in \eqref{eq:lambda_exp}. See Figure \ref{fig:phase} for an illustration of $\lambda_k$, $\gamma_k$, and $\gamma_k^{(0)}$.
Then we initialize Algorithm \ref{alg:tpi} with $\eta^{(0)} = [\bm{0}_{Nm,1}^\T, \gamma^{(0)\T}]^\T$.

We compare Algorithm \ref{alg:tpi} with an $\ell_1$ minimization approach. Wang and Chi \cite{Wang2016} adopted an approach tailored for the case where $A$ is the DFT matrix and $\lambda_k \approx 1$. They use a linear constraint $\sum_{k\in[n]}\gamma_k = n$ to avoid the trivial solution of all zeros. For fair comparison, we revise their approach to accommodate arbitrary $A$ and $\lambda$. The revised approach uses the alternating direction method of multipliers (ADMM) \cite{Boyd2010} to solve the following convex optimization problem: \footnote{In the noisy setting, one could replace the linear constraint $\diag(\gamma) Y = AX$ with an ellipsoid constraint $\norm{\diag(\gamma) Y - AX}_\rmF \leq \epsilon$. However, the parameter $\epsilon$ needs to be adjusted with noise levels. For fair comparison of robustness to noise, we use the linear constrained $\ell_1$ minimization in the noisy setting (similar to \cite{Wang2016}).}
\begin{align*}
\min_{\gamma,X} \quad & \norm{\vect(X)}_1 \\
\text{s.t.} \quad & \diag(\gamma) Y = AX, \\
& \gamma^{(0)*} \gamma = n.
\end{align*}
Here, $\gamma^{(0)}$ is the initial estimate of $\gamma$ defined in \eqref{eq:gamma0}, and used as initialization in our Algorithm \ref{alg:tpi} in this comparison.

\begin{figure}[htbp]%
\centering
\subfloat[]{
\begin{tikzpicture}[scale=0.6]

\begin{axis}[
width=3in,
height=2in,
xmin=0, xmax=7,
ymin=0, ymax=1.1,
axis on top,
xtick={0,1,2,3,4,5,6,7},
xticklabels={8,16,24,32,40,48,56,64}
]
\addplot [color=blue,solid,line width=2.0pt]
table {%
0 1
1 1
2 0.98
3 0.8
4 0
5 0
6 0
7 0
};
\addplot [color=red,dashed,line width=2.0pt]
table {%
0 1
1 1
2 1
3 1
4 0.17
5 0
6 0
7 0
};
\end{axis}

\end{tikzpicture}
\label{fig:sparsity_a}}
\subfloat[]{
\begin{tikzpicture}[scale=0.6]

\begin{axis}[
width=3in,
height=2in,
xmin=0, xmax=7,
ymin=0, ymax=1.1,
axis on top,
xtick={0,1,2,3,4,5,6,7},
xticklabels={8,16,24,32,40,48,56,64}
]
\addplot [color=blue,solid,line width=2.0pt]
table {%
0 1
1 1
2 0.06
3 0
4 0
5 0
6 0
7 0
};
\addplot [color=red,dashed,line width=2.0pt]
table {%
0 0.97
1 0.45
2 0
3 0
4 0
5 0
6 0
7 0
};
\end{axis}

\end{tikzpicture}
\label{fig:sparsity_b}}
\\
\subfloat[]{
\begin{tikzpicture}[scale=0.6]

\begin{axis}[
width=3in,
height=2in,
xmin=0, xmax=7,
ymin=0, ymax=1.1,
axis on top,
xtick={0,1,2,3,4,5,6,7},
xticklabels={8,16,24,32,40,48,56,64}
]
\addplot [color=blue,solid,line width=2.0pt]
table {%
0 1
1 1
2 0
3 0
4 0
5 0
6 0
7 0
};
\addplot [color=red,dashed,line width=2.0pt]
table {%
0 0.04
1 0
2 0
3 0
4 0
5 0
6 0
7 0
};
\end{axis}

\end{tikzpicture}
\label{fig:sparsity_c}}
\subfloat[]{
\begin{tikzpicture}[scale=0.6]

\begin{axis}[
width=3in,
height=2in,
xmin=0, xmax=7,
ymin=0, ymax=1.1,
axis on top,
xtick={0,1,2,3,4,5,6,7},
xticklabels={8,16,24,32,40,48,56,64}
]
\addplot [color=blue,solid,line width=2.0pt]
table {%
0 1
1 1
2 0
3 0
4 0
5 0
6 0
7 0
};
\addplot [color=red,dashed,line width=2.0pt]
table {%
0 0.04
1 0
2 0
3 0
4 0
5 0
6 0
7 0
};
\end{axis}

\end{tikzpicture}
\label{fig:sparsity_d}}
\caption{Sparsity case: The empirical success rates of truncated power iteration (blue solid line) and $\ell_1$ minimization (red dashed line). The $x$-axis represents $s_0$, and the $y$-axis represents the empirical success rate. (a) -- (d) are the results with $\sigma_W=0$, $0.1$, $0.2$, and $0.5$, respectively.}%
\label{fig:sparsity}%
\end{figure}
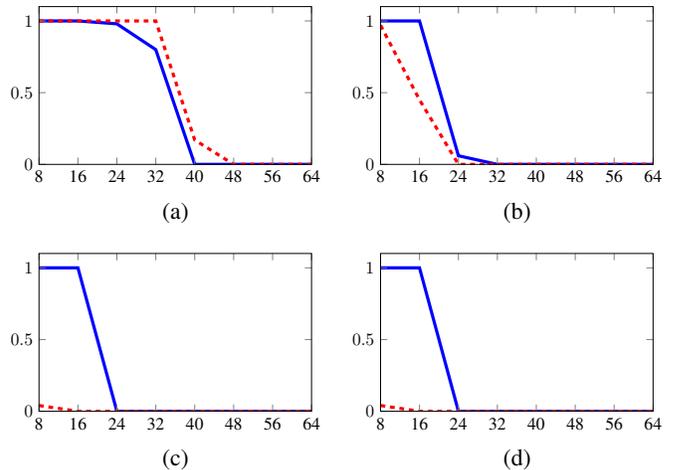

We conduct numerical experiments with the same four noise levels and criterion for successful recovery as in Section \ref{sec:exp_pi}. In the experiments, we set $n=128$, $m=256$, $N=16$, and $s_0 = 8, 16, 24, \dots, 64$. For each $s_0$, we repeat the experiments $100$ times and compute the empirical success rates, which are shown in Figure \ref{fig:sparsity}. In the noiseless case (Figure \subref*{fig:sparsity_a}), $\ell_1$ minimization achieves a slightly higher success rate near the phase transition. However, truncated power iteration is more robust against noise than $\ell_1$ minimization, which breaks down completely at the higher noise levels (Figures. \subref*{fig:sparsity_b} -- \subref*{fig:sparsity_d}).

Figure \subref*{fig:sparsity_a} clearly shows that truncated power iteration recovers $\eta$ successfully when $n=128$, $N=16$, and $s_0 = 32$. This suggests that truncated power iteration may succeed when $n$ and $N$ are (up to log factors) on the order of $s_0$ and $1$, respectively. However, while the scaling with the number of sensors $n$ agrees with Theorem \ref{thm:tpi_alt}, success with such small number of snapshots $N$ is not guaranteed by our current theoretical analysis.

\begin{figure}[htbp]%
\centering
\subfloat[]{
\begin{tikzpicture}[scale=0.6]

\begin{axis}[
width=3in,
height=2in,
xmin=0, xmax=7,
ymin=0, ymax=1.1,
axis on top,
xtick={0,1,2,3,4,5,6,7},
xticklabels={4,8,12,16,20,24,28,32}
]
\addplot [color=blue,solid,line width=2.0pt]
table {%
0 1
1 1
2 1
3 1
4 0.95
5 0.9
6 0.38
7 0
};
\addplot [color=red,dashed,line width=2.0pt]
table {%
0 1
1 1
2 1
3 1
4 1
5 0.9
6 0.37
7 0.01
};
\end{axis}

\end{tikzpicture}
\label{fig:wp_a}}
\subfloat[]{
\begin{tikzpicture}[scale=0.6]

\begin{axis}[
width=3in,
height=2in,
xmin=0, xmax=7,
ymin=0, ymax=1.1,
axis on top,
xtick={0,1,2,3,4,5,6,7},
xticklabels={4,8,12,16,20,24,28,32}
]
\addplot [color=blue,solid,line width=2.0pt]
table {%
0 0.94
1 0.98
2 0.91
3 0.78
4 0.48
5 0.06
6 0
7 0
};
\addplot [color=red,dashed,line width=2.0pt]
table {%
0 0.94
1 0.8
2 0.42
3 0.1
4 0.01
5 0
6 0
7 0
};
\end{axis}

\end{tikzpicture}
\label{fig:wp_b}}
\caption{Sparsity case: The empirical success rates of truncated power iteration (blue solid line) and $\ell_1$ minimization (red dashed line), with bad initial estimate of the phases. The $x$-axis represents $s_0$, and the $y$-axis represents the empirical success rate. (a) and (b) are the results for which $1/2$ and $3/4$ of $\{\varphi_k\}_{k=1}^n$ are initialized with random phases.}%
\label{fig:wrong_phase}%
\end{figure}
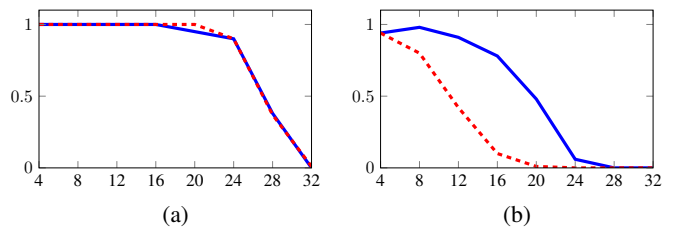

Next, we assume that only a subset of the phases $\{\varphi_k\}_{k=1}^n$ are available, and examine to what extent Algorithm \ref{alg:tpi} and $\ell_1$ minimization depend on a good initial estimate of $\gamma$. In the numerical results shown in Figure \ref{fig:wrong_phase}, we consider only the noiseless setting of BGPC with sparsity, and set $s_0 = 4, 8, 12, \dots, 32$. In Figures \subref*{fig:wp_a} and \subref*{fig:wp_b}, we replace $1/2$ and $3/4$ of $\{\varphi_k\}_{k=1}^n$ with random phases, respectively, and use the resulting bad estimate $\gamma^{(0)}$ in Algorithm \ref{alg:tpi} and $\ell_1$ minimization. As seen in Figure \ref{fig:wrong_phase}, truncated power iteration is less dependent on accurate initial estimate of $\gamma$.

We repeat the above experiments for the joint sparsity case, where we replace $\widetilde{\Pi}_{s_1}$ in Algorithm \ref{alg:tpi} with $\widetilde{\Pi}'_{s_1}$. We also replace the $\ell_1$ norm $\norm{\vect(X)}_1$ in the competing approach with a mixed norm:
\[
\norm{X}_{2,1} = \sum_{\ell\in[m]} \Bigl( \sum_{j\in[N]} |x_{\ell j}|^2 \Bigr)^{1/2},
\]
which is a well known convex method for the recovery of jointly sparse signals. The results for different noise levels and for inaccurate $\gamma^{(0)}$ are shown in Figures \ref{fig:jsparsity} and \ref{fig:js_wrong_phase}, respectively. In the joint sparsity case, truncated power iteration is robust against noise, but seems less robust against errors in the initial phase estimate. We conjecture that the failure of Algorithm \ref{alg:tpi} in the joint sparsity case is due to the restriction of $\widetilde{\Pi}'_{s_1}$. By projecting onto jointly sparse supports, the algorithm is likely to converge prematurely to an incorrect support. When compared to the results in Figures \subref*{fig:jwp_a} and \subref*{fig:jwp_b}, Figures \subref*{fig:jwp_c} and \subref*{fig:jwp_d} show that using $\widetilde{\Pi}_{s_1}$ instead of $\widetilde{\Pi}'_{s_1}$ in the first half of the iterations indeed improves the performance of Algorithm \ref{alg:tpi} in the joint sparsity case. In the rest of the experiments, we use $\widetilde{\Pi}_{s_1}$ during the first half of the iterations in Algorithm \ref{alg:tpi} for the joint sparsity case.

\begin{figure}[htbp]%
\centering
\subfloat[]{
\begin{tikzpicture}[scale=0.6]

\begin{axis}[
width=3in,
height=2in,
xmin=0, xmax=7,
ymin=0, ymax=1.1,
axis on top,
xtick={0,1,2,3,4,5,6,7},
xticklabels={8,16,24,32,40,48,56,64}
]
\addplot [color=blue,solid,line width=2.0pt]
table {%
0 1
1 1
2 1
3 1
4 0.95
5 0.92
6 0
7 0
};
\addplot [color=red,dashed,line width=2.0pt]
table {%
0 1
1 1
2 1
3 1
4 1
5 1
6 1
7 0.99
};
\end{axis}

\end{tikzpicture}
\label{fig:jsparsity_a}}
\subfloat[]{
\begin{tikzpicture}[scale=0.6]

\begin{axis}[
width=3in,
height=2in,
xmin=0, xmax=7,
ymin=0, ymax=1.1,
axis on top,
xtick={0,1,2,3,4,5,6,7},
xticklabels={8,16,24,32,40,48,56,64}
]
\addplot [color=blue,solid,line width=2.0pt]
table {%
0 1
1 1
2 1
3 0.02
4 0
5 0
6 0
7 0
};
\addplot [color=red,dashed,line width=2.0pt]
table {%
0 0.41
1 0.01
2 0
3 0
4 0
5 0
6 0
7 0
};
\end{axis}

\end{tikzpicture}
\label{fig:jsparsity_b}}
\\
\subfloat[]{
\begin{tikzpicture}[scale=0.6]

\begin{axis}[
width=3in,
height=2in,
xmin=0, xmax=7,
ymin=0, ymax=1.1,
axis on top,
xtick={0,1,2,3,4,5,6,7},
xticklabels={8,16,24,32,40,48,56,64}
]
\addplot [color=blue,solid,line width=2.0pt]
table {%
0 1
1 1
2 1
3 0
4 0
5 0
6 0
7 0
};
\addplot [color=red,dashed,line width=2.0pt]
table {%
0 0.28
1 0
2 0
3 0
4 0
5 0
6 0
7 0
};
\end{axis}

\end{tikzpicture}
\label{fig:jsparsity_c}}
\subfloat[]{
\begin{tikzpicture}[scale=0.6]

\begin{axis}[
width=3in,
height=2in,
xmin=0, xmax=7,
ymin=0, ymax=1.1,
axis on top,
xtick={0,1,2,3,4,5,6,7},
xticklabels={8,16,24,32,40,48,56,64}
]
\addplot [color=blue,solid,line width=2.0pt]
table {%
0 1
1 1
2 0.93
3 0
4 0
5 0
6 0
7 0
};
\addplot [color=red,dashed,line width=2.0pt]
table {%
0 0.36
1 0.01
2 0
3 0
4 0
5 0
6 0
7 0
};
\end{axis}

\end{tikzpicture}
\label{fig:jsparsity_d}}
\caption{Joint sparsity case: The empirical success rates of truncated power iteration (blue solid line) and mixed minimization (red dashed line). The $x$-axis represents $s_0$, and the $y$-axis represents the empirical success rate. (a) -- (d) are the results with $\sigma_W=0$, $0.1$, $0.2$, and $0.5$, respectively.}%
\label{fig:jsparsity}%
\end{figure}
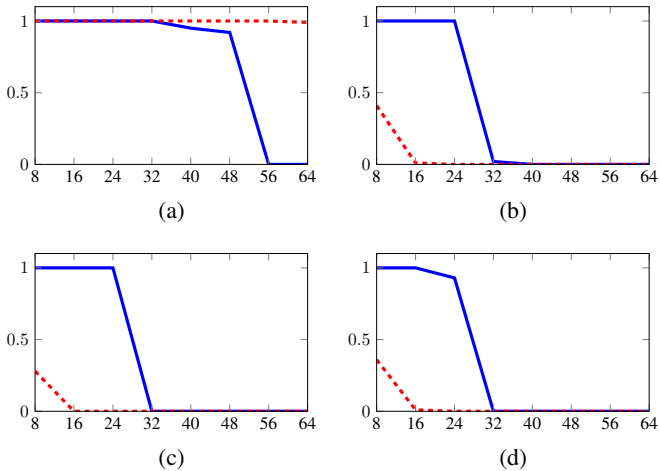

\begin{figure}[htbp]%
\centering
\subfloat[]{
\begin{tikzpicture}[scale=0.6]

\begin{axis}[
width=3in,
height=2in,
xmin=0, xmax=7,
ymin=0, ymax=1.1,
axis on top,
xtick={0,1,2,3,4,5,6,7},
xticklabels={4,8,12,16,20,24,28,32}
]
\addplot [color=blue,solid,line width=2.0pt]
table {%
0 1
1 1
2 0.97
3 0.91
4 0.71
5 0.49
6 0.35
7 0.09
};
\addplot [color=red,dashed,line width=2.0pt]
table {%
0 1
1 1
2 1
3 1
4 1
5 0.99
6 0.93
7 0.57
};
\end{axis}

\end{tikzpicture}
\label{fig:jwp_a}}
\subfloat[]{
\begin{tikzpicture}[scale=0.6]

\begin{axis}[
width=3in,
height=2in,
xmin=0, xmax=7,
ymin=0, ymax=1.1,
axis on top,
xtick={0,1,2,3,4,5,6,7},
xticklabels={4,8,12,16,20,24,28,32}
]
\addplot [color=blue,solid,line width=2.0pt]
table {%
0 0.37
1 0.27
2 0.07
3 0
4 0
5 0
6 0
7 0
};
\addplot [color=red,dashed,line width=2.0pt]
table {%
0 0.91
1 0.68
2 0.38
3 0.08
4 0.02
5 0
6 0
7 0
};
\end{axis}

\end{tikzpicture}
\label{fig:jwp_b}}
\\
\subfloat[]{
\begin{tikzpicture}[scale=0.6]

\begin{axis}[
width=3in,
height=2in,
xmin=0, xmax=7,
ymin=0, ymax=1.1,
axis on top,
xtick={0,1,2,3,4,5,6,7},
xticklabels={4,8,12,16,20,24,28,32}
]
\addplot [blue,solid,line width=2.0pt]
table {%
0 1
1 1
2 1
3 1
4 1
5 1
6 0.98
7 0.94
};
\addplot [red,dashed,line width=2.0pt]
table {%
0 1
1 1
2 1
3 1
4 1
5 0.99
6 0.93
7 0.57
};
\end{axis}

\end{tikzpicture}
\label{fig:jwp_c}}
\subfloat[]{
\begin{tikzpicture}[scale=0.6]

\begin{axis}[
width=3in,
height=2in,
xmin=0, xmax=7,
ymin=0, ymax=1.1,
axis on top,
xtick={0,1,2,3,4,5,6,7},
xticklabels={4,8,12,16,20,24,28,32}
]
\addplot [blue,solid,line width=2.0pt]
table {%
0 0.77
1 0.8
2 0.77
3 0.53
4 0.39
5 0.11
6 0.06
7 0
};
\addplot [red,dashed,line width=2.0pt]
table {%
0 0.91
1 0.68
2 0.38
3 0.08
4 0.02
5 0
6 0
7 0
};
\end{axis}

\end{tikzpicture}
\label{fig:jwp_d}}
\caption{Joint sparsity case: The empirical success rates of truncated power iteration with $\widetilde{\Pi}'_{s_1}$ (blue solid line) and mixed minimization (red dashed line), with bad initial estimate of the phases. The $x$-axis represents $s_0$, and the $y$-axis represents the empirical success rate. (a) and (b) are the results for which $1/2$ and $3/4$ of $\{\varphi_k\}_{k=1}^n$ are initialized with random phases. In (c) and (d), we repeat the experiments, but use $\widetilde{\Pi}_{s_1}$ instead of $\widetilde{\Pi}'_{s_1}$ in the first half of the iterations.}%
\label{fig:js_wrong_phase}%
\end{figure}

\begin{figure}[htbp]%
\centering
\subfloat[]{
\begin{tikzpicture}[scale=0.6]

\begin{axis}[
width=3in,
height=2in,
xmin=-0.5, xmax=9.5,
ymin=-0.5, ymax=3.5,
xtick={1,3,5,7,9},
xticklabels={16,32,48,64,80},
ytick={0,1,2,3},
yticklabels={64,128,192,256},
ultra thick
]
\addplot[plot graphics/node/.append style={yscale=-1,anchor=north west}] graphics [includegraphics cmd=\pgfimage,xmin=-0.5, xmax=9.5, ymin=3.5, ymax=-0.5] {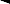};
\end{axis}

\end{tikzpicture}
\label{fig:tpi_pt_n}}
\subfloat[]{
\begin{tikzpicture}[scale=0.6]

\begin{axis}[
width=3in,
height=2in,
xmin=-0.5, xmax=15.5,
ymin=-0.5, ymax=3.5,
xtick={3,7,11,15},
xticklabels={32,64,96,128},
ytick={0,1,2,3},
yticklabels={2,4,6,8},
ultra thick
]
\addplot[plot graphics/node/.append style={yscale=-1,anchor=north west}] graphics [includegraphics cmd=\pgfimage,xmin=-0.5, xmax=15.5, ymin=3.5, ymax=-0.5] {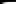};
\end{axis}

\end{tikzpicture}
\label{fig:tpi_pt_N}} \\
\subfloat[]{
\begin{tikzpicture}[scale=0.6]

\begin{axis}[
width=3in,
height=2in,
xmin=-0.5, xmax=7.5,
ymin=-0.5, ymax=3.5,
xtick={1,3,5,7},
xticklabels={32,64,96,128},
ytick={0,1,2,3},
yticklabels={64,128,192,256},
ultra thick
]
\addplot[plot graphics/node/.append style={yscale=-1,anchor=north west}] graphics [includegraphics cmd=\pgfimage,xmin=-0.5, xmax=7.5, ymin=3.5, ymax=-0.5] {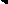};
\end{axis}

\end{tikzpicture}
\label{fig:js_tpi_pt_n}}
\subfloat[]{
\begin{tikzpicture}[scale=0.6]

\begin{axis}[
width=3in,
height=2in,
xmin=-0.5, xmax=15.5,
ymin=-0.5, ymax=3.5,
xtick={3,7,11,15},
xticklabels={32,64,96,128},
ytick={0,1,2,3},
yticklabels={2,4,6,8},
ultra thick
]
\addplot[plot graphics/node/.append style={yscale=-1,anchor=north west}] graphics [includegraphics cmd=\pgfimage,xmin=-0.5, xmax=15.5, ymin=3.5, ymax=-0.5] {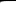};
\end{axis}

\end{tikzpicture}
\label{fig:js_tpi_pt_N}}
\caption{The empirical phase transition of truncated power iteration. Grayscale represents success rates, where white equals 1, and black equals 0.
(a) Sparsity case: The $x$-axis represents $s_0$, and the $y$-axis represents $n$. (b) Sparsity case: The $x$-axis represents $s_0$, and the $y$-axis represents $N$.
(c) Joint sparsity case: The $x$-axis represents $s_0$, and the $y$-axis represents $n$. (d) Joint sparsity case: The $x$-axis represents $s_0$, and the $y$-axis represents $N$.}%
\label{fig:tpi_pt}%
\end{figure}

Next, we plot the phase transitions for truncated power iteration. We fix $N=16$ and $m=2n$ and plot the empirical phase transition with respect to $n$ and $s_0$ (sparsity case in Figure \subref*{fig:tpi_pt_n}, and joint sparsity case in Figure \subref*{fig:js_tpi_pt_n}); we then fix $n=4s_0$ and $m=2n$ and plot the empirical phase transition with respect to $N$ and $s_0$ (sparsity case in Figure \subref*{fig:tpi_pt_N}, and joint sparsity case in Figure \subref*{fig:js_tpi_pt_N}). It is seen that, to achieve successful recovery, $n$ must scale linearly with $s_0$, but $N$ can be small compared to $s_0$ and $n$. On the one hand, the scaling law $n\gtrsim s_0$ in Theorem \ref{thm:tpi_alt} is confirmed by Figure \ref{fig:tpi_pt}; on the other hand, $N\gtrsim \sqrt{s_0}$ seems conservative and might be an artifact of our proof techniques. We have yet to come up with a theoretical guarantee that covers the more general sparsity case, or requires a less demanding sample complexity $N\gtrsim 1$. In Figures \subref*{fig:tpi_pt_N} and \subref*{fig:js_tpi_pt_N}, the success rates at smaller $s_0$ are lower than those at a larger $s_0$, because the number of sensors $n = 4s_0$ is too small to yield a high probability.

\subsection{Sparsity Case: Initialization} \label{sec:exp_init}

In this section, we examine the quality of the initialization produced by Algorithm \ref{alg:init} by comparing it with two different initializations: (i) the good initialization $\eta^{(0)} = [\bm{0}_{Nm,1}^\T, \gamma^{(0)\T}]^\T$ aided by side information on the phase in Section \ref{sec:exp_tpi}; and (ii) a baseline initialization $\eta^{(0)} = [\bm{0}_{Nm,1}^\T, \bm{1}_{n,1}^\T]^\T$. We use the same setting as in Section \ref{sec:exp_tpi}, except that $N=32$. We let $\sigma_W=0.1$, and claim the recovery is successful if the RSNR exceeds $20$dB. 
In the experiment for the joint sparsity case, for the reason mentioned in Section \ref{sec:exp_tpi}, we ignore the joint sparsity structure and estimate the support of different columns of $X$ independently in the initialization and during the first half of the iterations. Only in the second half of the iterations, we use the projection $\widetilde{\Pi}'_{s_1}$ onto jointly sparse supports.

\begin{figure}[htbp]%
\centering
\subfloat[]{
\begin{tikzpicture}[scale=0.6]

\begin{axis}[
width=3in,
height=2in,
xmin=0, xmax=7,
ymin=0, ymax=1.1,
axis on top,
xtick={0,1,2,3,4,5,6,7},
xticklabels={2,4,6,8,10,12,14,16}
]
\addplot [blue,solid,line width=2.0pt]
table {%
0 0.99
1 1
2 0.99
3 0.95
4 0.96
5 0.88
6 0.69
7 0.36
8 0.17
9 0.02
10 0
11 0
12 0
13 0
14 0
15 0
};
\addplot [red,dashed,line width=2.0pt]
table {%
0 0.3
1 0.28
2 0.22
3 0.14
4 0.04
5 0.02
6 0.02
7 0
8 0
9 0
10 0
11 0
12 0
13 0
14 0
15 0
};
\addplot [black, dash pattern=on 1pt off 3pt on 3pt off 3pt, line width=2.0pt]
table {%
0 1
1 1
2 1
3 1
4 1
5 1
6 1
7 0.99
8 1
9 0.97
10 0.94
11 0.72
12 0.02
13 0
14 0
15 0
};
\end{axis}

\end{tikzpicture}
\label{fig:init_s}}
\subfloat[]{
\begin{tikzpicture}[scale=0.6]

\begin{axis}[
width=3in,
height=2in,
xmin=0, xmax=7,
ymin=0, ymax=1.1,
axis on top,
xtick={0,1,2,3,4,5,6,7},
xticklabels={2,4,6,8,10,12,14,16}
]
\addplot [blue,solid,line width=2.0pt]
table {%
0 1
1 0.99
2 0.97
3 0.92
4 0.85
5 0.83
6 0.7
7 0.59
8 0.33
9 0.26
10 0.08
11 0.05
12 0.02
13 0
14 0
15 0
};
\addplot [red,dashed,line width=2.0pt]
table {%
0 0.03
1 0.01
2 0.03
3 0.01
4 0.02
5 0.02
6 0
7 0.01
8 0
9 0
10 0
11 0
12 0
13 0
14 0
15 0
};
\addplot [black, dash pattern=on 1pt off 3pt on 3pt off 3pt, line width=2.0pt]
table {%
0 1
1 1
2 1
3 1
4 1
5 1
6 1
7 1
8 1
9 1
10 1
11 1
12 1
13 0.99
14 0.47
15 0
};
\end{axis}

\end{tikzpicture}
\label{fig:init_js}}
\caption{The empirical success rates of truncated power iteration with the initialization in Algorithm \ref{alg:init} (blue solid line), with a baseline initialization $\eta^{(0)} = [\bm{0}_{Nm,1}^\T, \bm{1}_{n,1}^\T]^\T$ (red dashed line), and with the accurate initialization $\eta^{(0)} = [\bm{0}_{Nm,1}^\T, \gamma^{(0)\T}]^\T$ with side information in Section \ref{sec:exp_tpi} (black dash-dot line). The $x$-axis represents $s_0$, and the $y$-axis represents the empirical success rate. (a) is the result for the sparsity case, and (b) is the result for the joint sparsity case.}%
\label{fig:init}%
\end{figure}
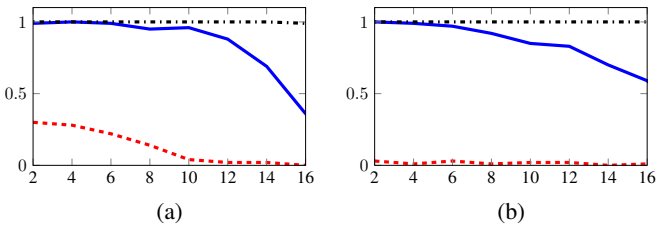

\begin{figure}[htbp]%
\centering
\subfloat[]{
\begin{tikzpicture}[scale=0.6]

\begin{axis}[
width=3in,
height=3in,
xmin=-0.5, xmax=15.5,
ymin=-0.5, ymax=7.5,
xtick={3,7,11,15},
xticklabels={8,16,24,32},
ytick={1,3,5,7},
yticklabels={64,128,192,256},
ultra thick
]
\addplot[plot graphics/node/.append style={yscale=-1,anchor=north west}] graphics [includegraphics cmd=\pgfimage,xmin=-0.5, xmax=15.5, ymin=7.5, ymax=-0.5] {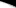};
\end{axis}

\end{tikzpicture}
\label{fig:tpi_pt_init_s}}
\subfloat[]{
\begin{tikzpicture}[scale=0.6]

\begin{axis}[
width=3in,
height=3in,
xmin=-0.5, xmax=15.5,
ymin=-0.5, ymax=7.5,
xtick={3,7,11,15},
xticklabels={8,16,24,32},
ytick={1,3,5,7},
yticklabels={64,128,192,256},
ultra thick
]
\addplot[plot graphics/node/.append style={yscale=-1,anchor=north west}] graphics [includegraphics cmd=\pgfimage,xmin=-0.5, xmax=15.5, ymin=7.5, ymax=-0.5] {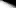};
\end{axis}

\end{tikzpicture}
\label{fig:tpi_pt_init_js}}
\caption{The empirical phase transition of truncated power iteration with the initialization in Algorithm \ref{alg:init}. The $x$-axis represents $s_0$, and the $y$-axis represents $n$. (a) is the result for the sparsity case, and (b) is the result for the joint sparsity case.}%
\label{fig:pt_init}%
\end{figure}

Figure \ref{fig:init} shows that, although the initialization provided by Algorithm \ref{alg:init} is not as good as the accurate initialization with side information, it is far better than the baseline. Figure \ref{fig:pt_init} shows the empirical phase transition with respect to $n$ and $s_0$, when Algorithm \ref{alg:init} is used to initialize truncated power iteration (sparsity case in Figure \subref*{fig:tpi_pt_init_s}, and joint sparsity case in Figure \subref*{fig:tpi_pt_init_js}). The results suggest that when $n$ scales linearly with $s_0$, Algorithm \ref{alg:init} can provide a sufficiently good initialization for truncated power iteration. For example, in \subref*{fig:tpi_pt_init_s}, the success rate is 1 when $n=256$ and $s_0=20$. Therefore, the sample complexity $n\gtrsim s_0^2$ in Theorem \ref{thm:init} could be overly conservative and an artifact of our analysis.

\subsection{Application: Inverse Rendering} \label{sec:inverse_render}

\begin{figure}[htbp]%
\centering
\subfloat[]{\includegraphics[width=0.9\columnwidth]{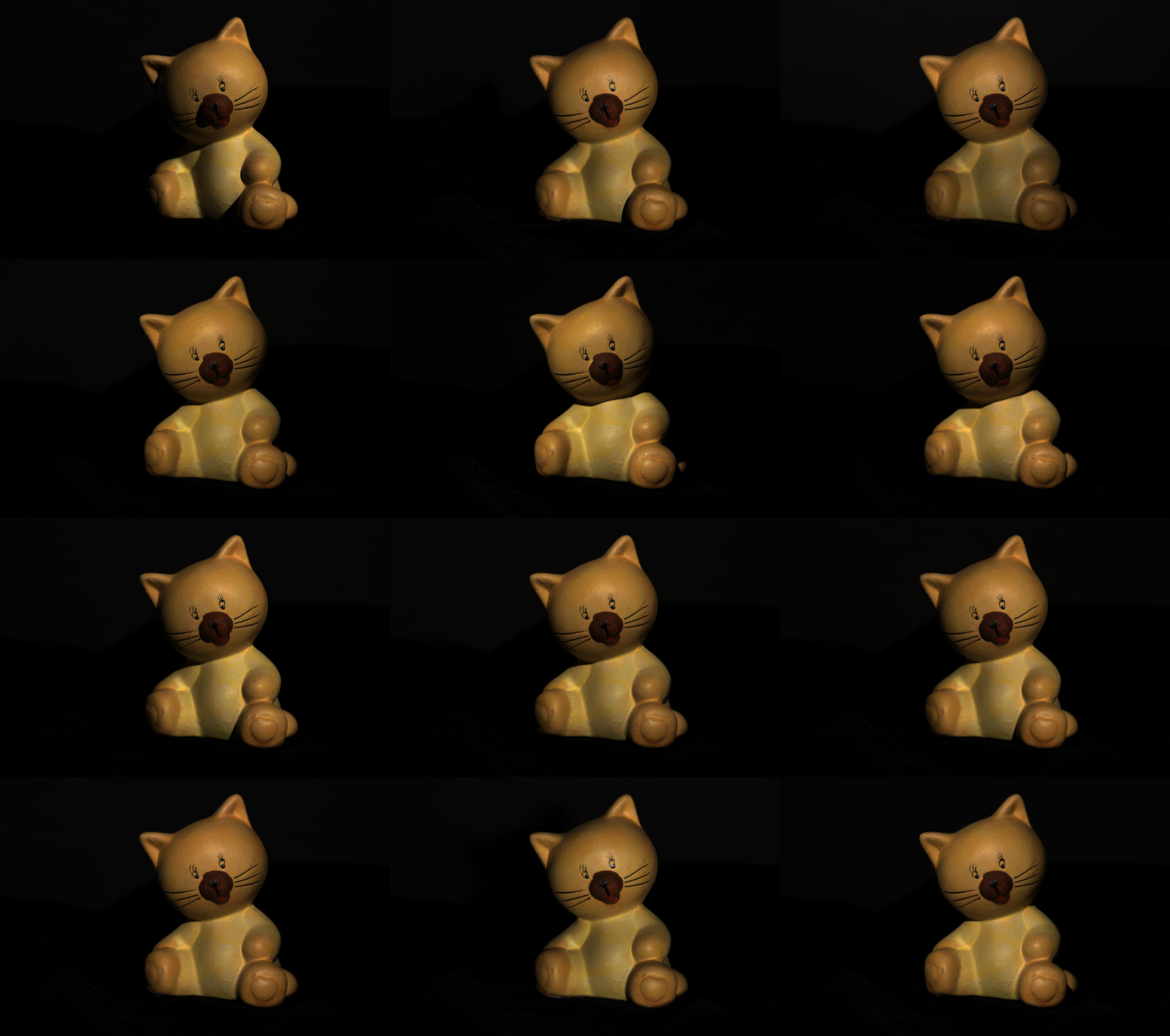}
\label{fig:cat_i}} \\
\subfloat[]{\includegraphics[width=0.4\columnwidth]{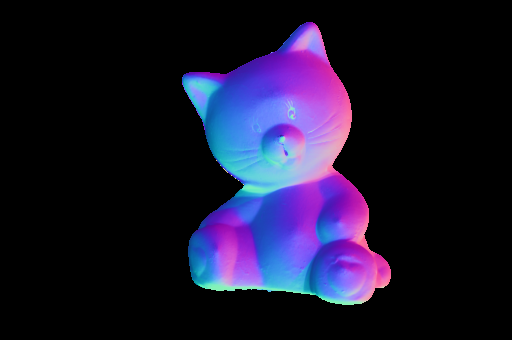}
\label{fig:cat_n}}
\subfloat[]{\includegraphics[width=0.4\columnwidth]{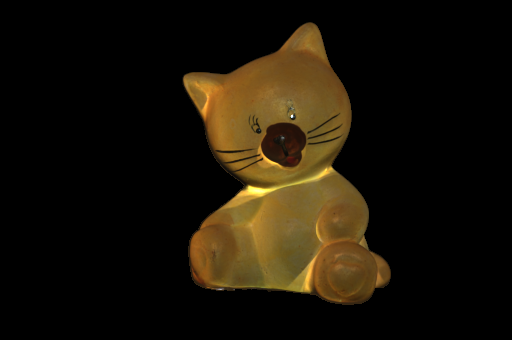}
\label{fig:cat_a}} \\
\subfloat[]{\includegraphics[width=0.9\columnwidth]{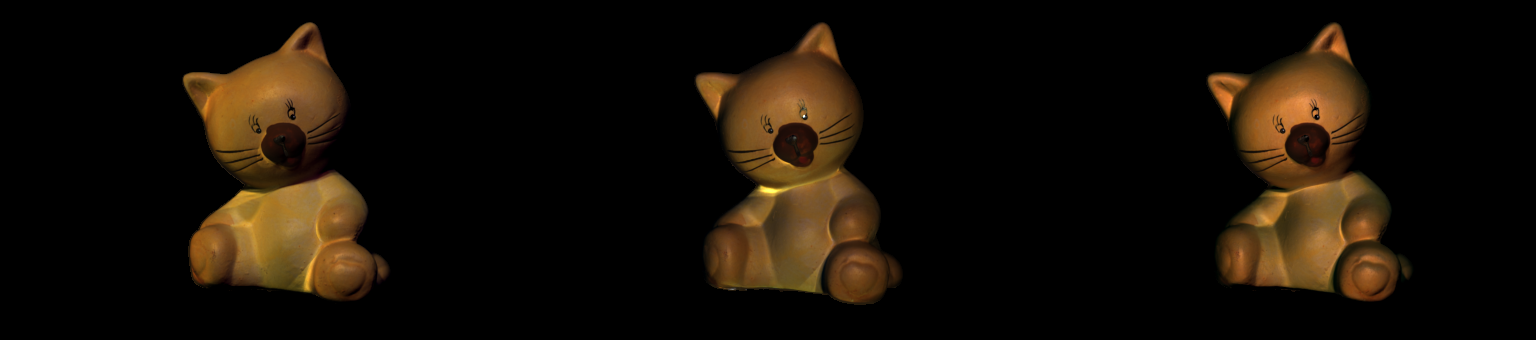}
\label{fig:cat_new}}
\caption{Inverse rendering and relighting. (a) We use 12 images of the object under different lighting conditions. (b) The surface normals. The three dimensions of the normal vectors are represented by the RGB channels of the color image. (c) The recovered albedo map. (d) Computed images of the object under new lighting conditions.}%
\label{fig:cat}%
\end{figure}

In this section, we apply the power iteration algorithm to the inverse rendering problem in computational relighting -- given images of an object under different lighting conditions (Figure \subref*{fig:cat_i}), and the surface normals of the object (Figure \subref*{fig:cat_n}), the goal is to recover the albedos (also known as reflection coefficients) of the object surface and the lighting conditions. In this problem, the columns of $Y=\diag(\lambda)AX \in \bbR^{n\times N}$ represent images under different lighting conditions, which are the products of the unknown albedo map $\lambda \in\bbR^n$ and the intensity maps of incident light under different conditions $AX$. For Lambertian surfaces, it is reasonable to assume that the intensity of incident light resides in a subspace spanned by the first $9$ spherical harmonics computed from the surface normals \cite{Nguyen2013}, which we denote by the columns of $A \in \bbR^{n\times 9}$. Then the columns of $X$ are the coordinates of the spherical harmonic expansion, which parameterize the lighting conditions. 
We can solve for $\lambda$ and $X$ using Algorithm \ref{alg:pi}. Our approach is similar to that of Nguyen et al. \cite{Nguyen2013}, which also formulates inverse rendering as an eigenvector problem. Despite the fact that the two approaches solve for the eigenvectors of different matrices, they yield identical solutions in the ideal scenario where the model is exact and the solution is unique.

In our experiment, we obtain $N=12$ color images and the surface normals of an object under different lighting conditions,\footnote{The images are downloaded from \emph{https://courses.cs.washington. edu/courses/csep576/05wi/projects/project3/project3.htm} on September 16, 2017. The surface normals are computed using the method described in the same webpage.} and we compute the first $m=9$ spherical harmonics. We apply Algorithm \ref{alg:pi} to each of the three color channels, and the albedo map recovered using 200 power iterations is shown in Figure \subref*{fig:cat_a}. We also compute new images of the object under new lighting conditions (Figure \subref*{fig:cat_new}).

\section{Conclusion} \label{sec:conclusion}

We formulate the BGPC problem as an eigenvector problem, and propose to solve BGPC with power iteration, and solve BGPC with a sparsity structure with truncated power iteration. We give theoretical guarantees for the subspace case with a near optimal sample complexity, and for the joint sparsity case with a suboptimal sample complexity. 
Numerical experiments show that both power iteration and truncated power iteration can recover the unknown gain and phase, and the unknown signal, using a near optimal number of samples. 
It is an open problem to obtain theoretical guarantees with optimal sample complexities, for truncated power iteration that solves BGPC with joint sparsity or sparsity constraints.


\appendix

\begin{IEEEproof}[Proof of Lemma \ref{lem:expB}]
We have
\begin{equation}
\label{eq:DstarD}
D^*D = I_N\otimes (A^*A),
\end{equation}
\begin{equation}
\label{eq:DstarEs}
D^*E_\rms = \begin{bmatrix}
\lambda_1 \overline{a_{1\cdot}} a_{1\cdot}^\T x_{\cdot 1} & \cdots & \lambda_n \overline{a_{n\cdot}} a_{n\cdot}^\T x_{\cdot 1} \\
\vdots & \ddots & \vdots \\
\lambda_1 \overline{a_{1\cdot}} a_{1\cdot}^\T x_{\cdot N} & \cdots & \lambda_n \overline{a_{n\cdot}} a_{n\cdot}^\T x_{\cdot N}
\end{bmatrix},
\end{equation}
\begin{equation}
\label{eq:EsstarEs}
E_\rms^*E_\rms = \begin{bmatrix}
|\lambda_1|^2 a_{1\cdot}^\T XX^* \overline{a_{1\cdot}} & & \\
& \ddots & \\
& & |\lambda_n|^2 a_{n\cdot}^\T XX^* \overline{a_{n\cdot}}
\end{bmatrix}.
\end{equation}
Under Assumption \ref{ass:A} and \ref{ass:X}, we have
\begin{equation}
\label{eq:EDstarD}
\bbE D^*D = I_{Nm},
\end{equation}
\begin{equation}
\label{eq:EDstarEs}
\bbE D^*E_\rms = \frac{1}{n} x \lambda^\T,
\end{equation}
\begin{align}
\bbE E_\rms^*E_\rms = \frac{1}{n}\norm{X}_\rmF^2 \diag([|\lambda_1|^2, \dots, |\lambda_n|^2]) \nonumber \\
= \frac{1}{n} \diag([|\lambda_1|^2, \dots, |\lambda_n|^2]). \label{eq:EEsstarEs}
\end{align}
Set $\alpha = \sqrt{n}$, we have
\[
\bbE B_\rms = \begin{bmatrix}
I_{Nm} & \frac{1}{\sqrt{n}} x \lambda^\T \\
\frac{1}{\sqrt{n}}  \overline{\lambda}x^*  & \diag([|\lambda_1|^2, \dots, |\lambda_n|^2])
\end{bmatrix},
\]
and
\begin{align*}
& \bbE \Omega_{T_\eta} B_\rms \Omega_{T_\eta}^* \\
& = \begin{bmatrix}
I_{Ns} & \frac{1}{\sqrt{n}} \Omega_{T_x} x \lambda^\T \\
\frac{1}{\sqrt{n}}  \overline{\lambda}x^* \Omega_{T_x}^*  & \diag([|\lambda_1|^2, \dots, |\lambda_n|^2])
\end{bmatrix} \\
& = P^*QP,
\end{align*}
where
\[
P = \diag([\bm{1}_{1,Ns},\lambda^\T]),
\]
\[
Q = \begin{bmatrix}
I_{Ns} & \frac{1}{\sqrt{n}} \Omega_{T_x} x \bm{1}_{n,1}^\T \\
\frac{1}{\sqrt{n}}  \bm{1}_{n,1} x^* \Omega_{T_x}^*  & I_n
\end{bmatrix}.
\]
The matrix $Q$ has eigenvalues $0, 1, 1, \dots, 1, 2$. The eigenvectors corresponding to $0$ and $2$ are $\mu = [(\Omega_{T_x} x)^\T, -\bm{1}_{n,1}^\T/\sqrt{n}]^\T/\sqrt{2}$ and $[(\Omega_{T_x} x)^\T, \bm{1}_{n,1}^\T/\sqrt{n}]^\top/\sqrt{2}$, respectively. Any vector orthogonal to these two vectors is an eigenvector of $Q$ corresponding to $1$. It follows that $ Q + \mu\mu^* - I_{Ns+n} $ is positive semidefinite.

Since $\mu$ is a null vector of $Q$, we have $P^{-1}\mu$ is a null vector of $P^*QP$ (note that $\Omega_{T_\eta}\eta = \sqrt{2}P^{-1}\mu$). Therefore, the smallest eigenvalue of the positive semidefinite matrix $P^*QP$ is $0$.

Next, we bound the largest eigenvalue of $P^*QP$, which satisfies
\begin{align}
\nonumber & \max_{\norm{z}_2\leq 1} \norm{P^*QPz}_2 \\
\nonumber & \leq \sqrt{1+\delta}\max_{\norm{Pz}_2\leq \sqrt{1+\delta}} \norm{QPz}_2 \\
\nonumber & = (1+\delta)\max_{\norm{z}_2\leq 1} \norm{Qz}_2 \\
\label{eq:max_amp} & \leq 2(1+\delta),
\end{align}
where the first inequality follows from Assumption \ref{ass:lambda}, and the second inequality follows from the largest eigenvalue of $Q$.

Next, we bound the second smallest eigenvalue of $P^*QP$, which satisfies
\begin{align}
\nonumber & \min_{z\perp P^{-1}\mu,~ \norm{z}_2\geq 1} \norm{P^*QP z}_2 \\
\nonumber & \geq \sqrt{1-\delta}\min_{Pz \perp (PP^*)^{-1}\mu,~ \norm{Pz}_2\geq \sqrt{1-\delta}} \norm{QPz}_2 \\
\nonumber & = (1-\delta)\min_{z\perp (PP^*)^{-1}\mu,~ \norm{z}_2\geq 1} \norm{Qz}_2 \\
\nonumber & \geq (1-\delta)\min_{z\perp (PP^*)^{-1}\mu,~ \norm{z}_2= 1} \norm{(I_{Ns+n} -\mu\mu^*)z}_2 \\
\nonumber & = (1-\delta)\min_{z\perp (PP^*)^{-1}\mu,~ \norm{z}_2= 1}\sqrt{1 - |\mu^*z|^2 } \\
\nonumber & = (1-\delta) \frac{|\mu^* (PP^*)^{-1} \mu|}{\norm{(PP^*)^{-1} \mu}_2} \\
\label{eq:min_amp} & \geq \frac{(1-\delta)^2}{1+\delta},
\end{align}
where the first and third inequalities follow from Assumption \ref{ass:lambda}, and the second inequality is due to the fact that $ Q + \mu\mu^* - I_{Ns+n} $ is positive semidefinite.

By \eqref{eq:max_amp} and \eqref{eq:min_amp}, all nonzero eigenvalues of $\bbE \Omega_{T_\eta} B_\rms \Omega_{T_\eta}^*$ reside in the interval $[\frac{(1-\delta)^2}{1+\delta},\, 2(1+\delta)]$. 
\end{IEEEproof}


\begin{IEEEproof}[Proof of Lemma \ref{lem:DstarD}]
We prove only the joint sparsity case. One can prove the subspace case by replacing $s$ with $m$ and getting rid of the union bound.

It is well-known that, for sufficiently large $n$, a Gaussian random matrix satisfies RIP \cite{Candes2005}. Here, we use a bound for real Gaussian random matrices \cite{Davidson2001}, and present its extension to complex Gaussian random matrices. Let $T\subset [m]$ denote an index set of cardinality $s$, i.e., $|T| = s < n$. Let $\widehat{A} \coloneqq [\re(A)\Omega_{T}^*, \im(A)\Omega_{T}^*]$.
By \cite[Theorem 2.13]{Davidson2001},
\[
\bbP\Bigl[ \norm{2\widehat{A}^*\widehat{A} - I_{2s}} \leq 3\Bigl(\sqrt{\frac{2s}{n}} + \varepsilon \Bigr) \Bigr] 
\geq 1 - 2 \exp\Bigl( - \frac{n \varepsilon^2}{2} \Bigr).
\]
Note also that
\begin{align*}
\Omega_{T} A^*A \Omega_{T}^* = &~ \Omega_{T}\re(A)^\T \re(A) \Omega_{T}^* \\
& + \sqrt{-1} \Omega_{T}\re(A)^\T \im(A) \Omega_{T}^* \\
& - \sqrt{-1}\Omega_{T}\im(A)^\T \re(A) \Omega_{T}^* \\
& + \Omega_{T}\im(A)^\T \im(A) \Omega_{T}^*.
\end{align*}
\begin{align*}
\norm{\Omega_{T} A^*A \Omega_{T}^* - I_s} \leq &~ \norm{\Omega_{T}\re(A)^\T \re(A) \Omega_{T}^* - I_s/2} \\
& + \norm{\Omega_{T}\re(A)^\T \im(A) \Omega_{T}^*} \\
& + \norm{\Omega_{T}\im(A)^\T \re(A) \Omega_{T}^*} \\
& + \norm{\Omega_{T}\im(A)^\T \im(A) \Omega_{T}^* - I_s/2} \\
\leq &~ 4\norm{\widehat{A}^* \widehat{A} - I_{2s}/2 }.
\end{align*}

It follows that
\begin{align*}
\bbP\Bigl[ \norm{\Omega_{T} A^*A \Omega_{T}^* - I_s} \leq 6\Bigl(\sqrt{\frac{s}{n}} + \varepsilon \Bigr) \Bigr] \\ 
\geq 1 - 2 \exp\Bigl( - \frac{n \varepsilon^2}{2} \Bigr).
\end{align*}

Therefore, there exist constants $C_1, c_1 >0$, such that
\begin{align*}
 & \bbP\Bigl[ \norm{\Omega_{T} A^*A \Omega_{T}^* - I_s} \leq C_1 \sqrt{\frac{s}{n}\log m},~\forall T~\text{s.t.}~|T| = s \Bigr] \\
& \geq 1 - 2 {m \choose s} \exp\Bigl( - \Bigl( \frac{C_1}{6}-1 \Bigr)^2 \frac{s}{2}\log m\Bigr) \\
& \geq 1 - m^{-c_1 s},
\end{align*}
where the first inequality follows from a union bound, and setting $\varepsilon = (\frac{C_1}{6}-1)\sqrt{\frac{s}{n}\log m}$; the second inequality follows from Stirling's approximation ${m \choose s}\leq \bigl(\frac{em}{s}\bigr)^s$. 

We obtain Lemma \ref{lem:DstarD} by applying the above bound to every diagonal block of the block diagonal matrix $\Omega_{T_x} D^*D \Omega_{T_x}^*$.
\end{IEEEproof} 


\begin{IEEEproof}[Proof of Lemma \ref{lem:EsstarEs}]
By a consequence of the Hanson-Wright inequality (see \cite[Theorem 2.1]{Rudelson2013}, and its complexification in \cite[Section 3.1]{Rudelson2013}), there exists an absolute constant $c_2'$ such that
\begin{align}
\bbP\Bigl[ \bigl|\sqrt{n}\norm{X^\T a_{k\cdot}}_2 - 1 \bigr| \leq \varepsilon \Bigr] \geq 1-2\exp\Bigl( -\frac{c_2'\varepsilon^2}{\norm{X}^2}\Bigr).
\label{eq:h-w}
\end{align}
Set $\varepsilon = C_2'\norm{X}\sqrt{\log n}$ for some $C_2'>0$, then by a union bound, there exists an absolute constant $c_2>0$ such that
\begin{align}
& \bbP\Bigl[ \bigl|\sqrt{n}\norm{X^\T a_{k\cdot}}_2 - 1 \bigr| \leq C_2'\norm{X}\sqrt{\log n},~ \forall k\in[n] \Bigr]  \nonumber\\
& \geq 1-n^{-c_2}. \label{eq:bdXa}
\end{align}
By Assumption \ref{ass:lambda},
\begin{align}
& \bbP\Bigl[ |\lambda_k|^2 \bigl | a_{k\cdot}^\T XX^* \overline{a_{k\cdot}} - \frac{1}{n} \bigr | \leq \frac{(2C_2'+C_2'^2)(1+\delta)}{n}  \nonumber\\
& \cdot \max\Bigl\{\norm{X}\sqrt{\log n}, \norm{X}^2\log n\Bigr\}, ~ \forall k\in[n] \Bigr]  \nonumber\\
& \geq 1-n^{-c_2}. \label{eq:H-Wineq}
\end{align}
The spectral norm $\norm{X}$ is bounded in Assumption \ref{ass:X}:
\begin{align*}
& \text{\textbf{Subspace case:}}~~\norm{X}^2 \leq (1+\theta)\max\{\frac{1}{N},\frac{1}{m}\},\\
& \text{\textbf{Joint sparsity case:}}~~\norm{X}^2 \leq (1+\theta)\max\{\frac{1}{N},\frac{1}{s_0}\}.
\end{align*}
Therefore, Lemma \ref{lem:EsstarEs} follows from \eqref{eq:EsstarEs}, \eqref{eq:EEsstarEs}, and \eqref{eq:H-Wineq}.
\end{IEEEproof}


\begin{IEEEproof}[Proof of Lemma \ref{lem:DstarEs}]
By \eqref{eq:DstarEs}, the columns of $D^*E_\rms$ are independent random vectors. Define
\[
\phi_k \coloneqq \begin{bmatrix}
\overline{a_{k\cdot}}a_{k\cdot}^\T x_{\cdot 1} \\
\overline{a_{k\cdot}}a_{k\cdot}^\T x_{\cdot 2} \\
\vdots \\
\overline{a_{k\cdot}}a_{k\cdot}^\T x_{\cdot N}
\end{bmatrix}.
\]
Then $D^*E_\rms = [\phi_1,\phi_2,\dots, \phi_n]\diag(\lambda)$. Next, we bound the spectral norm of the random matrix $\Phi-\bbE \Phi$, where $\Phi := [\phi_1,\phi_2,\dots, \phi_n]$, using matrix Bernstein inequality \cite[Theorem 1.6]{Tropp2011}. We need the following bounds to proceed:

\emph{1)} A bound on $\norm{\phi_k-\bbE \phi_k}_2$.

First, by \cite[Theorem 2.1 and Section 3.1]{Rudelson2013}, there exists a constant $c_3'$
\[
\bbP\Bigl[ \bigl|\sqrt{n}\norm{a_{k\cdot}}_2 - \sqrt{m}\bigr| \leq \varepsilon \Bigr] \geq 1-2\exp(-c_3'\varepsilon^2).
\]
By a union bound over all $k\in[n]$, there exists a constant $C_3'$ such that
\begin{align}
\nonumber & \bbP\Bigl[ \bigl|\sqrt{n}\norm{a_{k\cdot}}_2 - \sqrt{m}\bigr| \leq C_3'\sqrt{\log n},~  \forall k\in[n] \Bigr] \\
\nonumber & \geq 1-2 n\exp\bigl(-c_3'C_3'^2 \log n\bigr) \\
\label{eq:bdOa} & \geq 1-n^{-c_2}.
\end{align}
Note that
\[
\norm{\bbE \phi_k}_2 = \frac{1}{n} \norm{X}_\rmF = \frac{1}{n},
\]
\[
\norm{\phi_k}_2 \leq \norm{a_{k\cdot}}_2 \norm{X^\T a_{k\cdot}}_2.
\]
By \eqref{eq:bdXa} and \eqref{eq:bdOa}, there exists a constant $C_3''$, such that with probability at least $1-2n^{-c_2}$,
\begin{align*}
& \norm{\phi_k-\bbE \phi_k}_2 \\
\leq & \frac{C_3''}{n} \max\Big\{\sqrt{m},\sqrt{\log n}\Big\} \max\Big\{1,\sqrt{\frac{\log n}{N}},\sqrt{\frac{\log n}{m}}\Big\} \\
\leq & \frac{C_3''\sqrt{m}}{n} ,
\end{align*}
for all $k\in[n]$, where the second inequality uses the assumption that $\min\{N,m\}>\log n$.

\emph{2)} A bound on $\norm{\bbE [(\Phi-\bbE \Phi)^*(\Phi-\bbE \Phi)]}$.

One should observe that
\[
\bbE [(\phi_k-\bbE \phi_k)^*(\phi_k-\bbE \phi_k)] = \frac{m}{n^2},
\]
\[
\bbE [(\phi_k-\bbE \phi_k)^*(\phi_{k'}-\bbE \phi_{k'})] = 0,
\]
for $k\neq k'$. Therefore,
\[
\bbE [(\Phi-\bbE \Phi)^*(\Phi-\bbE \Phi)] = \frac{m}{n^2} I_n,
\]	
\[
\norm{\bbE [(\Phi-\bbE \Phi)^*(\Phi-\bbE \Phi)]} = \frac{m}{n^2}.
\]

\emph{3)} A bound on $\norm{\bbE [(\Phi-\bbE \Phi)(\Phi-\bbE \Phi)^*]}$.

Since $\{\phi_k\}_{k=1}^{n}$ are i.i.d. random vectors,
	\begin{align*}
	& \bbE [(\Phi-\bbE \Phi)(\Phi-\bbE \Phi)^*] \\
	= & \sum_{k=1}^n \bbE [(\phi_k-\bbE \phi_k)(\phi_k-\bbE \phi_k)^*] \\
	= & n \bbE [(\phi_1-\bbE \phi_1)(\phi_1-\bbE \phi_1)^*] \\
	= & n [\bbE (\phi_1\phi_1^*) - (\bbE \phi_1)(\bbE \phi_1)^*]\\
	= & \frac{1}{n} (X^\T \overline{X} \otimes I_m)
	\end{align*}
By Assumption \ref{ass:X}, in the subspace case,
\begin{align*}
\norm{\bbE [(\Phi-\bbE \Phi)(\Phi-\bbE \Phi)^*]} = \frac{1}{n}\norm{X^\T \overline{X}} \\
\leq \frac{1+\theta}{n} \max\{\frac{1}{N},\frac{1}{m}\}.
\end{align*}

Given the above bounds, we apply the matrix Bernstein inequality \cite[Theorem 1.6]{Tropp2011} as follows:
\begin{align*}
\bbP\Bigl[ \norm{\Phi-\bbE \Phi} \leq \varepsilon \Big| \norm{\phi_k-\bbE \phi_k}_2\leq R, ~ \forall k\in[n] \Bigr] \\
\geq 1 - (Nm+n)\exp\Bigl(-\frac{\varepsilon^2/2}{\sigma^2 + R\varepsilon/3} \Bigr),
\end{align*} 
where
\[
\sigma^2 = \max\Bigl\{ \frac{m}{n^2}, \frac{1+\theta}{nN}, \frac{1+\theta}{nm} \Bigr\},
\]
\[
R = \frac{C_3''\sqrt{m}}{n} .
\]
It follows that
\begin{align*}
& \bbP\Bigl[ \norm{\Phi-\bbE \Phi} \leq \varepsilon \Bigr] \\
& \geq 1 -  (Nm+n)\exp\Bigl(-\frac{\varepsilon^2/2}{\sigma^2 + R\varepsilon/3} \Bigr) - 2n^{-c_2},
\end{align*}
where the last term $2n^{-c_2}$ bounds the probability that $\norm{\phi_k-\bbE \phi_k}_2>R$ for some $k$.
Hence there exist constants $C_3, c_3 >0$ such that
\begin{align*}
& \bbP\Bigl[ \norm{\Phi-\bbE \Phi} \leq \frac{C_3}{\sqrt{1+\delta}} \max\Bigl\{ \sqrt{\frac{\log (Nm+n)}{nN}}, \\
&~~ \sqrt{\frac{\log(Nm+n)}{nm}},\frac{\sqrt{m}\log(Nm+n)}{n}\Bigr\}
\Bigr] \geq 1- n^{-c_3},
\end{align*}

Lemma \ref{lem:EsstarEs} follows from the above bound, and
\begin{align*}
\norm{\Omega_{T_x}D^*E_\rms-\bbE \Omega_{T_x}D^*E_\rms} = \norm{\Phi-\bbE \Phi}\norm{\diag(\lambda)} \\
\leq \sqrt{1+\delta} \norm{\Phi-\bbE \Phi}.
\end{align*}

\end{IEEEproof}


\begin{IEEEproof}[Proof of Lemma \ref{lem:DstarEs_alt}]
We introduce some notations for this proof. We use $B_p^n$ and $B_{S_p^{m,n}}$ to denote unit balls in $\bbC^n$ with $\ell_p$ norm, and in $\bbC^{m\times n}$ with Schatten $p$ norm, respectively. The projection on the support set $T$ is denoted by $\Pi_T$. For a set $\calA$ of matrices, $d_\mathrm{F}(\calA)$ and $d_\mathrm{op}(\calA)$ denote the radii of $\calA$ in the Frobenius norm and in the spectral norm, respectively. We use $\gamma_2(\calA, \norm{\cdot})$ the $\gamma_2$ functional of $\calA$, which is another way to quantify the size of $\calA$ \cite[Section 2.2]{Lee2015b}. These are key quantities in the upper bound of the supremum of an asymmetric second order process \cite[Theorem 2.3]{Lee2015b}, which we use to prove Lemma \ref{lem:DstarEs_alt}.

Note that
\begin{align}
& \max_{\begin{subarray}{c} T \subset [m] \\ |T| = s \end{subarray}}\norm{\Omega_{T_x}D^*E_\rms-\bbE \Omega_{T_x}D^*E_\rms}   \nonumber\\
& = \max_{\begin{subarray}{c} T \subset [m] \\ |T| = s \end{subarray}} 
\max_{\begin{subarray}{c} v \in B_2^{mN} \\ (I_N \otimes \Pi_T) v = v \end{subarray}} \max_{u \in B_2^n} |v^* \Phi u - \mathbb{E} v^* \Phi u|, \label{eq:alform1}
\end{align}
where $\Phi = D^*E_\rms$. Let ${z} = \sqrt{n} [a_{1\cdot}^*,\dots,a_{n\cdot}^*]^\T$. Then ${z}$ follows $\calCN(\bm{0}_{mn,1},I_{mn})$ and $v^* \Phi u$ is written as a quadratic form in ${z}$ as follows:
\begin{align}
& v^* \Phi u 
= \sum_{k=1}^n \sum_{j=1}^N u_k a_{k\cdot}^\T x_{\cdot j} v_{\cdot j}^* \overline{a_{k\cdot}}  \nonumber\\
& = {z}^* (\mathrm{diag}(u) \otimes \Pi_{T_0}) 
\Big(\frac{1}{n} I_n \otimes X V^*\Big) {z},  \label{eq:secondproc}
\end{align}
where $u = [u_1,\dots,u_n]^\T$, $v=[v_{\cdot 1}^\T,\dots,v_{\cdot N}^\T]^\T$, $V=[v_{\cdot 1},\dots, v_{\cdot N}]$, and $T_0 = \{i\in[m] | \norm{e_i^\T X}_2 > 0 \}$ denotes the row support of $X = [x_{\cdot 1},\dots,x_{\cdot N}]$. 

Let
\[
\mathcal{A} = \{ A_{u} | u \in B_2^n \},
\]
and
\[
\mathcal{B} = \{ B_{v} | v \in B_2^{mN}, ~ (I_N \otimes \Pi_T) v = v \},
\]
where $A_u$ and $B_v$ are left and right factors in the quadratic form in \eqref{eq:secondproc}, i.e.,
\[
A_{u} = \mathrm{diag}(u) \otimes \Pi_{T_0},
\]
and
\[
B_{v} = \frac{1}{n} I_n \otimes X V^*.
\]
Then \eqref{eq:alform1} is equivalent to
\[
\sup_{A_u \in \mathcal{A}} \sup_{B_v \in \mathcal{B}} 
|{z}^* A_u B_v {z} - \mathbb{E} {z}^* A_u B_v {z}|,
\]
which is a supremum of an asymmetric second order process. 
We use the result on suprema of asymmetric second order chaos processes by Lee and Junge \cite[Theorem 2.3]{Lee2015b}, which extends the original result by Krahmer et al. \cite{Krahmer2014} to asymmetric cases.

Next, we compute the key quantities, given as functions of $\mathcal{A}$ and $\mathcal{B}$, which we need to apply \cite[Theorem 2.3]{Lee2015b}. Let $A_u \in \mathcal{A}$. 
Since $|T_0| \leq s_0$, we have
\[
\norm{A_u}_\rmF = \sqrt{s_0} \norm{u}_2 \leq \sqrt{s_0}
\]
and the radius of $\mathcal{A}$ in the Frobenius norm satisfies
\[
d_\rmF(\mathcal{A}) \leq \sqrt{s_0}. 
\]
On the other hand, 
\[
\norm{A_u} = \norm{u}_\infty \leq 1, 
\]
which implies that the radius of $\mathcal{A}$ in the spectral norm satisfies
\[
d_{\mathrm{op}}(\mathcal{A}) \leq 1. 
\]
Moreover, for $A_u, A_u' \in \mathcal{A}$, we have
\[
\norm{A_{u} - A_{u'}} = \norm{u - u'}_\infty.
\]
Therefore, by the Dudley's inequality \cite{Ledoux2013}, 
\begin{align*}
\gamma_2(\mathcal{A},\norm{\cdot})
& \lesssim \int_0^\infty \sqrt{\log N(\mathcal{A},\norm{\cdot};t)} dt \\
& \leq \int_0^\infty \sqrt{\log N(B_2^n,\norm{\cdot}_\infty;t)} dt \\
& \lesssim \int_0^\infty \sqrt{\log N(B_1^n,\norm{\cdot}_2;t)} dt \\
& \lesssim \log^{3/2} n,
\end{align*}
where the third step follows from the entropy duality result by Artstein et al. \cite{Artstein2004} and the last step follows from Maurey's empirical method \cite{Carl1985} (also see \cite[Lemma 3.1]{Junge2017}).
Collecting the above estimates shows that the relevant quantities are given by
\begin{align*}
& \gamma_2(\mathcal{A},\norm{\cdot}) (d_{\mathrm{F}}(\mathcal{A})+\gamma_2(\mathcal{A},\norm{\cdot})) + d_{\mathrm{F}}(\mathcal{A}) d_{\mathrm{op}}(\mathcal{A}) \\
& \qquad \lesssim \max\{\sqrt{s_0} \log^{3/2} n,~ \log^3 n \}, \\
& d_{\mathrm{op}}(\mathcal{A})(\gamma_2(\mathcal{A},\norm{\cdot}) + d_{\mathrm{F}}(\mathcal{A})) \lesssim \max\{\sqrt{s_0},~ \log^{3/2} n \}, \\
& d_{\mathrm{op}}(\mathcal{A})^2 \leq 1.
\end{align*}

Next we consider the other set $\mathcal{B}$. Let $B_v \in \mathcal{B}$. Then
\begin{align*}
\norm{B_{v}}_{\mathrm{F}}
= \frac{1}{\sqrt{n}}\norm{X V^*}_{\mathrm{F}}
\leq \frac{1}{\sqrt{n}} \norm{X} \norm{V}_{\mathrm{F}}
= \frac{1}{\sqrt{n}} \norm{X}.
\end{align*}
Therefore
\[
d_{\mathrm{F}}(\mathcal{B}) \leq \frac{1}{\sqrt{n}} \norm{X}. 
\]
On the other hand,
\begin{align*}
\norm{B_{v}} = \frac{1}{n} \norm{X V^*} \leq \frac{1}{n} \norm{X} \norm{V},
\end{align*}
which implies
\[
d_{\mathrm{op}}(\mathcal{B}) \leq \frac{1}{n} \norm{X}. 
\]
Moreover, for $B_v, B_{v'} \in \mathcal{B}$, we have
\[
\norm{B_{v} - B_{v'}}
\leq \frac{1}{n} \norm{X} \norm{V - V'},
\]
where $V' = [v'_{\cdot 1},\dots,v'_{\cdot N}]$ and $v' = [v_{\cdot 1}'^\T,\dots,v_{\cdot N}'^\T]^\T$.
Therefore, 
\begin{align*}
& \gamma_2(\mathcal{B},\norm{\cdot}) \\
& \lesssim \frac{1}{n} \norm{X} \int_0^\infty \sqrt{\log N(\cup_{|T| = s} \Pi_T B_{S_2^{m,N}},\norm{\cdot}_{S_\infty^{m,N}};t)} dt \\
& \leq \frac{1}{n} \norm{X} \int_0^1 \sqrt{\log N(\cup_{|T| = s} \Pi_T B_{S_2^{m,N}},\norm{\cdot}_{S_\infty^{m,N}};t)} dt \\
& \leq \frac{1}{n} \norm{X} \int_0^1 \sqrt{\log \sum_{|T|=s} N(\Pi_T B_{S_2^{m,N}},\norm{\cdot}_{S_\infty^{m,N}};t)} dt \\
& \leq \frac{1}{n} \norm{X} \int_0^1 \sqrt{s \log m + \log N(B_{S_2^{s,N}},\norm{\cdot}_{S_\infty^{s,N}};t)} dt \\
& \leq \frac{1}{n} \norm{X} \Big( \sqrt{s \log m} \\
& \qquad + \int_0^1 \sqrt{\log N(B_{S_2^{s,N}},\norm{\cdot}_{S_\infty^{s,N}};t)} dt
\Big) \\
& \lesssim \frac{1}{n} \norm{X} \sqrt{s+N} \log(sN+m),
\end{align*}
where the last step follows from Lemma \ref{lemma:entint}.
Therefore, the parameters for $\mathcal{B}$ are estimated as
\begin{align*}
& \gamma_2(\mathcal{B},\norm{\cdot}) (d_{\mathrm{F}}(\mathcal{B})+\gamma_2(\mathcal{B},\norm{\cdot})) + d_{\mathrm{F}}(\mathcal{B}) d_{\mathrm{op}}(\mathcal{B}) \\
& \quad \lesssim \frac{1}{n^2} \norm{X}^2 ((s+N) \log^2(sN+m) \\
& \quad + \sqrt{s+N} \sqrt{n} \log(sN+m)), \\
& d_{\mathrm{op}}(\mathcal{B})(\gamma_2(\mathcal{B},\norm{\cdot}) + d_{\mathrm{F}}(\mathcal{B})) \\
& \quad \lesssim \frac{1}{n^2} \norm{X}^2 (\sqrt{s+N} \log(sN+m) + \sqrt{n}), \\
& d_{\mathrm{op}}(\mathcal{B})^2 \leq \frac{1}{n^2} \norm{X}^2.
\end{align*}

According to \cite[Theorem 2.3]{Lee2015b}, the optimal upper bound is obtained as the geometric mean of the dominant parameters for the two sets. More precisely, the suprema is (up to an absolute constant) no larger than
\begin{align*}
\frac{s_0^{1/4} (s+N)^{1/4} (\sqrt{n}+\sqrt{s+N})^{1/2}}{n} \\
\cdot \norm{X} \log^3 n \log(sN+m)
\end{align*}
with probability $1 - n^{-c_3}$. By Assumptions \ref{ass:lambda} and \ref{ass:X},
\[
|\lambda_k| \leq \sqrt{1+\delta} ,
\]
\[
\norm{X} \leq \max\Bigl\{\sqrt{\frac{1+\theta}{N}},~\sqrt{\frac{1+\theta}{s_0}} \Big\},
\]
which completes the proof.
\end{IEEEproof}

\begin{lemma}
\label{lemma:entint}
\[
\int_0^\infty \sqrt{\log N(B_{S_2^{m,N}}, tB_{S_\infty^{m,N}})} dt 
\lesssim \sqrt{m+N} \log(mN).
\]
\end{lemma}

\begin{IEEEproof}[Proof of Lemma \ref{lemma:entint}]
First, by the dual entropy result by Artstein et al. \cite{Artstein2004}, we have
\[
\log N(B_{S_2^{m,N}}, tB_{S_\infty^{m,N}}) \lesssim \log N(B_{S_1^{m,N}}, tB_{S_2^{m,N}}).
\]

Then we approximate the $S_1$ ball as a polytope using a trick proposed by Junge and Lee \cite{Junge2017}. 
Let $R$ be the set of all rank-1 matrices in the unit sphere of $S_2^{m,N}$.
Then $B_{S_1^{m,N}}$ is the absolute convex hull of $R$.
We construct an $\epsilon$-net $\Delta_m$ of the sphere $S^{m-1}$. Then
\[
|\Delta_m| \leq \Big(1+\frac{2}{\epsilon}\Big)^m.
\]
For an arbitrary $f \in S^{m-1}$, we have a sequence $\{f_l\}_{l=1}^\infty \subset \Delta_m$ such that
\[
f = \sum_{l=1}^\infty \alpha_l f_l,
\]
and
\[
\sum_{l=1}^\infty |\alpha_l| \leq \frac{1}{1-\epsilon}.
\]
The existence of such a sequence follows from the optimality of the construction of the net.
Similarly we construct an $\epsilon$-net $\Delta_N \subset S^{N-1}$ of $S^{N-1}$. Then
\[
|\Delta_N| \leq \Big(1+\frac{2}{\epsilon}\Big)^N.
\]
For an arbitrary $g \in S^{N-1}$, we have a sequence $\{g_k\}_{k=1}^\infty \subset \Delta_N$ such that
\[
g = \sum_{k=1}^\infty \beta_k g_k
\]
and
\[
\sum_{k=1}^\infty |\beta_k| \leq \frac{1}{1-\epsilon}.
\]
Therefore,
\[
fg^* = \sum_{l,k=1}^\infty \alpha_l \beta_k f_l g_k^*
\]
and
\[
\sum_{l,k=1}^\infty |\alpha_l| |\beta_k|
\leq \Big(\frac{1}{1-\epsilon}\Big)^2.
\]
We can choose $\epsilon$ so that
\[
\Big(\frac{1}{1-\epsilon}\Big)^2 \leq 2
\]
and
\[
1+\frac{2}{\epsilon} \leq 8.
\]
Let $\Delta_{m,N} = \Delta_m \times \Delta_N$. Then
\[
\log(|\Delta_{m,N}|) \leq (m+N) \log 8
\]
and
\[
B_{S_1^{m,N}} \subset 2 \mathrm{absconv}(\Delta_{m,N}).
\]

Now, it suffices to compute
\[
\int_0^\infty \sqrt{\log N(2 \mathrm{absconv}(\Delta_{m,N}), tB_{S_2^{m,N}})} dt.
\]
Then use a change of variable and get
\begin{align*}
& \int_0^\infty \sqrt{\log N(2 \mathrm{absconv}(\Delta_{m,N}), tB_{S_2^{m,N}})} dt \\
& = 2 \int_0^\infty \sqrt{\log N(\mathrm{absconv}(\Delta_{m,N}), tB_{S_2^{m,N}})} dt.
\end{align*}
Let $\Delta_{m,N} = \{q_1,\dots,q_M\}$, where $M = |\Delta_{m,N}|$.
Define linear mapping $Q: \ell_1^M \to \ell_2^{mN}$ by $Q(e_i) = \mathrm{vec}(q_i)$ for $i=1,\dots,M$.
Since $\norm{\mathrm{vec}(q_i)}_2 = \norm{q_i}_{S_2} = 1$ for all $i$, we have
\[
\norm{Q: \ell_1^M \to \ell_2^{mN}} = 1.
\]
Note
\begin{align*}
& \int_0^\infty \sqrt{\log N(\mathrm{absconv}(\Delta_{m,N}), tB_{S_2^{m,N}})} dt \\
& = \int_0^\infty \sqrt{\log N(Q(B_1^M), tB_{\ell_2^{mN}})} dt
\end{align*}
By a version of Maurey's empirical method (see for example \cite[Proposition 3.2]{Junge2017}), we have
\begin{align*}
\int_0^\infty \sqrt{\log N(Q(B_1^M), tB_{\ell_2^{mN}})} dt
\lesssim \sqrt{\log M} \log (mN) \\
\lesssim \sqrt{m+N} \log (mN).
\end{align*}
This completes the proof.
\end{IEEEproof}


\begin{IEEEproof}[Proof of Lemma \ref{lem:DstarEn}]
Bear in mind that the columns of $\Psi\coloneqq D^* E_\rmn$, which we denote by $\{\psi_k\}_{k=1}^n$, are independent random vectors with zero mean: 
\[
\psi_k \coloneqq \begin{bmatrix}
\overline{a_{k\cdot}} w_{k1}\\
\overline{a_{k\cdot}} w_{k2}\\
\vdots\\
\overline{a_{k\cdot}} w_{kN}
\end{bmatrix}.
\]
We bound $\norm{D^* E_\rmn}$ using the matrix Bernstein inequality \cite[Theorem 1.6]{Tropp2011}. We need the following bounds:

\emph{1)} A bound on $\norm{\psi_k}_2$.

Since
\[
\norm{\psi_k}_2 \leq \norm{a_{k\cdot}}_2 \norm{w_{k\cdot}}_2
\]
By \eqref{eq:bdOa}, and $m>\log n$,
\[
\norm{\psi_k}_2 \leq (C_3'+1) \sqrt{\frac{m}{n}} \times \sqrt{N}\max_{k\in[n], j\in[N]}|w_{kj}|,
\]
with probability at least $1-n^{-c_2}$.

\emph{2)} A bound on $\norm{\bbE \Psi^*\Psi}$.

Since
\[
\bbE \Psi^*\Psi = \frac{m}{n} \diag([\norm{w_{1\cdot}}_2^2,\norm{w_{2\cdot}}_2^2,\dots,\norm{w_{k\cdot}}_2^2]),
\]
we have
\[
\norm{\bbE \Psi^*\Psi} = \frac{m}{n} \max_{k\in[n]}\norm{w_{k\cdot}}_2^2 \leq \frac{mN}{n} \max_{k\in[n], j\in[N]}|w_{kj}|^2.
\]

\emph{3)} A bound on $\norm{\bbE \Psi\Psi^*}$.

Since
\[
\bbE \Psi\Psi^* = \sum_{k\in[n]} \frac{1}{n} \diag([|w_{k1}|^2,|w_{k2}|^2,\dots,|w_{kN}|^2])\otimes I_m,
\]
we have
\[
\norm{\bbE \Psi\Psi^*} = \frac{1}{n} \max_{j\in[N]} \sum_{k\in[n]}|w_{kj}|^2 \leq \max_{k\in[n], j\in[N]}|w_{kj}|^2.
\]

Given the above bounds, we completes the proof using the matrix Bernstein inequality (similar to the proof of Lemma \ref{lem:DstarEs}). There exist constants $C_4, c_4 >0$ such that
\begin{align*}
& \norm{D^* E_\rmn} = \norm{\Psi} \leq C_4 \max\Bigl\{\sqrt{\log (Nm+n)}, \\
& \qquad\qquad \sqrt{\frac{Nm}{n}} \log (Nm+n) \Bigr\} \max_{k\in[n], j\in[N]}|w_{kj}|,
\end{align*}
with probability at least $1-n^{-c_4}$.
\end{IEEEproof}


\begin{IEEEproof}[Proof of Lemma \ref{lem:DstarEn_alt}]
Note that
\begin{align*}
\max_{\begin{subarray}{c} T \subset [m] \\ |T| = s \end{subarray}}\norm{\Omega_{T_x}D^*E_\rmn}   = \max_{\begin{subarray}{c} T \subset [m] \\ |T| = s \end{subarray}} 
\max_{\begin{subarray}{c} v \in B_2^{mN} \\ (I_N \otimes \Pi_T) v = v \end{subarray}} \max_{u \in B_2^n} |v^* \Psi u|,
\end{align*}
where
\begin{align*}
& \Psi = D^* E_\rmn \\
& = \begin{bmatrix}
I_N \otimes \overline{a_{1\cdot}} & \dots & I_N \otimes \overline{a_{n\cdot}}
\end{bmatrix}
\begin{bmatrix}
w_{1\cdot} & & \\
& \ddots & \\
& & w_{n\cdot}
\end{bmatrix}.
\end{align*}
Let ${z} = \sqrt{n} [a_{1\cdot}^*,\dots,a_{n\cdot}^*]^\T$. Then ${z}$ is a standard Gaussian vector, and
\[
v^* \Psi u
= \frac{1}{\sqrt{n}} (\mathbf{1}_{1,n} \otimes v^*) (E_\rmn \otimes I_m) (\mathrm{diag}(u) \otimes I_m) {z}.
\]
Let
\[
q_{u,v} \coloneqq \frac{1}{\sqrt{n}} (\mathrm{diag}(u)^* \otimes I_m) (E_\rmn^* \otimes I_m) (\mathbf{1}_{n,1} \otimes v).
\]
The $L_2$ metric is given by
\begin{align*}
& d((u,v),(u',v')) \\
& = \sqrt{\mathbb{E} (q_{u,v}^* {z} - q_{u',v'}^* {z})^2} \\
& = \norm{q_{u,v} - q_{u',v'}}_2.
\end{align*}
Indeed,
\begin{align*}
& d((u,v),(u',v')) \\
& \leq d((u,v),(u,v')) + d((u,v'),(u',v')) \\
& \leq \norm{\mathrm{diag}(u-u')}_\infty \norm{E_{\mathrm{n}}} \norm{v}_2 \\
& \quad + \norm{\mathrm{diag}(u')}_\infty \norm{E_{\mathrm{n}}} \norm{v-v'}_2 \\
& \leq \norm{\mathrm{diag}(u-u')}_\infty \norm{E_{\mathrm{n}}} \\
& \quad + \norm{E_{\mathrm{n}}} \norm{v-v'}_2.
\end{align*}

Let $\Gamma_s = \{v\in B_2^{mN}:~T \subset [m],~|T| = s,~(I_N \otimes \Pi_T) v = v \}$.
By Dudley's theorem (see e.g., \cite[Theorem 11.17]{Ledoux2013}), we have
\begin{align*}
& \bbE \sup_{\begin{subarray}{c} T \subset [m] \\ |T| = s \end{subarray}} \sup_{\begin{subarray}{c} v \in B_2^{mN} \\ (I_N \otimes \Pi_T) v = v \end{subarray}} \sup_{u \in B_2^n} v^* \Psi u \\
& \leq 24 \int_0^\infty \sqrt{\log N(\Gamma_s \times B_2^n, d(\cdot); \epsilon) d\epsilon} \\
& \leq 24 \norm{E_\rmn} \Bigl( \int_0^\infty \sqrt{\log N(\Gamma_s, \norm{\cdot}_2; \epsilon) d\epsilon} \\
& \quad + \int_0^\infty \sqrt{\log N(B_2^n, \norm{\cdot}_\infty; \epsilon) d\epsilon} \Bigr) \\
& \leq 24 \norm{E_\rmn} \Bigl( \int_0^\infty \sqrt{\log N(\Gamma_s, \norm{\cdot}_2; \epsilon) d\epsilon} \\
& \quad + \int_0^\infty \sqrt{\log N(B_1^n, \norm{\cdot}_2; \epsilon) d\epsilon} \Bigr) \\
& \lesssim \norm{E_\rmn} (\sqrt{s \log m} + \sqrt{Ns} + \log^{3/2} n)
\end{align*}
By an extension of Dudley's inequality to moments \cite[Section 8.9, Page 263]{Foucart2013},
\begin{align*}
& \Bigl(\bbE \sup_{\begin{subarray}{c} T \subset [m] \\ |T| = s \end{subarray}} \sup_{\begin{subarray}{c} v \in B_2^{mN} \\ (I_N \otimes \Pi_T) v = v \end{subarray}} \sup_{u \in B_2^n} | v^* \Psi u |^p \Bigr)^{1/p}\\
& \lesssim \norm{E_\rmn} (\sqrt{s \log m} + \sqrt{Ns} + \log^{3/2} n) \sqrt{p}
\end{align*}
By a variation of Markov's inequality \cite[Proposition 7.11]{Foucart2013}, there exist absolute constants $C_4, c_4 > 0$ such that
\begin{align*}
& \sup_{\begin{subarray}{c} T \subset [m] \\ |T| = s \end{subarray}} \sup_{\begin{subarray}{c} v \in B_2^{mN} \\ (I_N \otimes \Pi_T) v = v \end{subarray}} \sup_{u \in B_2^n} | v^* \Psi u | \\
& \leq C_4 \norm{E_\rmn} (\sqrt{s \log m} + \sqrt{Ns} + \log^{3/2} n) \sqrt{\log n}, 
\end{align*}
with probability at least $1 - n^{-c_4}$.

Therefore, Lemma \ref{lem:DstarEn_alt} follows from: 
\[
\norm{E_\rmn} = \max_{k \in [n]} \norm{w_{k\cdot}}_2 \leq \sqrt{N}\max_{k\in[n],j\in[N]}|w_{kj}|.
\]

\end{IEEEproof}


\begin{IEEEproof}[Proof of Lemma \ref{lem:EsstarEn}]
If assumptions \ref{ass:A} -- \ref{ass:X} are satisfied, then by \eqref{eq:bdXa}, 
\[
\norm{y_{k\cdot}}_2 \leq \frac{(C_2' + 1)\sqrt{1+\delta}}{\sqrt{n}} \max\Bigl\{1, \norm{X}\sqrt{\log n} \Bigr\}
\]
for all $k\in[n]$, with probability at least $1-n^{-c_2}$.

Since
\[
E_\rms^*E_\rmn = \diag([y_{1\cdot}^*w_{1\cdot},y_{2\cdot}^*w_{2\cdot},\dots,y_{n\cdot}^*w_{n\cdot}]),
\]
there exist constants $C_5 = (C_2' + 1)\sqrt{(1+\delta)(1+\theta)} > 0$ such that
\begin{align*}
& \norm{ E_\rms^*E_\rmn } \\
\leq & \max_k\norm{y_{k\cdot}}_2 \times \sqrt{N}\max_{k\in[n], j\in[N]}|w_{kj}| \\
\leq & \frac{C_5}{\sqrt{1+\theta}}\sqrt{\frac{N}{n}} \max\Bigl\{1, \norm{X}\sqrt{\log n} \Bigr\} \max_{k\in[n], j\in[N]}|w_{kj}|,
\end{align*}
with probability at least $1-n^{-c_2}$. Therefore, Lemma \ref{lem:EsstarEn} follows from Assumption \ref{ass:X}.
\end{IEEEproof}


\begin{IEEEproof}[Proof of Lemma \ref{lem:EnstarEn}]
Lemma \ref{lem:EnstarEn} follows from
\[
E_\rmn^*E_\rmn = \diag([\norm{w_{1\cdot}}_2^2,\norm{w_{2\cdot}}_2^2,\dots, \norm{w_{n\cdot}}_2^2]).
\]
\end{IEEEproof}


\begin{IEEEproof}[Proof of Lemma \ref{lem:ineq_square}]
We prove these inequalities using the Hoeffding's inequality.

For all $j\in[N]$, $\ell\in[m]$, and $k\in[n]$,
\begin{align*}
& \left| |\overline{a_{k\ell}}a_{k\cdot}^\T x_{\cdot j}|^2 - \bbE |\overline{a_{k\ell}}a_{k\cdot}^\T x_{\cdot j}|^2 \right| \\
& \leq |a_{k\ell}|^2 |a_{k\cdot}^\T x_{\cdot j}|^2 + \frac{1}{n^2} (\norm{x_{\cdot j}}_2^2 + |x_{\ell j}|^2) \\
& \leq C_6' \frac{\log (nm)}{n} \cdot \frac{\norm{x_{\cdot j}}_2^2\log (nN)}{n} + \frac{2\norm{x_{\cdot j}}_2^2}{n^2} \\
& \leq \frac{(C_6'+2) \norm{x_{\cdot j}}_2^2 \log^2 (nmN)}{n^2} ,
\end{align*}
where the third line is true with probability at least $1-n^{-c_6'}$ for some absolute constant $c_6'$. We show this by applying a Chernoff bound and a union bound to $|a_{k\ell}|^2$, and applying the Hanson-Wright inequality \eqref{eq:h-w} and a union bound to $|a_{k\cdot}^\T x_{\cdot j}|^2$. 
Then it follows from the Hoeffding's inequality and a union bound, that there exist absolute constants $C_6, c_6>0$ such that for all $j\in[N]$ and $\ell\in[m]$ we have \eqref{eq:ineq1}.

Similarly, for all $j\in[N]$, $\ell\in[m]$, and $k\in[n]$,
\begin{align*}
|a_{k\ell}|^2 |a_{k\cdot}^\T\, x_{\cdot j}| \leq C_6' \frac{\log (nm)}{n} \cdot \frac{\norm{x_{\cdot j}}_2\sqrt{\log (nN)} }{\sqrt{n}},
\end{align*}
with probability at least $1-n^{-c_6'}$. By the Hoeffding's inequality and a union bound, we have \eqref{eq:ineq2}. Here we use the following facts: By Assumption \ref{ass:X}, $\norm{x_{\cdot j}}\geq \sqrt{\frac{1-\theta}{N}}$. By Assumption \ref{ass:W}, $\max_{k\in[n],j\in[N]} |w_{kj}| \leq \frac{C_W}{\sqrt{nN}}$.

For $\ell\in[m]$ and $k\in[n]$,
\begin{align*}
|a_{k\ell}|^2 - \bbE |a_{k\ell}|^2 \leq C_6' \frac{\log (nm)}{n},
\end{align*}
with probability at least $1-n^{-c_6'}$. By the Hoeffding's inequality and a union bound, we have \eqref{eq:ineq3}.
\end{IEEEproof}


\end{document}